\documentclass[prd,aps,twocolumn,nofootinbib,floatfix,longbibliography, superscriptaddress]{revtex4-1}
\usepackage{graphicx}
\usepackage{bm}
\usepackage{mathtools}
\usepackage{hyperref}
\usepackage{amsmath}
\usepackage{mathrsfs}
\usepackage{color,xcolor}

\usepackage{hyperref}
\hypersetup{
 pdftitle={Constraining Inputs to Realistic Kilonova Simulations}, %
 pdfauthor={Ristic}, %
 colorlinks=true,
 citecolor={blue},
}

\newcommand\editedthree[1]{{\color{black}#1}}
\newcommand\editedToo[1]{{\color{black}#1}}
\newcommand\edited[1]{{#1}}

\usepackage{soul}

\newcommand\rurl[1]{%
  \href{https://#1}{\nolinkurl{#1}}%
}

\newcommand{\apjl}{Astrophysical Journal Letters}
\newcommand{\apjs}{Astroph. J. S.}
\newcommand{\pasp}{PASP}

\newcommand{\mnras}{Monthly Notices of the Royal Astronomical Society} 
\newcommand{\aap}{Astronomy and Astrophysics}

\newcommand{\araa}{Annual Reviews of Astronomy and Astrophysics}

\def\RIT{Center for Computational Relativity and Gravitation, Rochester Institute of Technology, Rochester, New York
 14623, USA}
\def\XCP{Computational Physics Division, Los Alamos National Laboratory, Los Alamos, NM, 87545, USA}
\def\CTA{Center for Theoretical Astrophysics, Los Alamos National Laboratory, Los Alamos, NM, 87545, USA}
\def\CCSS{Computer, Computational, and Statistical Sciences Division, Los Alamos National Laboratory, Los Alamos, NM,
 87545, USA}
\def\TD{Theoretical Division, Los Alamos National Laboratory, Los Alamos, NM 87545, USA}
\def\JINA{Joint Institute for Nuclear Astrophysics---Center for the Evolution of the Elements, USA}
\def\UA{The University of Arizona, Tucson, AZ 85721, USA}
\def\NM{Department of Physics and Astronomy, The University of New Mexico, Albuquerque, NM 87131, USA}

\def\UR{Department of Physics and Astronomy, University of Rochester, Rochester, NY 14627, USA}
\def\CO{Observatories of the Carnegie Institution for Science, Pasadena, CA 91101, USA}

\graphicspath{{figures/}}

\begin{document}

\title{Constraining inputs to realistic kilonova simulations through comparison to observed \emph{r}-process abundances}

\author{M.\ Ristic}
\affiliation{\RIT}
\author{E.\ M.\ Holmbeck}
\affiliation{\CO}
\affiliation{Hubble Fellow}
\affiliation{\JINA}
\author{R.\ Wollaeger}
\affiliation{\CTA}
\affiliation{\CCSS}
\author{O.\ Korobkin}
\affiliation{\CTA}
\affiliation{\CCSS}
\author{E.\ Champion}
\affiliation{\UR}
\author{R.\ O'Shaughnessy}
\affiliation{\RIT}
\author{C.~L. Fryer}
\affiliation{\CTA}
\affiliation{\CCSS}
\affiliation{\UA}
\affiliation{\NM}
\author{C.~J. Fontes}
\affiliation{\CTA}
\affiliation{\XCP}
\author{M.~R. Mumpower}
\affiliation{\JINA}
\affiliation{\CTA}
\affiliation{\TD}
\author{T.~M. Sprouse}
\affiliation{\JINA}
\affiliation{\CTA}
\affiliation{\TD}

\date{\today}

\begin{abstract}
Kilonovae, one source of electromagnetic emission associated with neutron star mergers, are powered by the decay of
radioactive isotopes in the neutron-rich merger ejecta. Models for kilonova emission consistent with the electromagnetic counterpart to GW170817 predict characteristic abundance patterns, determined by the
relative balance of different types of material in the outflow. Assuming the observed source is prototypical, this
inferred abundance pattern in turn must match \emph{r}-process abundances deduced by other means, such as what is observed in the solar system. 
We report on 
analysis comparing the input mass-weighted elemental compositions adopted in our radiative
transfer simulations to the mass fractions of elements in the Sun, \editedToo{as a practical prototype for the
  potentially universal abundance signature from neutron-star mergers.} We characterise the extent to which our parameter inference results depend on our assumed composition for the
dynamical and wind ejecta and examine how the new results compare to previous work. We find that \edited{a dynamical ejecta composition calculated using the FRDM2012 nuclear mass and FRLDM fission models with extremely neutron-rich ejecta ($Y_{\rm{e}} = 0.035$) along with moderately neutron-rich ($Y_{\rm e} = 0.27$) wind ejecta composition yields a  wind-to-dynamical mass ratio of \editedthree{$M_{\rm{w}}/M_{\rm{d}}$ = 0.47}} \edited{which best matches} the observed AT2017gfo kilonova light curves while also producing the best-matching abundance of neutron-capture elements in the solar system.
\end{abstract}

\maketitle

\section{Introduction}
\label{sec:intro}

For nearly half a century, neutron star binaries have been known to exist in nature, stemming from the first detection
of a binary pulsar system \cite{hulsetaylor}. Shortly thereafter, the general relativistic prediction of gravitational
radiation from a compact object binary was measured in the same system, implying the possibility of neutron star binary
coalescence \cite{1982ApJ...253..908T}. Recently, neutron star mergers were confirmed as astrophysical sources of both
gravitational wave and electromagnetic emission with the detection of the binary neutron star merger GW170817 and its transient electromagnetic counterpart AT2017gfo
\cite{LIGO-GW170817-bns, LIGO-GW170817-kilonova, LIGO-GW170817-mma, LIGO-GW170817-astro, Tanvir_2017}.

Around the same time as the first pulsar binary detection, compact object mergers involving neutron stars, either binary neutron star (BNS) or black-hole-neutron-star (BHNS), were predicted to be candidates for rapid neutron capture (\emph{r}-process) nucleosynthesis \cite{1974ApJ...192L.145L, 1976ApJ...210..549L, 1982ApL....22..143S, 1989Natur.340..126E}. 
The nuclei synthesized in the \edited{immediate aftermath of the} post-merger ejecta were thought to \edited{be} heavy ($A > 140$), \edited{with a sizeable fraction of} radioactive isotopes having short lifetimes due to their instability \cite{2015IJMPD..2430012R}.
As these nuclei decay, they release energy into the surrounding matter which would be emitted as ultraviolet, optical, and infrared thermal radiation once the ejecta becomes optically thin \cite{LiLX1998, Kulkarni2005, short-grb-Metzger-EMCounterpartViaRProcess-2010}. This thermal emission is now commonly referred to as a kilonova \cite{Metzger_LRR_2019} and serves as the bridge between the \emph{r}-process elements produced by neutron star mergers and their resultant electromagnetic emission \cite{Roberts2011, Goriely2011, short-grb-GWCoincidenceEM-MetzgerBerger2011, Korobkin_2012, 2021RvMP...93a5002C}.
\edited{Aside from the transient electromagnetic kilonova emission (including a gamma-ray burst \cite{2017ApJ...848L..13A, 2017ApJ...848L..15S}), \emph{r}-process material ejecta from neutron star binary mergers like GW170817 could produce another observable signature: relic \emph{r}-process abundances such as in ancient, metal-poor stars and in our solar system.}

\edited{Modeling kilonova light curves from merger events as viewed from the solar system by a distant observer requires the ejecta mass, velocity, composition, morphology, and viewing angle to be known, or otherwise supplied as model inputs. 
It has been conclusively demonstrated that the multiband light curves of AT2017gfo are poorly fit with single-component models, i.e., models consisting of a single type of ejecta described by fixed velocity, mass, and composition \cite{Cowperthwaite2017}.
Instead, the AT2017gfo light curve is better fit by two (or even three) components describing multiple types of ejecta: generally, a high-opacity ``dynamical" component, and a low-opacity ``wind" component \cite{Tanvir_2017, Troja_2017, Cowperthwaite2017, 2017ApJ...851L..21V}.}

\edited{In two-component models of kilonovae, the low-opacity wind component typically includes elements only up to the ``second \emph{r}-process peak" at $A\sim 130$, while the higher-opacity dynamical component includes even heavier elements \cite{2017Natur.551...67P, 2017ApJ...848L..18N, 2019PhRvD.100b3008M}.
While these composition trends set a general opacity scale (see e.g., \cite{2020MNRAS.496.1369T}), the full details of the composition effects on electromagnetic emission depend on the components' nuclear physics considerations as well as their physical parameterizations described, in part, by the components' masses $M_{\rm{d}}, M_{\rm{w}}$ and velocities $v_{\rm{d}}, v_{\rm{w}}$, where ``d" and ``w" refer to the dynamical and wind components, respectively.}
Previous studies of kilonovae have highlighted the importance of nuclear physics inputs on \emph{r}-process nucleosynthesis and the resultant effect on observed kilonova emission \cite{2021ApJ...918...44B, 2021ApJ...906...94Z}. \edited{In this work, we build on previous studies by considering the effect that nuclear physics uncertainties have on parameter inference from kilonova light curves.}

This \edited{work} presents two-component nucleosynthetic yield constraints assuming \emph{r}-process contribution exclusively from neutron star mergers and electromagnetic constraints assuming all neutron star mergers \edited{are phenomenologically similar to GW170817}. We investigate the effects of comparing elemental abundances from kilonova simulations to solar \emph{r}-process abundances under the assumption that the second ($A\sim130$) and third ($A\sim 195$) \emph{r}-process peaks follow universal behavior, \editedToo{which is justified by the robustness of the \emph{r}-process pattern observed among metal-poor stars (e.g. \cite{1994ApJ...431L..27S, 2002A&A...387..560H, 2003ApJ...591..936S, 2007ApJ...660L.117F, 2013A&A...550A.122S, PhysRevLett.114.192501})}. We use this comparison to create a parameter estimation prior driven by explicit consideration of \emph{r}-process elemental abundances in kilonova ejecta to gauge the effects on recovered ejecta properties\edited{; i.e., the masses and velocities of the ejecta components}. As kilonova models improve in complexity and more observations become available for parameter estimation purposes, we can use more representative simulation abundances to hone this prior in future studies.

In this work, we will assess the extent to which our assumptions about the ejected material are simultaneously
consistent with both types of aforementioned observations. Specifically, we will examine whether the abundances produced by our \edited{nucleosynthesis} simulations realistically match the \emph{r}-process abundances observed in the Sun \edited{while simultaneously reproducing the AT2017gfo light curve as well}.
In Section \ref{sec:sim_setup}, we discuss the \edited{radiative transfer,} atomic, and nuclear physics codes used to calculate the \edited{surrogate light curves,} line-binned opacities, and ejecta compositions, respectively, considered in this work. In Section \ref{sec:rprocess_prior}, we describe our method of comparing mass-weighted \emph{r}-process abundances from our simulations with the solar abundance pattern. Section \ref{sec:pe} describes our parameter estimation framework and the effects of the \emph{r}-process prior introduced in this work. In Section \ref{sec:discussion} we discuss whether the inclusion of the \emph{r}-process prior makes a substantial difference in the parameter estimation process compared to prior work.

\editedToo{Our proof-of-concept analysis provides two key new approaches to multimessenger inference of BNS mergers.  On the one hand, we provide a method to   quantitatively  assess the hypothesis of a universal \emph{r}-process origin in BNS mergers with observations of kilonovae, by requiring consistent predictions for  the ejecta's electromagnetic
  emission and its asymptotic impact on  \emph{r}-process abundances.  On the other hand, if the universal origin of \emph{r}-process abundances is from binary neutron star mergers, our method can sharply refine our inferences about the ejected material.
While in our proof-of-concept calculation we presently employ the Sun's \emph{r}-process abundances as a prototype for
a pristine, universal abundance signature from BNS mergers, our method should ideally be applied to ongoing and future
efforts to disentangle the BNS merger's natal abundance contribution, for example from isolated metal-poor stars or
from abundance principal-component correlations. 
}

\section{Methods}
\label{sec:methods}

\subsection{Simulation Setup}
\label{sec:sim_setup}

\begin{figure}
\includegraphics[width=\columnwidth]{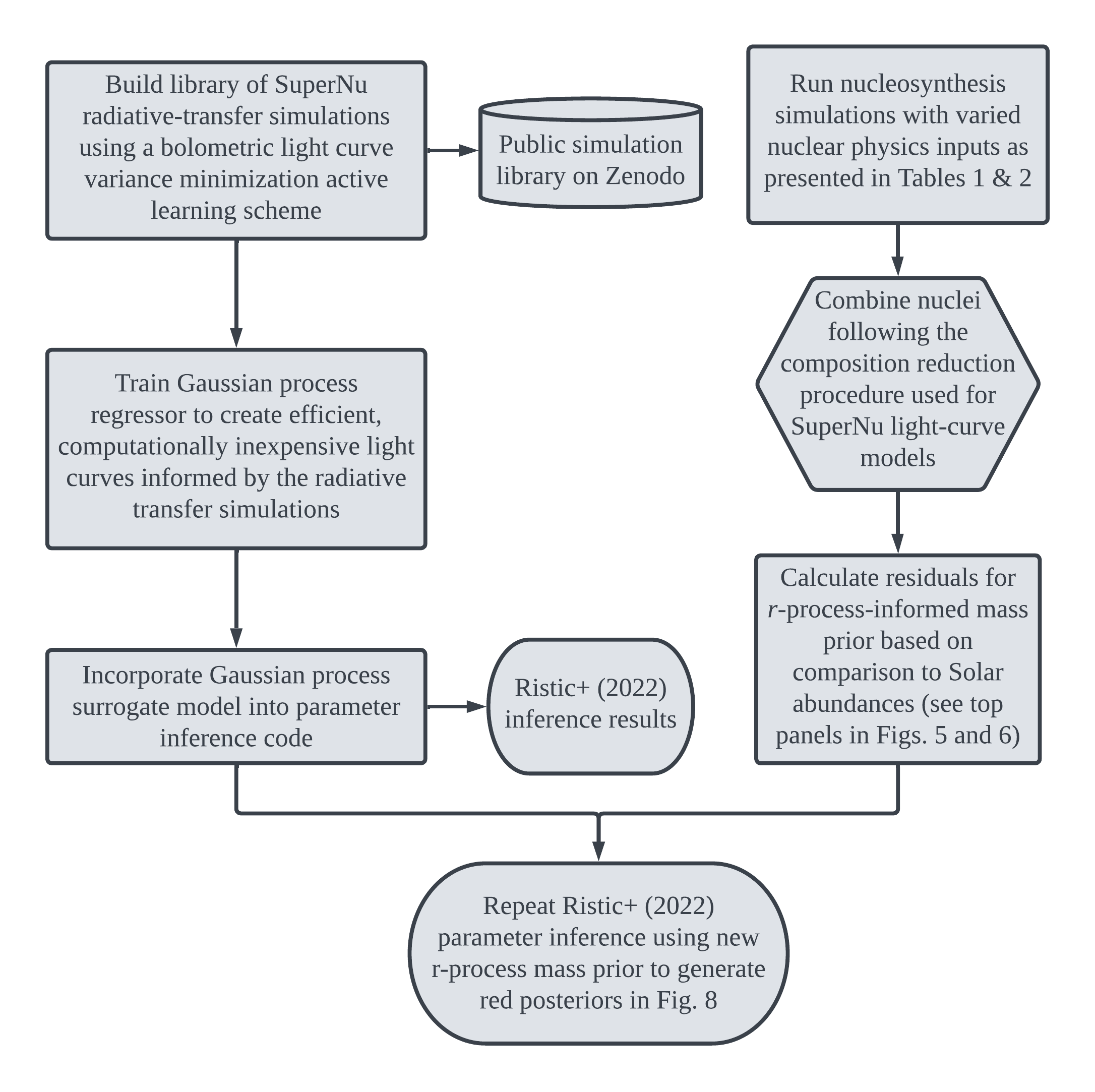}
\caption{\edited{Unified Modeling Language (UML) diagram describing the large-scale steps taken in creating the \emph{r}-process mass prior and using it during parameter inference to generate the red posteriors in Figure \ref{fig:corner_plot}. Per standard UML definition, rectangles represent processes, cylinders represent databases, hexagons represent data preparation steps and ovals represent terminators, or final products.}
}
\label{fig:flowchart}
\end{figure}

\edited{

The aforementioned model abundances are not sufficient to create a direct link to kilonova electromagnetic emission on their own. However, they restrict which radioactive isotopes can plausibly exist and determine the radioactive heating rates powering the kilonova at different times. In this section, we describe the details of our ejecta compositions, the relevant thermalization efficiencies, and the composition-dependent ejecta opacity effects which constitute our kilonova emission model. \editedToo{Figure \ref{fig:flowchart} schematically shows the subsequent process for using these models to perform kilonova parameter inference.}

Throughout this work, we assume a two-component kilonova model composed of a lanthanide-rich ``dynamical" ejecta component and a lanthanide-poor ``wind" ejecta component. Each of our two ejecta components, dynamical and wind, is described by a fixed morphology and elemental composition. The morphologies are fixed to torus-shaped and peanut-shaped for the dynamical and wind ejecta, respectively (as defined in Ref. \cite{2021ApJ...910..116K}). The wind component compositions, contributing to elements around and between the first ($A\sim 80$) and second ($A\sim 130$) \emph{r}-process peaks, are fixed in this study and are described by the H5 and H1 tracers in Ref. \cite{2014MNRAS.443.3134P} for the ``wind1" and ``wind2" models, respectively. We consider two different wind models with lower (wind1) and higher (wind2) neutron-richness to gauge the effects of lighter and heavier element contributions in our comparison to solar \emph{r}-process abundances. The dynamical ejecta compositions, composed of the elements from the second to the third \emph{r}-process peak and beyond, are varied as described in Table \ref{tbl:residuals}; the dynamical ejecta composition used in our previous study, marked with the (*) label, is described by the model B tracer in \cite{2014MNRAS.439..744R}.}

\edited{
We use the models from our previous study (see Ref. \cite{Ristic22}) using \texttt{SuperNu}, a Monte Carlo code for simulation of time-dependent radiation transport with matter in local thermodynamic equilibrium \cite{SuperNu}. Our light-curve simulations use radioactive power sources calculated from decaying the \emph{r}-process composition from the \texttt{WinNet} code \cite{2012ApJ...750L..22W}. The contributions from these power sources are weighted by thermalization efficiencies first presented in \cite{Barnes_2016} (see Ref. \cite{Wollaeger2018} for a detailed description of the adopted nuclear heating). We utilize detailed opacity calculations from the tabulated binned opacities generated using the Los Alamos suite of atomic physics codes \cite{2015JPhB...48n4014F,2020MNRAS.493.4143F}. Our tabulated binned opacities are not calculated for all elements; therefore, we produce opacities for representative proxy elements by combining pure-element opacities of nuclei with similar atomic properties \cite{2020MNRAS.493.4143F}. Our final \texttt{SuperNu} outputs are simulated kilonova spectra evaluated at 1024 equally log-spaced wavelength bins from $0.1 \ \mu\text{m}$ to $12.8 \ \mu\text{m}$ across 54 viewing angles spaced equally in $\cos \theta$ for $-1 \leq \cos \theta \leq 1$. These spectra are then post-processed into light curves assuming a source distance of $10$ parsecs.
}

\edited{
Our \texttt{SuperNu} simulations require discrete mass and average outflow velocity parameter inputs for the ejecta components. To sample our parameter space continuously during parameter inference, we require a continuous mapping of ejecta parameter inputs to kilonova light curve outputs in the form of a Gaussian process surrogate model. 
We built our surrogate model training library of $\sim450$ kilonova light curve simulations using iterative simulation placement guided by Gaussian process variance minimization. In other words, we placed new simulations in regions of parameter space where our interpolation root-mean-square (RMS) uncertainty was largest. For simulation placement purposes, we only examined the uncertainty on the entire bolometric light curve rather than uncertainty at individual simulation times (see Ref. \cite{Ristic22} for a full discussion on the creation of the simulation library).
}

\edited{
Using Gaussian process regression interpolation in conjunction with our simulation library \cite{Ristic22}, we created a continuous mapping of our four ejecta parameter simulation inputs ($M_{\rm{d}}$, $v_{\rm{d}}$, $M_{\rm{w}}$, $v_{\rm{w}}$) to a scalar output $M_{AB, \lambda}$ at some time $t$, angle $\theta$, and wavelength $\lambda$. Because of the substantial dynamic range of our many outputs, we interpolate in AB magnitudes using the LSST \textit{grizy} and 2MASS \textit{JHK} bands as our reference wavelengths. Our Gaussian process uses a squared-exponential kernel and a white noise (diagonal) kernel. Unless otherwise noted, we quantify the performance of our interpolation with the RMS difference between our prediction and the true value.
}

\edited{
Combining our surrogate light curves and parameter inference code, we generated posteriors for the ejecta parameters of GW170817 given our model assumptions \cite{Ristic22}. We perform four-dimensional Monte Carlo integration of the electromagnetic likelihood over our model's four parameters using the EM\_PE package \footnote{\texttt{\rurl{github.com/markoris/EM\_PE}}} to provide the likelihood and the RIFT adaptive Gaussian mixture model integrator to perform the integration \cite{Wofford22}. The parameter estimation in this work, discussed in Section \ref{sec:pe}, follows the same methodology as in Ref. \cite{Ristic22}, with the additional composition-based prior constraint described in Section \ref{sec:rprocess_prior}.
}

\edited{In creating our surrogate model light-curve training library, we considered only one dynamical ejecta composition and one wind ejecta composition, indicated by the (*) label in Table \ref{tbl:residuals}. \editedToo{With regard to constructing a composition-informed mass prior,} the wind compositions in this work are unchanged from previous studies \cite{Wollaeger2018, 2021ApJ...910..116K, Ristic22} while new considerations for dynamical ejecta compositions with different nuclear physics inputs were included in addition to the original dynamical composition used in prior studies. As a result of its ejection process, dynamical ejecta typically has a much lower electron fraction $Y_{\rm{e}} \equiv (n_{\rm{p}})/(n_{\rm{p}} + n_{\rm{n}})$, where $n_{\rm{p}}$ is the number of protons and $n_{\rm{n}}$ is the number of neutrons in the ejecta. A low electron fraction results in a higher availability of neutrons for capture during \emph{r}-process nucleosynthesis and thus generally the creation of heavier elements such as lanthanides and actinides \cite{1957PASP...69..201C, RevModPhys.29.547}. Due to the dynamical ejecta's dominance on the elemental abundance pattern compared to the relatively minimal contribution of the wind ejecta, we only present newly calculated compositions for dynamical ejecta in this work.}

The compositions presented in this work were generated using two nuclear network codes: \texttt{WinNet} and \texttt{PRISM}. The dynamical and wind ejecta compositions considered here and in previous work (i.e. \cite{Wollaeger2018, Ristic22}) were generated using \texttt{WinNet}. \edited{Specifically, these are the wind1 and wind2 compositions and the dynamical ejecta model using the Panov+ (2010) fission model in Table \ref{tbl:residuals}.}

\begin{table*}[ht]
\begin{center}
{\color{black}
\begin{tabular}{c@{\hskip8mm} c@{\hskip4mm} c@{\hskip4mm} c@{\hskip4mm} c@{\hskip4mm} c@{\hskip4mm}} 
\hline
Mass Model & Fission Model & $Y_{\rm{e}}$ & Wind Comp. &  Mass Ratio & Min. Residual \\
 & & $\left(\frac{n_{\rm{p}}}{n_{\rm{p}}+n_{\rm{n}}}\right)$ & & ($M_{\rm{w}}/M_{\rm{d}}$) & $(r_{\rm{min}}$) \\ [0.5ex]
\hline\hline
FRDM2012 & FRLDM & 0.035 & wind2 & 0.47 & 1257  \\
FRDM2012 & 50/50 & 0.180 & wind2 & 0.35 & 1849  \\
FRDM2012 & FRLDM & 0.035 & wind1 & 0.20 & 2001  \\
HFB24 & FRLDM & 0.035 & wind2 & 0.52 & 2470  \\
FRDM2012 (*) & Panov+ (2010) & 0.035 & wind2 & 0.24 & 2550  \\
HFB24 & FRLDM & 0.180 & wind2 & 0.19 & 2722  \\
HFB24 & FRLDM & 0.035 & wind1 & 0.10 & 2872  \\
FRDM2012 & 50/50 & 0.180 & wind1 & 17.07 & 3323  \\
FRDM2012 & Panov+ (2010) & 0.035 & wind1 & 17.69 & 4048  \\
HFB24 & FRLDM & 0.180 & wind1 & 12.66 & 4192  \\ \hline 
\end{tabular}
}
\end{center}
\caption{Wind-to-dynamical mass ratios \edited{sorted by increasing} minimum residuals for each dynamical composition considered. Mass ratios were
 determined by calculating the mean mass ratio of the bottom 2nd percentile of all residuals so as to eliminate outliers.
 The residuals were calculated as in Equation \ref{eq:r} and the minimum residual was identified as the smallest residual 
 across all the mass pairs considered for a given composition. The two \edited{wind1} and \edited{wind2} trajectories are described in detail in Ref. \cite{Wollaeger2018}. 
 The two nuclear
 mass models considered are FRDM2012 and HFB24 \cite{2016ADNDT.109....1M, 2014EPJA...50...43P}. The two nuclear
 fission models considered in our study are FRLDM \cite{2020PhRvC.101e4607M} and ``50/50," a simple symmetric
 assumption that fission yields split into two identical nuclei. The fission rates for the simulations performed in previous work, labeled Panov+ (2010), are taken from Ref. \cite{2010A&A...513A..61P}. The (*) indicates the compositions used in creating
 the surrogate light curves used during parameter estimation (see Section \ref{sec:discussion}). \editedToo{The reported $Y_{\rm e}$ values describe the neutron-richness of just the dynamical ejecta component.}
}
\label{tbl:residuals}
\end{table*}

\begin{table*}[ht]
\begin{center}
{\color{black}
\begin{tabular}{c@{\hskip8mm} c@{\hskip4mm} c@{\hskip4mm} c@{\hskip4mm} c@{\hskip4mm}} 
\hline
Mass Model & Fission Model & Wind Comp. &  Mass Ratio & Min. Residual \\
 & & & ($M_{\rm{w}}/M_{\rm{d}}$) & $(r_{\rm{min}}$) \\ [0.5ex]
\hline\hline
FRDM2012 & FRLDM & wind2 & 0.32 & 1467  \\
FRDM2012 & 50/50 & wind2 & 0.34 & 1809  \\
FRDM2012 & FRLDM & wind1 & 0.04 & 1854  \\
HFB27 & FRLDM & wind2 & 0.81 & 2433  \\
HFB27 & 50/50 & wind2 & 0.81 & 2548  \\
FRDM2012 & 50/50 & wind1 & 0.01 & 3242  \\
HFB27 & FRLDM & wind1 & 0.04 & 4080  \\
HFB27 & 50/50 & wind1 & 46.89 & 4609 \\ \hline 
\end{tabular}
}
\end{center}
\caption{Same as Table~\ref{tbl:residuals} except considering dynamical ejecta compositions derived from the $Y_{\rm e}$ distribution presented in Figure 5 of \cite{2022arXiv221107637K}. We also consider the HFB27 mass model in place of HFB24.
}
\label{tbl:yedist}
\end{table*}

The varied dynamical ejecta compositions new to this work were generated using \texttt{PRISM}. \texttt{PRISM} is a single-zone nuclear reaction network code that evolves an initial seed abundance of nuclei along a time-temperature-density thermodynamic trajectory, while allowing full flexibility with the input nuclear data \cite{2015APS..DNP.EA097S}. In this work, we use state-of-the-art nuclear reaction and decay rates that are calculated to be self-consistent with the nuclear mass model. Following from the thermodynamic trajectories of dynamical ejecta from neutron star merger simulations presented in Ref. \cite{Korobkin_2012}, all of our \texttt{PRISM} runs begin in nuclear statistical equilibrium at a temperature of 10 GK in the thermodynamic trajectory. \edited{All dynamical ejecta models presented in Table \ref{tbl:residuals} with a fission model different from Panov+ (2010) were generated using \texttt{PRISM}.}

\begin{figure}
\includegraphics[width=\columnwidth]{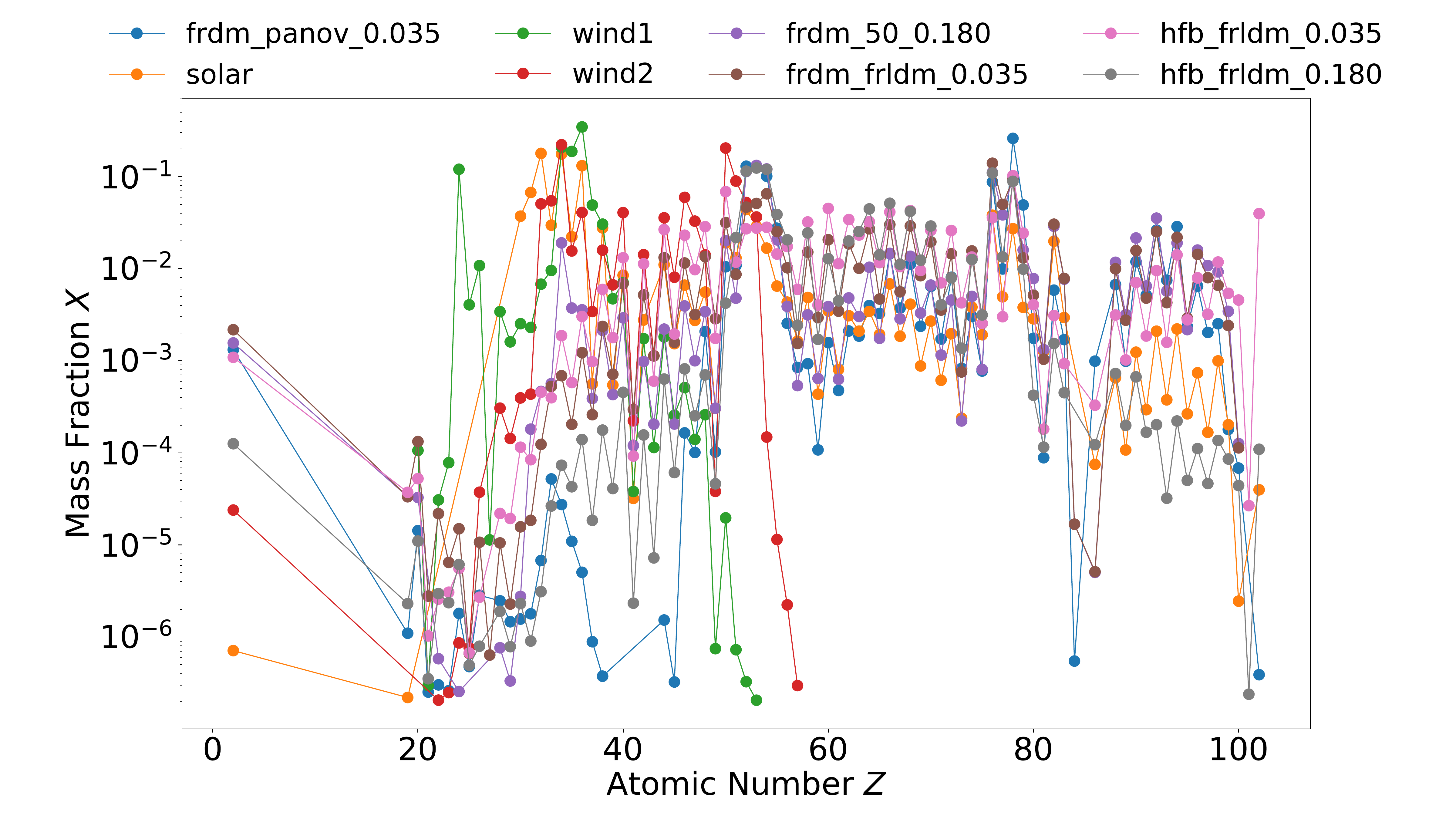}
\caption{\edited{Unscaled mass-fractions $X$ as a function of atomic number $Z$ for all single-$Y_e$ compositions considered in this work. The labels pertaining to the dynamical ejecta compositions considered in this work indicate the nuclear mass model, fission model, and electron fraction $Y_{\rm e}$ used to generate the respective composition. The remaining labels indicate the solar and wind compositions. The wind1 and wind2 compositions do not extend to higher atomic numbers $Z$ due to their higher electron fractions $Y_{\rm e} = 0.37$ and $Y_{\rm e} = 0.27$, respectively.}
}
\label{fig:abundances}
\end{figure}

\edited{The mass fractions of all the compositions considered in this work are shown in Figure \ref{fig:abundances}. Figure \ref{fig:abundances} highlights the main difference between our wind ejecta compositions; the wind1 composition has very low mass fractions at the second $r$-process peak ($A\sim130$) while a significant portion of the wind2 composition consists of elements around this peak.}

\subsection{Ejecta Prior Implied by \emph{r}-process Observations}
\label{sec:rprocess_prior}

We \edited{seek} to compare the combined mass fractions $X_{\rm{sim}}$ to the seemingly universal pattern of elements between the 2nd and 3rd \emph{r}-process peaks (the ``main'' \emph{r}-process, \cite{2021RvMP...93a5002C}) \edited{observed among some of the oldest stars}. This ``\emph{r}-process universality" has been noted for iron-poor (or ``metal-poor") stars that show enhancements in the main \emph{r}-process elements relative to their iron content \edited{in excess of ten times the equivalent Solar ratio}.
However, \edited{observationally derived abundances in} metal-poor stars are necessarily elemental since the abundances are derived from atomic transitions in stellar spectra \edited{that are overall insensitive to atomic mass number}.
Except for a handful of elements, the detailed \emph{isotopic} distributions of \emph{r}-process elements in metal-poor stars is observationally unknown.
As a proxy for a representative example of the universal \emph{r}-process, we use the well-studied solar isotopic abundance pattern $X_{\odot}$, relying on previously-published projections of the high-$A$ elements into different neutron-capture process contributions. Specifically, the \emph{r}-process fractions presented in Ref.\ \cite{Arlandini1999} are used in conjunction with the total abundances from Ref. \cite{Sneden2008} to isolate the contribution to the solar abundances by the \emph{r}-process.

\begin{figure}[tp]
\includegraphics[width=\columnwidth]{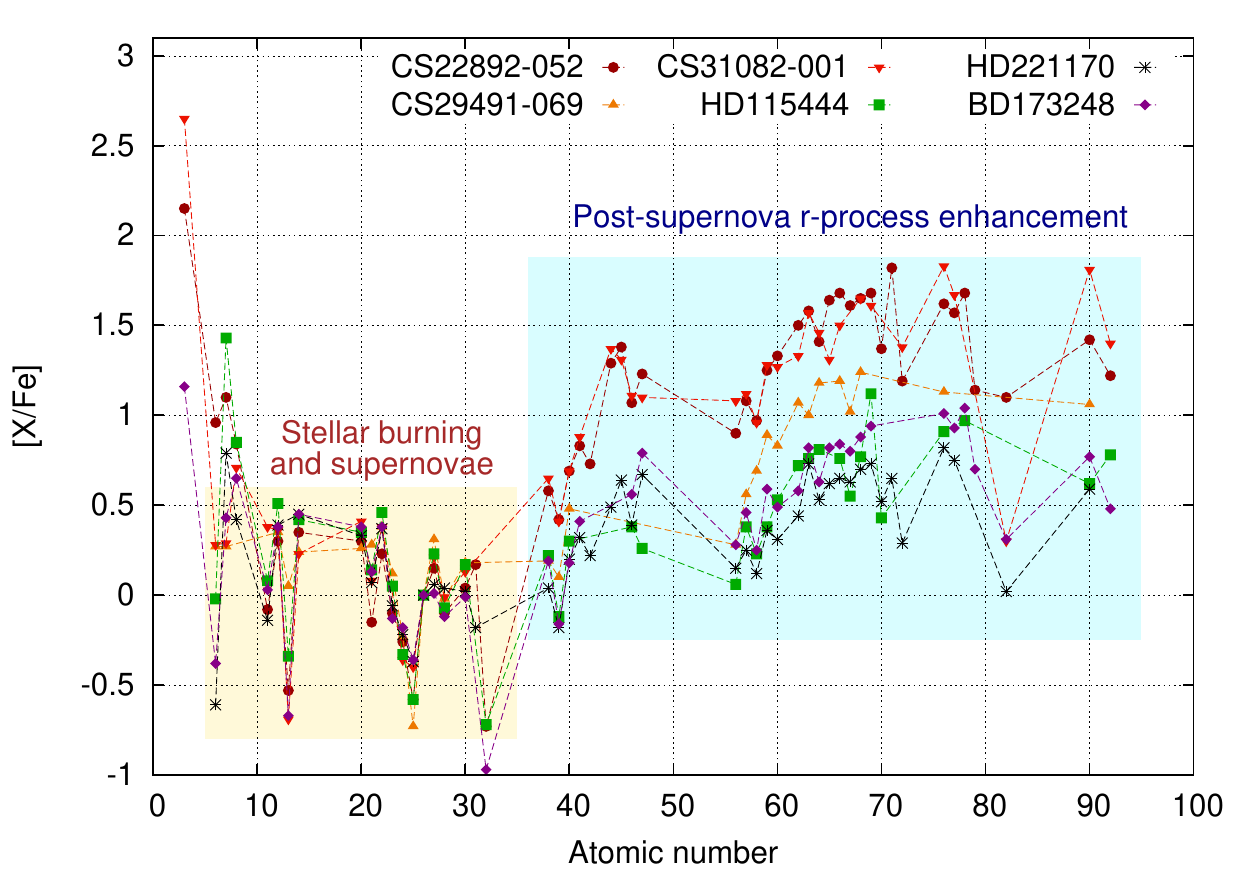}
\caption{Elemental mass fraction ratios relative to iron of a sample of \emph{r}-process enriched metal-poor stars. 
$[X/\textrm{Fe}] > 0$ implies enhanced abundance of element $X$ compared to the solar system with respect to iron. 
We assume all elements that are significantly enhanced compared to iron to have been introduced post-supernova, \edited{exclusively from neutron star mergers.} 
The region of enhanced elements ($Z\ge37$, highlighted in blue) is the focus of our comparison to solar composition.
The iron-peak elements and supernova \emph{r}-process are not strongly enhanced compared to solar (highlighted in yellow).
Stellar elemental abundances obtained from \texttt{JINAbase} \cite{2018ApJS..238...36A} 
with the respective stars reported in references 
\cite{SNE03,HAY09,HIL02,WES00,IVA06,COW02}.}
\label{fig:metal_poor_stars}
\end{figure}

Figure \ref{fig:metal_poor_stars} shows the [X/Fe] abundances of six metal-poor stars \editedToo{with \emph{r}-process enhancements}.
The ``[X/Fe]" notation means that each elemental ratio $\log\epsilon(\textrm{X/Fe})$ is compared to the same elemental ratio in the solar abundance 
pattern.\footnote{Definition: 
$[X/{\rm Fe}] := \log\epsilon({\rm X/Fe}) - \log\epsilon({\rm X/Fe})_\odot$, where 
$\log\epsilon({\rm X}) := \log_{10}\left(Y_{\rm X}/Y_{\rm H}\right) = \log_{10}\left(Y_{\rm X}\right) + 12$,
where $Y_{\rm X}$ is the \emph{abundance} (mole fraction) of the element X.}
Stars with $[{\rm X/Fe}]>0$ are considered ``enhanced" in that element relative to the solar system abundance.
For many metal-poor stars, elements with $Z\geq 37$ have an enhanced abundance compared to \editedToo{the Sun}.
\edited{For this work, guided by the enhancement seen in elements $Z \geq 37$ in Figure \ref{fig:metal_poor_stars} and with the assumption that iron was created during supernova nucleosynthesis, we assume that elements with $Z\geq 37$ originate \emph{exclusively} from neutron star mergers.}
The trends of elements with $Z< 37$ are less clear; they are not uniformly enhanced in stars that are enhanced with the $Z\geq 37$ elements, likely pointing to multiple (non-merger) origins for these elements.

In our \texttt{SuperNu} simulations, we adopt a two-component compositional model and vary the mass ratio of the two components: the dynamical ($M_{\rm{d}}$) and wind ($M_{\rm{w}})$ ejecta masses. Each component has a fixed isotopic abundance, computed via nucleosynthesis network \cite{2012ApJ...750L..22W}. 
Due to the fixed nature of the compositions, we weight each component's composition, represented by mass fractions $X_{\rm{d}}$ and $X_{\rm{w}}$, by the mass of the respective ejecta component, dynamical $M_{\rm{d}}$ and wind $M_{\rm{w}}$, to introduce composition variation as a function of component mass in the combined simulation mass fraction $X_{\rm{sim}} = (M_{\rm{d}}X_{\rm{d}} + M_{\rm{w}}X_{\rm{w}})/(M_{\rm{d}}+M_{\rm{w}})$.
For every isotope, the combined mass fraction $X_{\rm{sim}}$ is simply the mass-weighted sum of its mass fractions in the constitutive components. We varied our dynamical and wind component masses over a grid between $-3 \leq \log({M_{\rm{d,w}}/M_{\odot})} \leq -1$, encompassing the most realistic ejecta masses predicted by numerical relativity simulations of neutron star mergers \cite{2014MNRAS.443.3134P, Rosswog_2017, 2018ApJ...869L..35R, 2019MNRAS.482.3373F, 2019PhRvD.100b3008M, 2021ApJ...906...98N}.

\edited{To account for isotopes of actinides with short decay timescales,} we rescale the solar mass fractions of actinides $X_{\odot, Ac}$ to what they would have been at 1 day to better match the kilonova-timescale mass fractions used in our simulations. The rescaling is achieved by setting the 1-day solar actinide mass fractions to values that would decay to present day values after 4.5 Gyr. The rescaled solar mass fractions are also mapped into a subset of representative elements used for \texttt{SuperNu} light-curve modeling as described in Section \ref{sec:sim_setup}. Hereafter, any mention of the solar mass-fraction pattern $X_\odot$ refers to the 1-day rescaled and mapped mass fractions using data from Refs. \cite{Arlandini1999} and \cite{Sneden2008}.

To get $X_\odot$ and $X_{\rm{sim}}$ on the same relative scale, we introduce an offset \edited{$C_{{\rm scale},Z}$} that shifts $X_\odot$ down to comparable values for $X_{\rm{sim}}$ by matching the two mass-fraction values at some element $Z$. To minimize how the choice of \edited{$C_{{\rm scale},Z}$} affects our results, we integrate over the range of possible values of \edited{$C_{{\rm scale}, Z}$} introduced by scaling $X_\odot$ and $X_{\rm{sim}}$ to matching values at different elements $Z$. \edited{This integration marginalizes over our uncertainty in $C_{{\rm scale}, Z}$, adopting a Gaussian prior on $\log(C_{{\rm scale}, Z})$ with mean $\mu = 0.08$ and variance $\sigma^2 = 0.695$}. After integrating over \edited{$C_{{\rm scale}, Z}$}, we are left with a single required choice of a new $C$ value that sets the constrained scale at some element $Z_{\rm{choice}}$ such that $\log X_{{\rm sim}, Z_{\rm{choice}}} = \log X_{\odot, Z_{\rm{choice}}} - \log C$. We chose $Z_{\rm{choice}}=46$ as it is one of the elements present in all the dynamical, wind, and solar mass fractions. Shifting $X_\odot$ to be on the same relative scale as $X_{\rm{sim}}$ and choosing a specific value of $C$ at some $Z_{\rm{choice}}$ are both done solely for the purpose of calculating well-behaved residuals. \edited{While the scaling by $C$ removes the constraint that $\sum_Z X_{\odot, Z} = 1$, this has no detrimental effects on our analysis as we are interested exclusively in our compositions' relative abundances}.

\begin{figure}[ht!]
    \centering
    \includegraphics[width=\columnwidth]{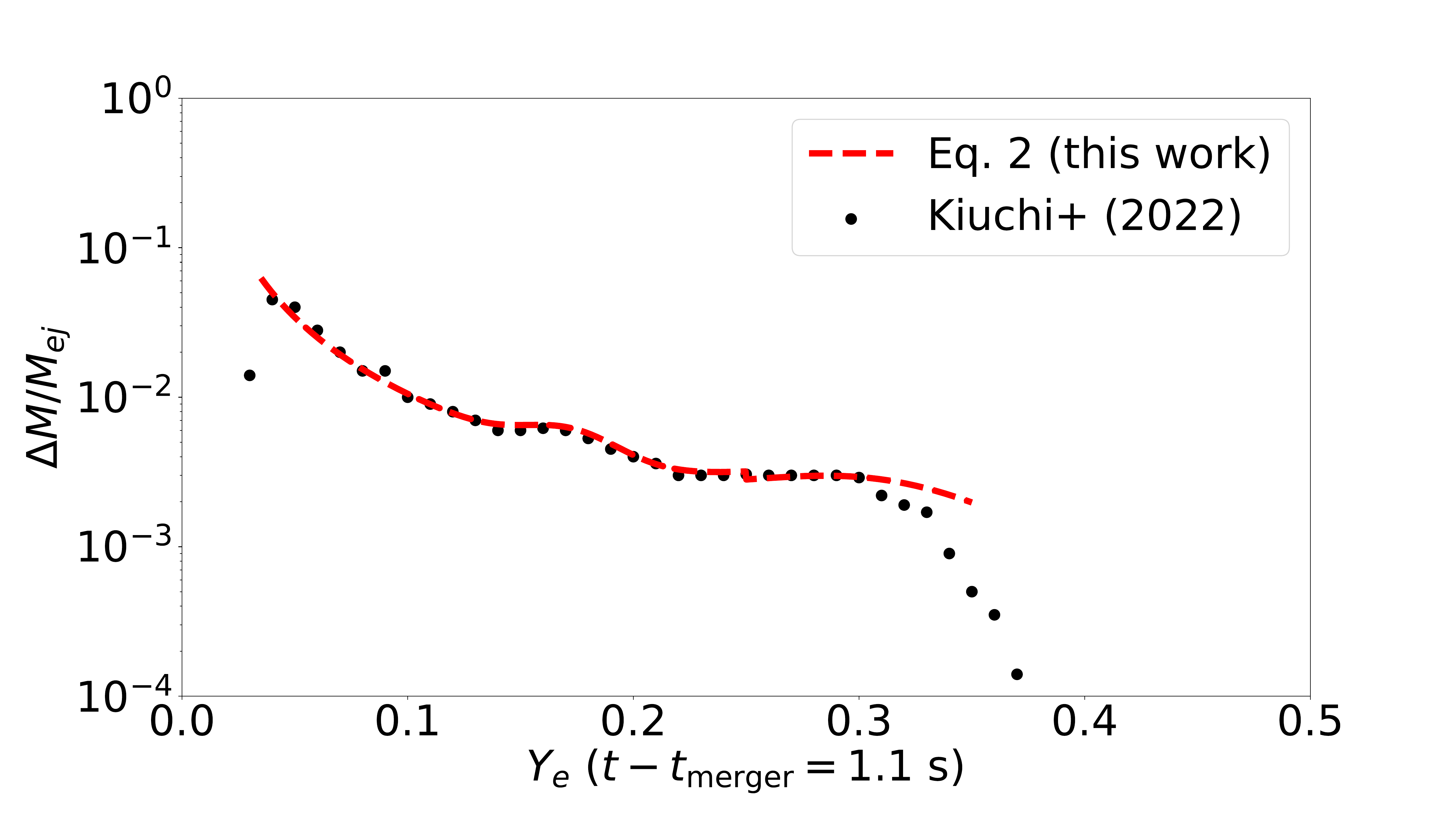}
    \caption{\editedthree{Recreated $Y_{\rm e}$ distribution for dynamical ejecta as presented in Fig. 5 of Ref. \cite{2022arXiv221107637K}. The analytic fit to the distribution, presented in Equation~\ref{eq:yedist_analytic}, is overlaid as a red line. The analytic fit was used to generate the $Y_e$-distribution compositions presented in Table~\ref{tbl:yedist}.}}
    \label{fig:kiuchi_fit}
\end{figure}

\edited{As part of our assumption that elements $Z \geq 37$ were synthesized exclusively in neutron star mergers (see Figure \ref{fig:metal_poor_stars}),} we only consider elements from $Z = 37$ up to and including $Z = 103$ when computing the residual $r(M_{\rm{d}}, M_{\rm{w}})$ for all available elements in the solar abundance pattern $Z \in Z_\odot$ given a simulation with component masses $M_{\rm{d}}, M_{\rm{w}}$:
\begin{multline}
\label{eq:r}
r(M_{\rm{d}}, M_{\rm{w}}) = \\ \sum_{Z=37}^{103} \frac{\left(\log{X_{\odot, Z}} - \log{C} - \log{X_{\rm{sim}, Z}}\right)^2}{2\sigma^2} \\
- \frac{N}{2\sigma^2}\frac{\left(\log{\overline{X}_{\rm{sim}}}-\log{\overline{X}_\odot}\right)^2}{1 + \sigma^2/\left(N\sigma_C^2\right)}
\end{multline}
where $r(M_{\rm{d}}, M_{\rm{w}})$ is the residual for the given dynamical and wind mass pair used to calculate $X_{\rm{sim}} = (M_{\rm{d}}X_{\rm{d}} + M_{\rm{w}}X_{\rm{w}})/(M_{\rm{d}}+M_{\rm{w}})$, $Z$ is the element's atomic number, $\log{X_{\odot, Z}}$ is the decimal logarithm of the solar mass fraction of element $Z$, $\log{C}$ is the decimal logarithm of the offset matching $X_\odot$ to $X_{\rm{sim}}$ at $Z=46$, $\log{X_{\rm{sim, Z}}}$ is the decimal logarithm of the simulation mass fraction of element $Z$ in both components (if present), $\sigma$ is the uncertainty on $\log{X_{\rm{sim}}}$, $\sigma_C$ is the uncertainty introduced by integrating out \edited{$C_{{\rm scale}, Z}$}, $\overline{X}_\odot$ is the average solar mass fraction across all elements, $\overline{X}_{\rm{sim}}$ is the average simulation mass fraction across all elements, and N is the total number of elements considered.

\begin{figure}[tp!]
\includegraphics[width=.99\columnwidth]{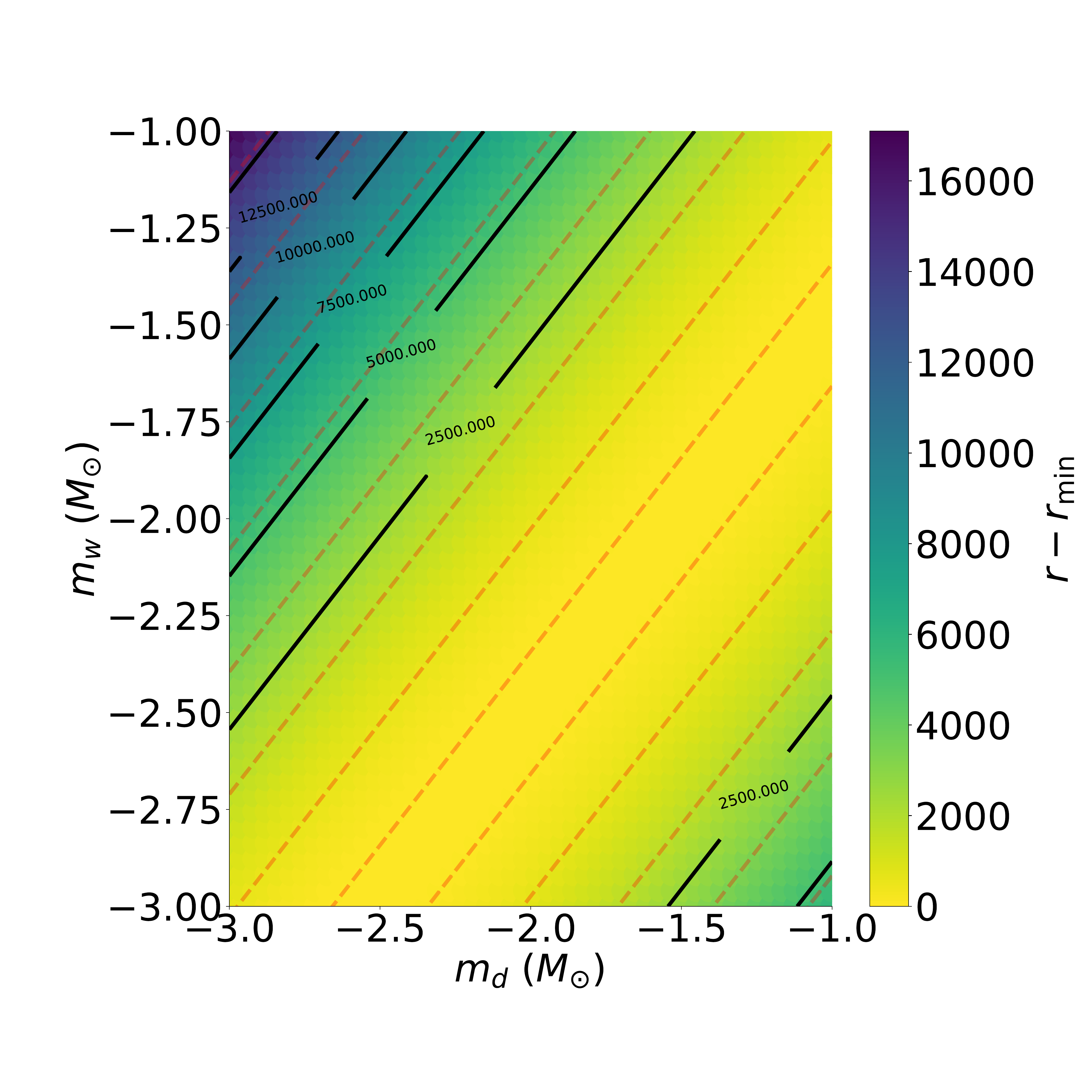}
\includegraphics[width=.99\columnwidth]{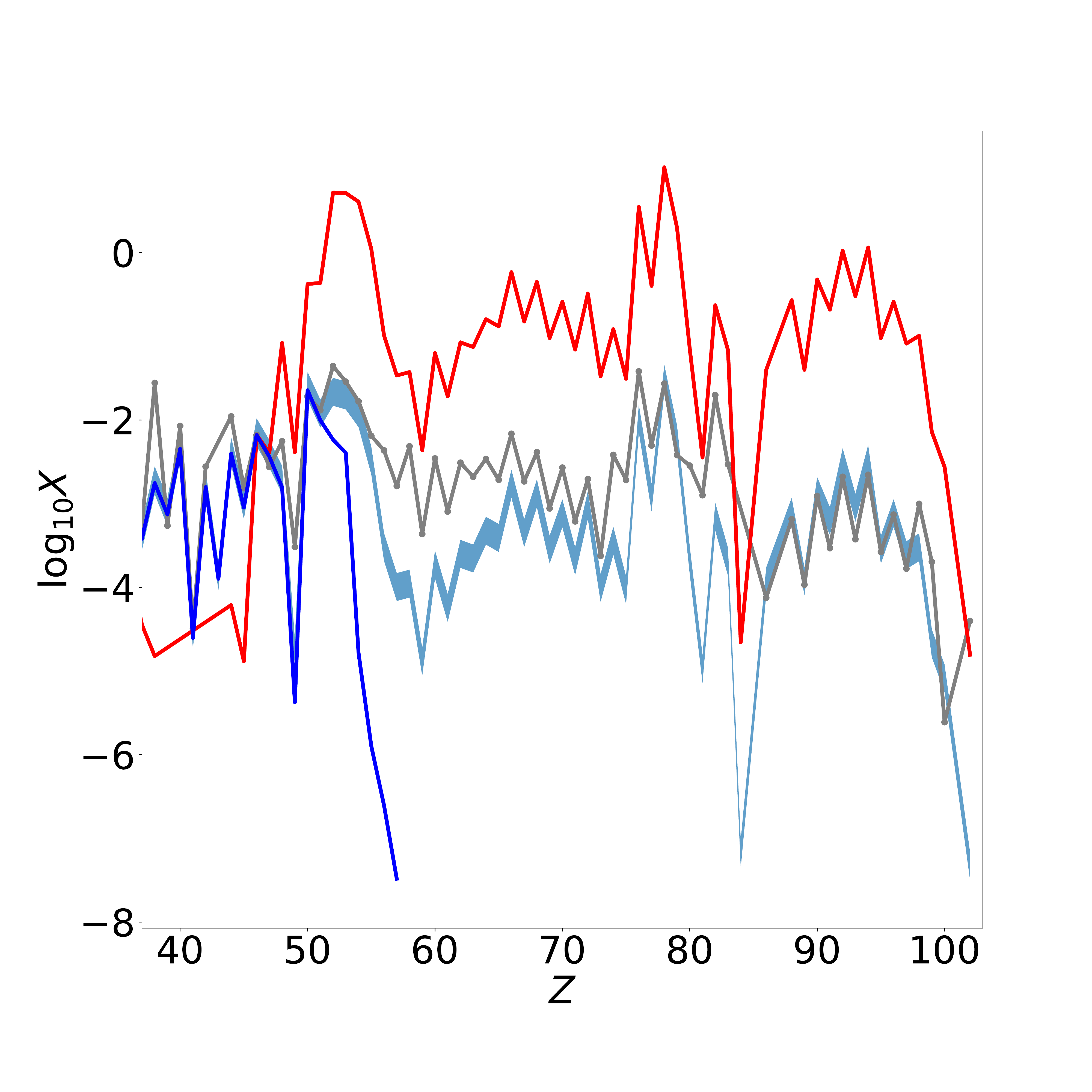}
\caption{\textit{Top}: 2-D distribution of residuals \edited{calculated as in Equation \ref{eq:r} by comparing $\log X_{\odot}$ to $\log X_{\rm{sim}}$ using the dynamical and wind compositions matching those in \cite{Ristic22}, represented by the (*) label in Table \ref{tbl:residuals}}. \edited{The residual grid is composed of} 50 mass values equally log-spaced between $-3 \leq \log{M_{\odot}} \leq -1$ for both dynamical and wind mass. \edited{The black lines indicate contours of equally-spaced residual values. The dashed red lines indicate a wind-to-mass ratio of 1 and serve purely as a visual aid.} \textit{Bottom}: Mass fractions of individual ejecta components compared to the \editedthree{best-fit mass fraction $X_{\rm sim}$ obtained from comparison to $X_\odot$}. The red and blue lines are the initial unweighted dynamical ($X_{\rm d}$) and wind ($X_{\rm w}$) ejecta mass fractions, respectively, scaled \edited{by $C$} to match the solar mass fraction at $Z=46$. The gray line is the solar mass fraction $X_\odot$ and the blue shaded region is the $90\%$ confidence interval for all the mass-weighted mass fractions $\log{X_{\rm{sim}}}$. The dynamical ejecta mass fraction only exceeds $\log{X} = 0$ due to the scale matching at $Z=46$.}
\label{fig:Ristic_2022_comps}
\end{figure}

\begin{figure}[tp!]
\includegraphics[width=.99\columnwidth]{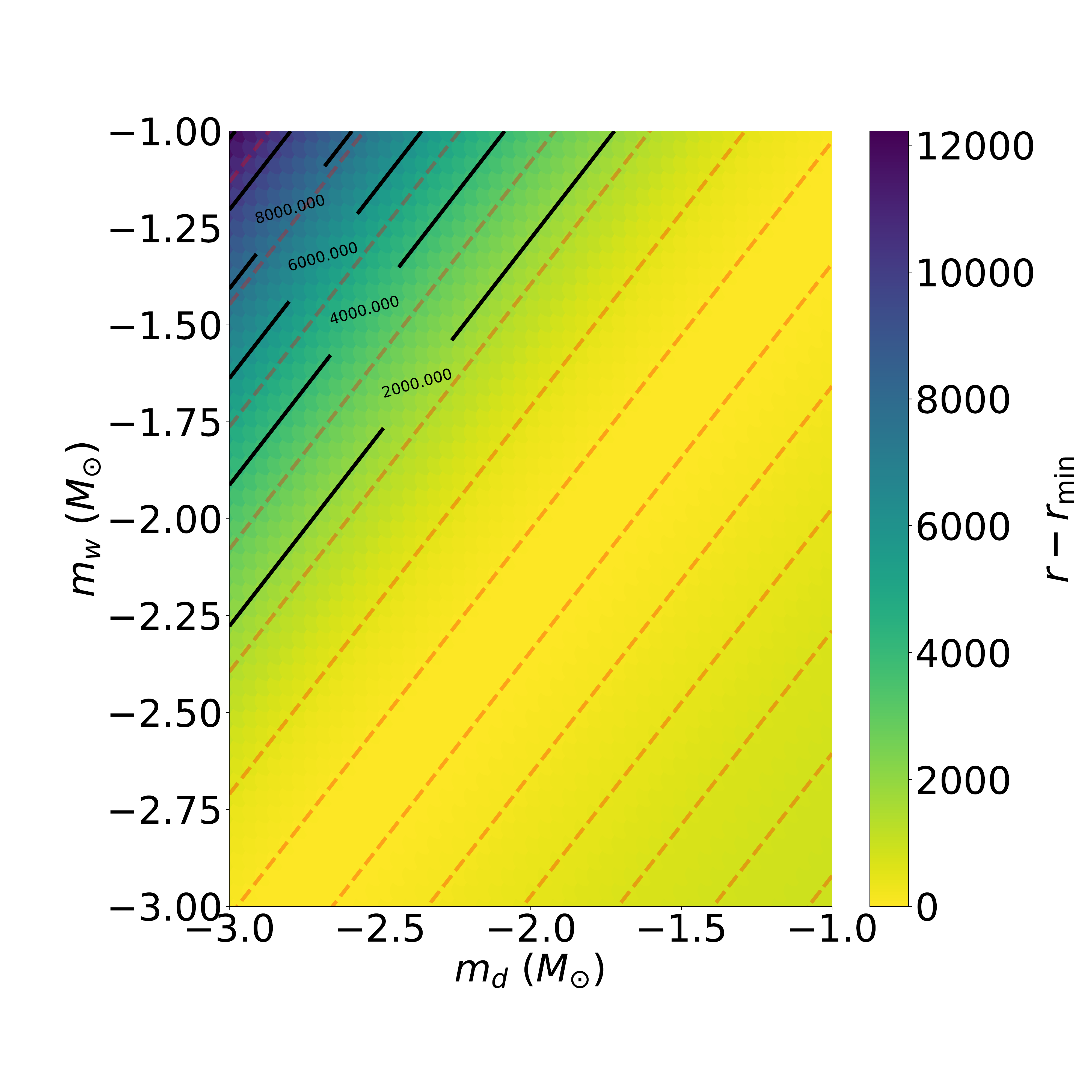}
\includegraphics[width=.99\columnwidth]{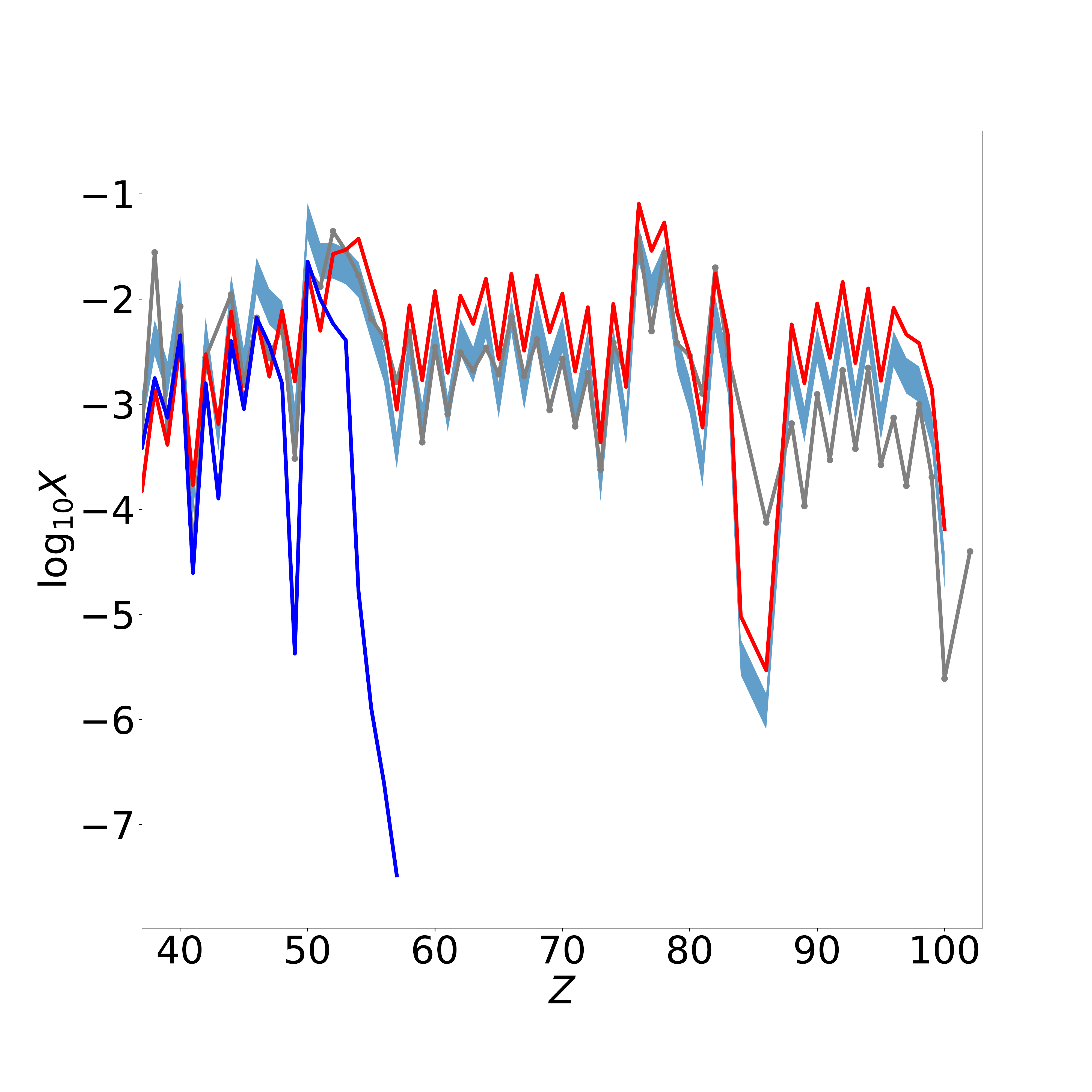}
\caption{\textit{Top}: Same 2-D distribution as described in Figure \ref{fig:Ristic_2022_comps} except with the compositions which yielded the lowest residual in comparison to the solar abundance pattern $X_\odot$ (top row of Table \ref{tbl:residuals}). \textit{Bottom}: Mass fractions of individual ejecta components compared to the \editedthree{best-fit mass fraction $X_{\rm sim}$ obtained from comparison to $X_\odot$}. The line colors represent the same quantities as in Figure \ref{fig:Ristic_2022_comps}. The minimum residual was identified as the smallest residual across all the mass pairs considered for a given composition $X_{\rm sim}$.}
\label{fig:new_comps}
\end{figure}

\edited{We compute the residual between each of our composition models $X_{\rm{sim}}$ from Table \ref{tbl:residuals} and the solar mass-fraction pattern $X_{\odot}$ using Equation \ref{eq:r}. We consider component mass weights across a log-spaced grid with $-3 \leq \log({M_{\rm{d,w}}/M_{\odot})} \leq -1$, with 50 masses for each component resulting in a total of 2500 residuals per composition model. \editedthree{The scaled residual values $r - r_{\rm min}$ are shown in the top panel of Figures \ref{fig:Ristic_2022_comps} and \ref{fig:new_comps}, with \emph{r} representing the residual calculated for each mass pair and $r_{\rm{min}}$ the lowest residual for all mass pairs considered for a given model.}

\editedthree{Guided by numerical-relativity simulations which suggest a distribution of $Y_{\rm e}$ values in neutron-star ejecta (e.g., \cite{2019PhRvD.100b3008M, 2021ApJ...906...98N, 2021PhRvD.104l4012M, 2022arXiv221107637K}), we also analyze a selection of compositions derived from the $Y_{\rm e}$ distribution for dynamical ejecta presented in Figure 5 of Ref. \cite{2022arXiv221107637K}. To calculate the mass-weights for each $Y_{\rm e}$ value, we recreate the $Y_{\rm e}$ distribution with a piece-wise analytic fit

\begin{equation}
\begin{split}
\Delta M / M &= 0.05C(Y_{\rm e}/0.04)^{-1.7} \\
&+ \frac{A}{\sqrt{(2\pi\sigma_1^2)}}e^{-0.5((Y_{\rm e}-0.17)/\sigma_1)^2} \\
&+ \frac{B}{\sqrt{(2\pi\sigma_2^2)}} e^{-0.5((Y_{\rm e}-0.3)/\sigma_2)^2} \\
\text{with }
C &= \begin{cases} 1 & Y_{\rm e} \leq 0.25 \\
 (0.04/Y_{\rm e})^{0.1} & Y_{\rm e} > 0.25,\end{cases}
\end{split}
\label{eq:yedist_analytic}
\end{equation}

with fit parameters $A = 10^{-4}$, $B = 2\times10^{-4}$, $\sigma_1 = 0.02$, $\sigma_2 = 0.05$. A comparison of the fit to the Ref. \cite{2022arXiv221107637K} $Y_{\rm e}$ distribution is shown in Figure~\ref{fig:kiuchi_fit}. Our fit begins to deviate from the distribution at $Y_{\rm e} = 0.3$; this is of little concern as less than $3\%$ of the total ejecta mass is described by $Y_{\rm e} > 0.3$. 

For each $Y_{\rm e}$-distribution composition, we run 10 nucleosynthesis simulations, evenly spaced between $0.035 \leq Y_{\rm e} \leq 0.35$. The mass-weight for each single-$Y_{\rm e}$ composition is calculated using Equation~\ref{eq:yedist_analytic}. Once all the weights are calculated, they are normalized such that their net contribution describes the total ejecta mass. The final $Y_{\rm e}$-distribution composition is the weighted sum of the abundances from the single-$Y_{\rm e}$ nucleosynthesis simulations. After the net $Y_{\rm e}$-distribution composition is calculated, we repeat the same methodology as for the models presented in Table~\ref{tbl:residuals} to calculate mass ratios and residuals for each composition model.}
}

\subsection{Parameter Inference}
\label{sec:pe}

As in many previous applications of Bayesian inference to infer parameters of kilonovae
\cite{gwastro-mergers-em-CoughlinGPKilonova-2020,2018MNRAS.480.3871C,2019MNRAS.489L..91C,2017ApJ...851L..21V,2017Natur.551...75S, 2021arXiv210101201B, Nicholl21, 2022MNRAS.516.1137L},
we seek to compare the observed magnitudes $x_i$ at evaluation points $i$ (denoting a combination of band and time) to a
continuous model that makes predictions $m(i|{\bm \theta})$ which depend on some model parameters $\theta$. Bayes
theorem expresses the posterior probability $p({\bm\theta})$ in terms of a prior probability $p_{\rm
 prior}({\bm\theta})$ for the model parameters $\bm\theta$ and a likelihood ${\cal L}(\theta)$ of all observations,
given the model parameters, as 
\begin{equation}
p({\bm \theta}) = \frac{{\cal L}({\bm \theta}) p_{\rm prior}({\bm \theta})}{
 \int d {\bm \theta} {\cal L}({\bm \theta}) p_{\rm prior}({\bm \theta})
} \;\;.
\end{equation}
Unless otherwise noted, for simplicity we assume that the source sky location, distance, and merger time are known.
We adopt a uniform prior on the ejecta velocity $v/c\in[0.05,0.3]$ and the two-dimensional prior discussed in Section \ref{sec:rprocess_prior} on the ejecta masses $m/M_\odot \in [0.001,0.1]$. 

\edited{We assume the observations have Gaussian-distributed magnitude errors with presumed known observational (statistical) uncertainties $\sigma_i$, convolved with
some additional unknown systematic uncertainty $\sigma$, so that our log-likelihood is 
\begin{equation}
    \ln \mathcal{L}(\bm{\theta}) = -0.5 \sum_{i=1}^n \left [ \frac {(x_i - m(i|\bm{\theta}))^2} {\sigma_i^2 + \sigma^2} + \ln(2 \pi (\sigma_i^2 + \sigma^2)) \right ]
\end{equation}
where the sum is taken over every data point in every band used in the analysis. For inference using our Gaussian process surrogate models, we set $\sigma$ to the estimated Gaussian process model error.}
For a full discussion of our parameter inference considerations, see Ref. \cite{Ristic22}.

\section{Results}
\label{sec:discussion}
For our two-component models, assuming a single source like GW170817 dominates the observed solar \emph{r}-process abundances, the inferred abundances from such mergers only depend on the mass ratio $M_{\rm{w}}/M_{\rm{d}}$. In other words, since in our study we emphasize only the
relative and not absolute \emph{r}-process abundances, motivated by considerable uncertainty in the binary neutron star merger rate, we
therefore only use and constrain the abundance \emph{ratios}. The relative abundances from a single channel
only depend on the relative proportions of this channel; for our two-component model, this is simply dependent on $M_{\rm{w}}/M_{\rm{d}}$. 
Thus for each set of initial assumptions---the composition of the dynamical ejecta (\edited{represented by the electron fraction} $Y_{\rm{e}}$), the presumed nuclear mass
and fission model, and other details---our comparison with solar abundances necessarily constrains $M_{\rm{w}}/M_{\rm{d}}$ narrowly
around a preferred value unique to that model. We note that the abundances we are considering are effectively frozen out for the processes we're interested in at times later than $\mathcal O(1)$ second.

Tables~\ref{tbl:residuals} and \ref{tbl:yedist} provide a list of models and their preferred $M_{\rm{w}}/M_{\rm{d}}$, in the sense that they minimize the residual mismatch with the solar abundances as calculated in Equation \ref{eq:r}.
With a few exceptions, most models prefer $M_{\rm{w}}/M_{\rm{d}}$ \editedthree{substantially lower than order unity}. In other words, most of our
abundance comparisons suggest \editedthree{that less} wind than dynamical ejecta would be required for GW170817-like mergers to reproduce the solar \emph{r}-process abundances.
These results are \editedthree{at odds} with those found from other contemporary modeling \cite{2019MNRAS.489L..91C, Nicholl21, 2021arXiv211215470A, 2020ApJ...889..171K} as well as numerical relativity results \cite{2014MNRAS.443.3134P, 2016MNRAS.460.3255R, PhysRevD.95.044045, 2019ARNPS..69...41S, 2021ApJ...906...98N}\editedthree{, which typically predict more post-merger (i.e. wind) mass ejection}. 

\editedthree{One particularly interesting result is that the lowest-residual single-$Y_{\rm e}$ composition in Table~\ref{tbl:residuals} agrees with $X_\odot$ better than the $Y_{\rm e}$-distribution composition with similar nuclear physics inputs. This result suggests that a single-$Y_{\rm e}$ approximation is adequate for computational simplicity in the context of nucleosynthesis calculations.}

\edited{
The top panel of Figure \ref{fig:Ristic_2022_comps} shows the mass-pair residuals for the composition and morphology assumptions considered in previous work \cite{Ristic22}, denoted by the (*) label in Table \ref{tbl:residuals}. The yellow stripe indicating the lowest residual region highlights the best-fitting ratio of wind-to-dynamical mass implied by the calculated residuals. This corresponds to the ``Mass Ratio ($M_{\rm{w}}$/$M_{\rm{d}}$)" value recorded in Table \ref{tbl:residuals}. 

The top panel of Figure \ref{fig:new_comps} shows the best-fitting wind-to-dynamical mass ratio for the lowest residual composition model presented in this work: dynamical ejecta with a composition characterized by the FRDM 2012 mass model, FRLDM fission model, electron fraction $Y_{\rm{e}} = 0.035$ and wind ejecta corresponding to the wind2 model. The bottom panels of Figures \ref{fig:Ristic_2022_comps} and \ref{fig:new_comps} show the solar mass-fraction pattern $X_{\odot}$ in gray, the dynamical and wind ejecta fixed composition mass-fraction patterns $X_{\rm{d}}$ and $X_{\rm{w}}$ in red and blue, respectively, and the 90\% confidence interval for the mass-fraction pattern of the relevant composition model $X_{\rm{sim}}$ as the light-blue shaded region. The 90\% confidence interval was calculated for the spread of possible mass-fraction patterns which arose when scaling the fixed component compositions $X_{\rm{d}}\ (X_{\rm{w}})$ by the component mass $M_{\rm{d}}\ (M_{\rm{w}})$.} \editedthree{Based on the good agreement shown in Figure~\ref{fig:new_comps}, we adopt the associated two-dimensional likelihood versus $M_d$ and $M_w$ as a prior constraint on ejecta masses. In other words, we assume GW170817 was produced from a representative member of a single population of kilonovae, which alone are responsible for the solar $r$-process abundance).}

\begin{figure}[tp!]
\includegraphics[width=\columnwidth]{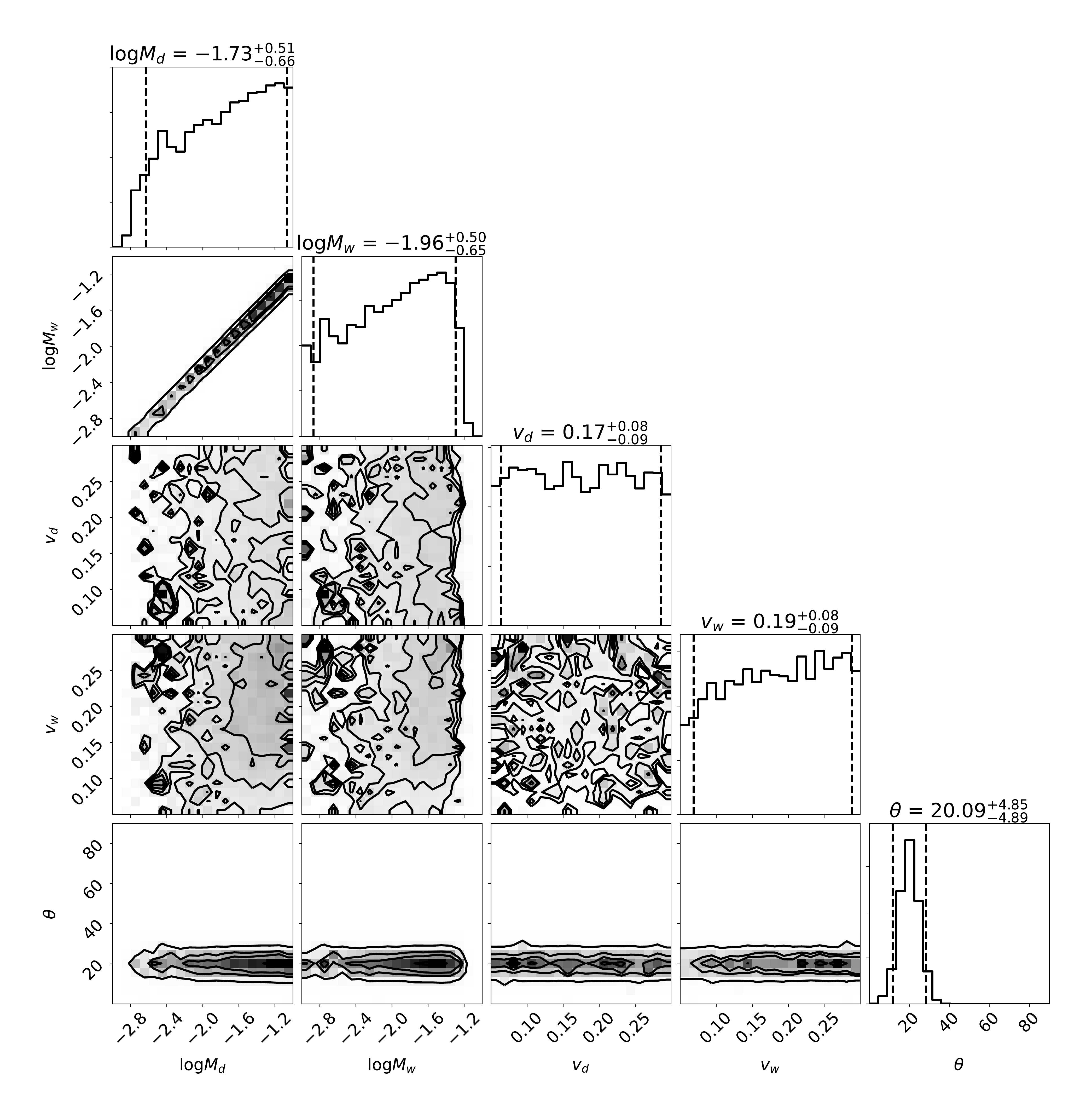}
\caption{Posterior distributions created using only the 2-D \emph{r}-process prior presented in Figure \ref{fig:new_comps}
 with no electromagnetic information provided during sampling besides constraints on the opening angle (see Ref. \cite{Troja_2020} and references therein). \editedthree{We apply a scale factor in our likelihood calculation during inference to prevent underflow from the large residual values.} Note the recovery of the yellow band of lowest-residual mass pairs from Figure \ref{fig:new_comps} in the $M_{\rm{w}}$ vs. $M_{\rm{d}}$ panel as well as the flat velocity posteriors stemming from the lack of velocity constraints introduced by our mass-focused prior. Residual small-scale substructure in the one- and two-dimensional  marginal distributions reflects sample size artifacts.}
\label{fig:corner_plot_prior}
\end{figure}

\begin{figure}[tp!]
\includegraphics[width=\columnwidth]{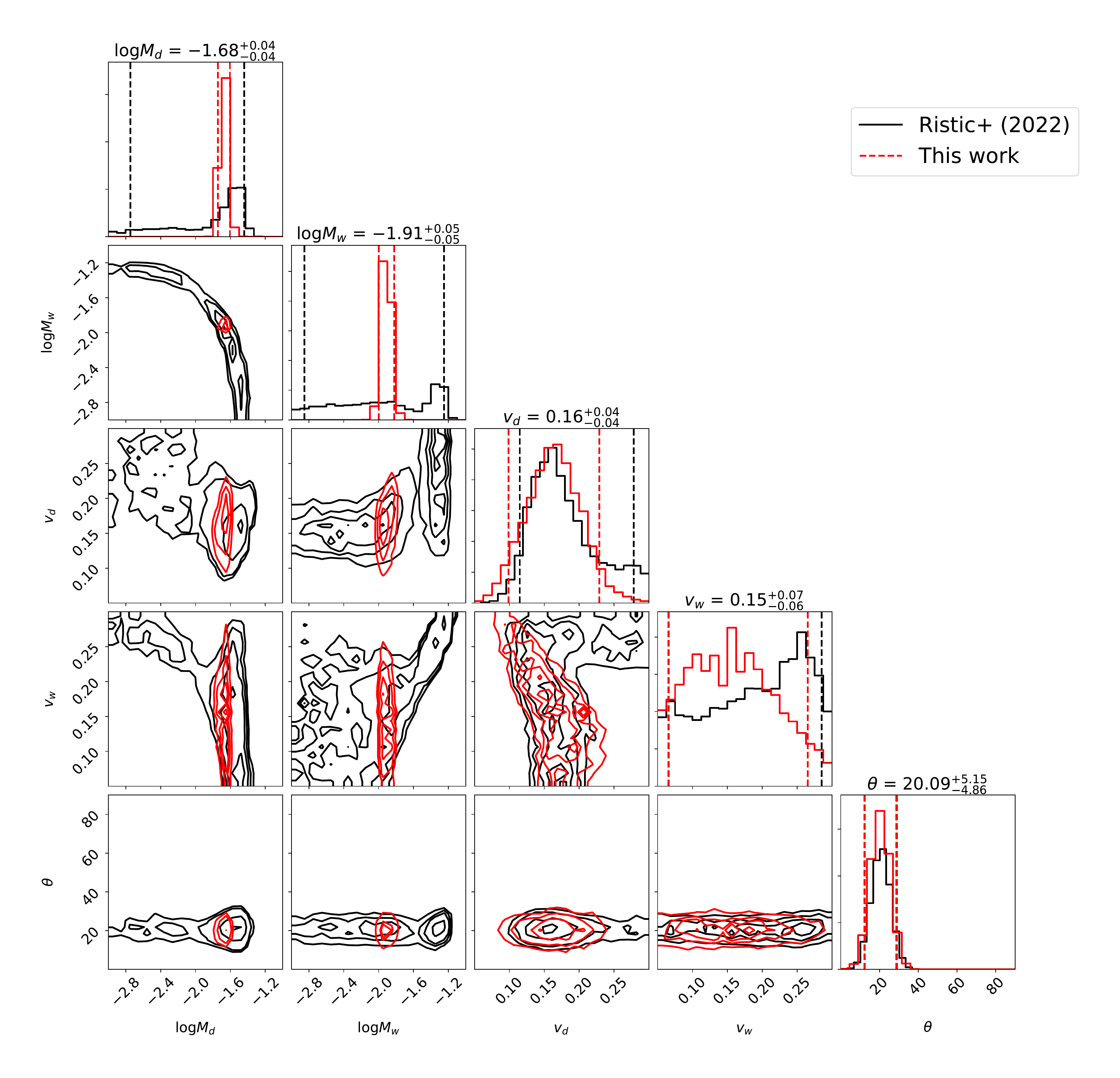}
\caption{Posterior distributions for samples generated when using the $grizyJHK$ bands considered in \cite{Ristic22} (black) and samples generated using the same bands along with the \emph{r}-process prior from Figure \ref{fig:new_comps} (red). \editedthree{The values reported at the top of each posterior distribution represent the inference results from this study.} \editedthree{The composition prior effect is most evident in the wind mass posterior shift to lower ejecta mass.}}
\label{fig:corner_plot}
\end{figure}

\begin{figure}[tp!]
\includegraphics[width=\columnwidth]{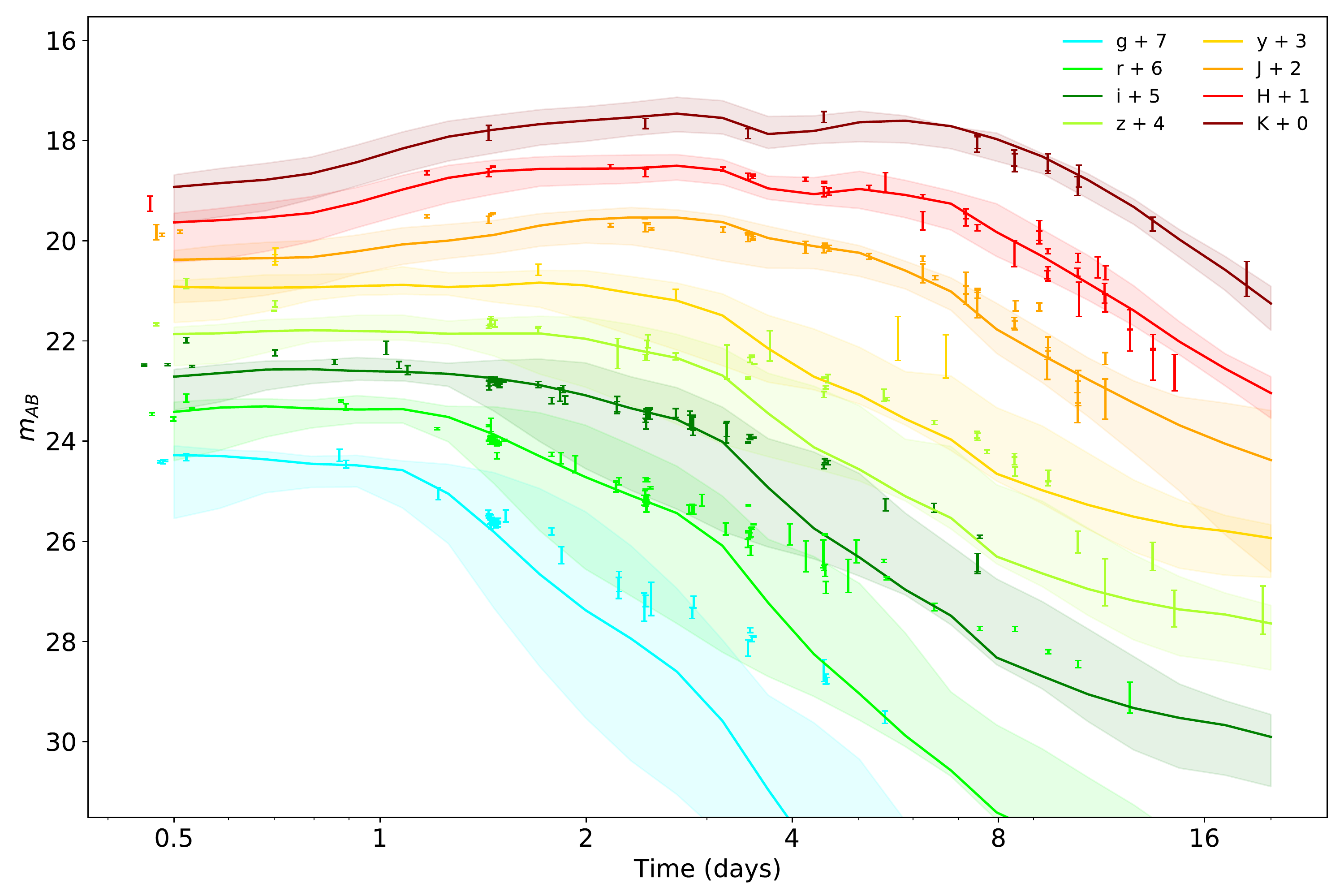}
\caption{\edited{Broadband light-curve predictions for the ejecta parameters recovered in Figure \ref{fig:corner_plot}. The inclusion of the composition-based mass prior (top panel of Figure \ref{fig:new_comps}) \editedthree{reduces model prediction uncertainty} compared to previous predictions \cite{Ristic22}.}}
\label{fig:lc_plot}
\end{figure}

For each set of initial assumptions, the inferred constraint on $M_{\rm{w}}/M_{\rm{d}}$ therefore also strongly constrains the
ingredients powering the associated kilonova. For example, Figure \ref{fig:corner_plot_prior} shows the results of
inferring the parameters of GW170817, using \emph{only} our prior constraints on $M_{\rm{w}}/M_{\rm{d}}$ from the top panel of Figure \ref{fig:new_comps} (and weak constraints on the binary orientation relative to our line of sight). 
Figure \ref{fig:corner_plot} shows how these constraints propagate into joint electromagnetic inference. The solid
black contours show inferences derived without using constraints on $M_{\rm{w}}/M_{\rm{d}}$; the red contours show inferences
supplemented with this insight, for a specific set of initial assumptions.
\editedthree{Figure~\ref{fig:lc_plot} shows the light curves associated with the recovered posterior distributions presented in Figure~\ref{fig:corner_plot}. The inclusion of the composition prior results in much tighter model uncertainties compared to the light-curve fits in Ref. \cite{Ristic22}.}

Each set of our input assumptions about ejecta composition and physics makes a prediction about \emph{r}-process abundances.
As shown by the last column in Tables~\ref{tbl:residuals} and \ref{tbl:yedist}, some of our input assumptions fit better than others. 
Given substantial systematic uncertainties associated in the many assumptions in our study, we approach these nominal
residuals with considerable cautions. However, the minimum residuals presented in Tables~\ref{tbl:residuals} and \ref{tbl:yedist} suggest that the wind2 model is a notably better fit to the solar mass-fraction pattern, consistent with similar findings in previous studies \cite{Evans_2017}.
The nearly distinct separation of the two wind models' lowest residuals implies that the wind1 model is less indicative of \emph{r}-process nucleosynthesis from neutron star mergers; \edited{however, our work neglects consideration of lighter $r$-process elements ($Z \leq 37$) which disfavors compositions with higher $Y_{\rm e}$ like wind1}. More importantly, the separation between the models also implies that new models for the wind ejecta composition need to be considered in comparison to the wind2 model. The results of Tables~\ref{tbl:residuals} and \ref{tbl:yedist} indicate the need for further studies involving updated wind ejecta composition modeling informed by GRMHD disk simulations \cite{2019PhRvD.100b3008M}.

\edited{The results of Tables~\ref{tbl:residuals} and \ref{tbl:yedist} also depend strongly on the assumption that neutron star mergers are the \emph{dominant} \emph{r}-process mechanism for the creation of elements with $Z \geq 37$ (see Figure \ref{fig:metal_poor_stars} and Section \ref{sec:rprocess_prior}), which may not be the case; see, e.g., Ref. \cite{2019ApJ...882...40J} and references therein.}

Our method as stated also assumes that a narrow distribution of mergers in total mass $M_{\rm{tot}} = M_1 + M_2$ dominates nucleosynthesis yields. While self-evidently consistent with the binary neutron star population inferred from the merging Galactic neutron star binaries, this assumption could even still hold for a wider binary neutron star population as suggested by gravitational wave observations, if ejecta are (as expected) suppressed for the most massive mergers with large $M_{\rm{tot}}$. 

Another caveat that presents limitations to our results is that we only
incorporate very specific wind1 and wind2 compositions. There
can be a broad variety of compositions permitted for electron fractions
${Y_{\rm{e}}>0.20}$ due to varying hydrodynamic conditions. 
An extensitve study of these compositions, along with the tests of how much
they can be considered ``representative" of their respective components, is
beyond the scope of this work.

Our results can further be improved by incorporating the observed higher variability 
of the lighter \emph{r}-process abundances between the first and the second 
peak, compared to the universal pattern between the second and third
\emph{r}-process peaks.
The lighter \emph{r}-process as observed in metal-poor stars, exhibits
variation on the order of 1 dex, while the ``strong'' \emph{r}-process
pattern varies by only about 0.3 dex~\cite{2021RvMP...93a5002C}.
An investigation with more accurate numbers based on careful statistical analysis 
of observations will be the subject of future studies~\cite{2022A&A...663A..70F}.

Finally, we note that we cannot ignore the bias introduced by the dynamical ejecta composition of the surrogate kilonova light curves presented in \cite{Ristic22}. While the constraints imposed by the \emph{r}-process abundance prior indeed shift the recovered parameters as in Figure \ref{fig:corner_plot}, there still remains some contribution to the parameter estimation stemming from the surrogate models having been trained on a different dynamical ejecta composition. In other words, our surrogate light curves were trained using the ejecta compositions in Table \ref{tbl:residuals} labeled with (*). Although the primary contribution to the parameter inference in this work comes from the prior discussed in Section \ref{sec:rprocess_prior}, some bias from the surrogates' original compositions is unavoidable.

\section{Conclusion}
\label{sec:conclusion}

We have presented an approach for incorporating nuclear physics-based composition effects as a prior for our kilonova
parameter inference framework.
\editedToo{Identifying a self-consistent electromagnetic and isotopic signature from kilonova models enables us to make
  sharper conclusions about any specific kilonova's ejecta.  
Moreover, our calculations provide a Bayesian evidence, assessing how well both observations can be fit independently and
together.  Our approach may therefore provide a new avenue to directly test whether BNS mergers are the primary source
of \emph{r}-process enrichment, using  any potential  star as a prototype (e.g., the Sun or a metal-poor, \emph{r}-process enriched
example).}

\editedToo{While our self-consistent approach will remain impactful as pertinent inputs improve, of course the
  quantitative numerical results shown here are merely illustrative, given substantial systematics.  Our calculations
  rely on physical models of BNS mergers, ejecta, and the \emph{r}-process abundance signature in solar and
  metal-poor stars, all of which have substantial and widely-investigated systematics.  To highlight one important
  potential source of systematics in this approach, }
we considered a range of models with varying nuclear physics inputs, and, given the assumptions discussed above, the best-fitting model appears to be the one with FRDM2012 nuclear mass, FRLDM fission, $Y_{\rm{e}} = 0.035$ (extreme neutron richness in dynamical ejecta), moderately neutron-rich wind ejecta, \edited{producing inferred dynamical and wind ejecta masses of \editedthree{$M_{\rm{d}} \sim 0.021$ and $M_{\rm{w}} \sim 0.012$ and corresponding to a relatively low mass ratio: $M_{\rm{w}}/M_{\rm{d}} = 0.47$.}} \edited{Our consideration of additional dynamical ejecta compositions, when compared to solar abundances, has indicated that the mass ratio between the two ejecta components is larger than what was implied by previous inference \editedthree{($M_{\rm{w}}/M_{\rm{d}} = 0.24$)}}. However, our conclusions should be taken with care, since the number of input compositions considered were quite limited.

\edited{We have also shown that the inferred mass ratio stemming from a comparison of $r$-process elemental abundances is highly sensitive to the input nuclear physics. For our preferred wind2 model, variations in the dynamical ejecta composition can change the recovered wind-to-dynamical ejecta mass ratio by a factor of \editedthree{$\sim4.3$}. For the wind1 model, the inferred mass ratio can change by a factor of \editedthree{$\sim4689$}, although this is largely due to the assumptions made in this work.}
\editedToo{While we focused on selected nuclear physics systematics, we have identified several areas meriting further
  study, including propagating different choices for nuclear physics uncertainties into our
  parameter inferences; examining whether more complex composition  (e.g., $Y_{\rm e}$), angular, or velocity distributions in
the outflow can mimic these effects; and employing better prototypes than the Sun for a potential \emph{r}-process signature.}

\edited{Even allowing for extremely conservative systematic uncertainties on our inputs (e.g., assuming $M_{\rm{w}}/M_{\rm{d}}$'s optimal value is well-localized between $0.1$ and $10$), these prior abundance constraints should still provide useful insight into kilonova ejecta modeling. For instance, this framework of kilonova surrogates with abundance priors can be used as a constraint to identify merger simulations that produce consistent properties.}

\edited{With the introduction of this composition-based prior, we are able to continue using our existing kilonova surrogate model and parameter inference frameworks while updating our inference priors to match contemporary results in the literature. The ability to update our mass prior using the underlying properties of kilonova models, without requiring expensive simulations (outside of nuclear network outputs), allows us to inexpensively and rapidly update our parameter inference results.}

In this work, we have considered fiducial initial conditions for the outflow, including composition, without allowing for correlations induced by the fact that both the composition and outflows are initialized by binary neutron star mergers. In future work, we will explore self-consistent initialization from merger properties, in particular exploring the effects of binary mass ratio and neutron star remnant lifetime, which should have significant impact on the ejecta amount and composition \cite{2023ApJ...942...39F, 2022arXiv220209437V, 2022MNRAS.510.2804K}.

\section{Acknowledgments}
ROS, MR, and EMH acknowledge support from NSF AST 1909534. ROS acknowledges support from NSF AST 2206321. EMH acknowledges additional support for this work by NASA through the NASA Hubble Fellowship grant HST-HF2-51481.001, awarded by the Space Telescope Science Institute, which is operated by the Association of Universities for Research in Astronomy, Inc., for NASA, under contract NAS5-26555.
CJF, CLF, MRM, OK, RW, and TMS were supported by the US Department of Energy through the Los Alamos National Laboratory. Los Alamos National Laboratory is operated by Triad National Security, LLC, for the National Nuclear Security Administration of U.S. Department of Energy (Contract No. 89233218CNA000001). Research presented in this article was supported by the Laboratory Directed Research and Development program of Los Alamos National Laboratory under project number 20190021DR. This research used resources provided by the Los Alamos National Laboratory Institutional Computing Program, which is supported by the U.S. Department of Energy National Nuclear Security Administration under Contract No.\ 89233218CNA000001.

\begin{table}[ht!]
\begin{tabular}{ccc}
\hline
$Z_{\rm removed}$ & $M_w/M_d$ & Residual \\
\hline
37 & 0.32 & 1466  \\
38 & 0.31 & 1354  \\
39 & 0.33 & 1464  \\
40 & 0.31 & 1463  \\
41 & 0.32 & 1438  \\
42 & 0.32 & 1466  \\
43 & 0.32 & 1467  \\
44 & 0.32 & 1461  \\
45 & 0.32 & 1463  \\
47 & 0.32 & 1466  \\
48 & 0.32 & 1464  \\
49 & 0.32 & 1458  \\
50 & 0.27 & 1466  \\
51 & 0.29 & 1464  \\
52 & 0.33 & 1444  \\
53 & 0.33 & 1460  \\
54 & 0.35 & 1466  \\
55 & 0.33 & 1466  \\
56 & 0.34 & 1454  \\
57 & 0.34 & 1412  \\
58 & 0.33 & 1464  \\
59 & 0.32 & 1465  \\
60 & 0.33 & 1466  \\
61 & 0.33 & 1466  \\
62 & 0.33 & 1466  \\
63 & 0.33 & 1467  \\
64 & 0.33 & 1464  \\
65 & 0.33 & 1456  \\
66 & 0.33 & 1466  \\
67 & 0.33 & 1463  \\
68 & 0.33 & 1466  \\
69 & 0.32 & 1459  \\
70 & 0.33 & 1466  \\
71 & 0.32 & 1467  \\
72 & 0.32 & 1464  \\
\hline
\end{tabular}
\quad
\begin{tabular}{ccc}
\hline
$Z_{\rm removed}$ & $M_w/M_d$ & Residual \\
\hline
73  & 0.33 & 1463 \\
74  & 0.33 & 1464 \\
75  & 0.34 & 1428 \\
76  & 0.37 & 1464 \\
77  & 0.33 & 1459 \\
78  & 0.36 & 1464 \\
79  & 0.33 & 1464 \\
80  & 0.34 & 1447 \\
81  & 0.34 & 1406 \\
82  & 0.35 & 1439 \\
83  & 0.33 & 1459 \\
84  & 0.32 & 1467 \\
85  & 0.32 & 1467 \\
86  & 0.38 & 1112 \\
87  & 0.32 & 1467 \\
88  & 0.31 & 1437 \\
89  & 0.3 & 1413 \\
90  & 0.32 & 1443 \\
91  & 0.31 & 1433 \\
92  & 0.32 & 1447 \\
93  & 0.31 & 1450 \\
94  & 0.32 & 1455 \\
95  & 0.31 & 1452 \\
96  & 0.31 & 1425 \\
97  & 0.31 & 1416 \\
98  & 0.31 & 1436 \\
99  & 0.28 & 1454 \\
100 & 0.26 & 1312  \\
101 & 0.32 & 1467  \\
102 & 0.36 & 1332  \\
103 & 0.32 & 1467  \\
La  & 0.44 & 1372 \\
Ac  & 0.14 & 775 \\
La+Ac & 0.1 & 651  \\
\\
\hline
\end{tabular}
\caption{\editedthree{Sensitivity study results using a leave-one-out approach with specific elements, the lanthanides, the actinides, and both lanthanides and actinides. In each case, the elements in the $Z$ column are removed from consideration during the residual calculation. Interesting cases include $Z=86$ which significantly reduces the calculated residual and yields the largest ratio of $M_w/M_d$, as well as $Z=100$ which yields the lowest ratio. We find that we are particularly sensitive to the actinides in our compositions, as the mass ratio dramatically changes when they are removed from the residual calculation.}}
\label{tbl:sens}
\end{table}

\appendix

\section{Sensitivity Study}

\editedthree{Due to the variability in mass ratios as nuclear physics assumptions change (see Tables~\ref{tbl:residuals} and \ref{tbl:yedist}), we conduct a sensitivity study to identify if certain elements have significant impact on the recovered mass ratio. We modify our best-fitting case with mass model FRDM2012, FRLDM fission, a fixed-$Y_{\rm e}$ of 0.035 and a wind2 composition by choosing specific elements to be omitted during the residual calculation. The results are shown in Table~\ref{tbl:sens}. Most notably, the inclusion (or lack thereof) of actinides strongly influences the preferred mass ratio, dropping it from an average of $M_w/M_d \approx 0.3$ to $M_w/M_d \approx 0.1$. We also note elements of interest $Z=86$ and $Z=100$ which, when removed during the residual calculation, yield the highest and lowest mass ratios, respectively. Our sensitivity to the actinides indicates that our choice of mass ratio during inference should be used with caution. Specifically, the mass prior can provide additional parameter constraints when used self-consistently with the input model nuclear physics, but can have detrimental effects when applied incorrectly. Despite some sensitivity to the actinide composition, our central conclusion that $M_w/M_d < 1$ remains robust.
}

\section{Derivation of Marginalized \\ Abundance Likelihood}
\editedthree{
We assume abundance measurement uncertainties are uncorrelated and individually normally distributed in $\log X$.  To
marginalize considerable uncertainty in the overall normalization of $X$, we marginalize over uncertainty in this
normalization, yielding the effective residual  Eq. (\ref{eq:r}).  In this appendix, we briefly outline the derivation
of this marginal log likelihood.

We consider a signal $y=f_A +F_B \lambda_B +\epsilon$ generated by superposing two known models $f_A$ and $F_B \lambda_B$ with parameters $\lambda_A, \lambda_B$ on top of Gaussian noise $\epsilon$.
In this expression,  $y$ is an N-dimensional vector (i.e., our log
abundance data), $\epsilon$ is normally distributed with mean zero and inverse covariance $\gamma$, and $F \lambda$
are linear operations transforming model parameters $\lambda$ into predictions for $y$.   We want to marginalize out the
impact of the linear model $F_B$ (here, the average abundance).   For this model, 
the log likelihood has the form (up to an overall normalization constant)
\begin{align}
-2 \ln (L p_B(\lambda_B)) &= (f+F_B \lambda_B - y)^T \gamma (f+F_B\lambda_B -y) \nonumber\\&+ \lambda_B \Gamma_B \lambda_B
\end{align}
where we use a normal prior for $\lambda_B $ with mean zero and inverse covariance $\Gamma_B$.  To marginalize over
$\lambda_B$, we complete the square in this gaussian log-likelihood versus $\lambda_B$, then integrate over $\lambda_B$,
arriving at (up to an overall normalization constant)
\begin{align}
-2 \ln L_{\rm marg} &= (f-y)^T G  (f-y) \\
G&= \left[\gamma - \gamma F_B(F_B^T\gamma F_b+\Gamma_B)^{-1} F_B^T\gamma\right]
\end{align}
In our case, $F_B=(1,1,1,1\ldots 1)\equiv \textbf{v} $ is a unit column vector (the same prediction for all element abundances, set to
agree with the  parameter $\lambda_B$),
$\gamma=\textbf{1}/\sigma^2$ is the assumed-common uncertainty in each (log) abundance measurement, and
$\Gamma_B=1/\sigma_B^2$ is our assumed uncertainty in the common matching value.  As  a result, $F^T_B\gamma F_B =
N/\sigma^2$ (a constant, not a matrix); $F^T_B\gamma F_B + \Gamma_B = N/\sigma^2 + 1/\sigma_B^2$; and the combination
appearing in the second term of  $G$ has the form
\begin{align}
\gamma F_B(F_B^T\gamma F_b+\Gamma_B)^{-1} F_B^T\gamma 
= \frac{1}{\sigma^4} {\bf v} {\bf v}^T \frac{1}{\sigma_B^{-2} + N/\sigma^2}.
\end{align}

}

\bibliography{bibliography,thesis,LIGO-publications,gw-astronomy-mergers-ns-gw170817}

\begin{thebibliography}{90}%
\makeatletter
\providecommand \@ifxundefined [1]{%
 \@ifx{#1\undefined}
}%
\providecommand \@ifnum [1]{%
 \ifnum #1\expandafter \@firstoftwo
 \else \expandafter \@secondoftwo
 \fi
}%
\providecommand \@ifx [1]{%
 \ifx #1\expandafter \@firstoftwo
 \else \expandafter \@secondoftwo
 \fi
}%
\providecommand \natexlab [1]{#1}%
\providecommand \enquote  [1]{``#1''}%
\providecommand \bibnamefont  [1]{#1}%
\providecommand \bibfnamefont [1]{#1}%
\providecommand \citenamefont [1]{#1}%
\providecommand \href@noop [0]{\@secondoftwo}%
\providecommand \href [0]{\begingroup \@sanitize@url \@href}%
\providecommand \@href[1]{\@@startlink{#1}\@@href}%
\providecommand \@@href[1]{\endgroup#1\@@endlink}%
\providecommand \@sanitize@url [0]{\catcode `\\12\catcode `\$12\catcode
  `\&12\catcode `\#12\catcode `\^12\catcode `\_12\catcode `\%12\relax}%
\providecommand \@@startlink[1]{}%
\providecommand \@@endlink[0]{}%
\providecommand \url  [0]{\begingroup\@sanitize@url \@url }%
\providecommand \@url [1]{\endgroup\@href {#1}{\urlprefix }}%
\providecommand \urlprefix  [0]{URL }%
\providecommand \Eprint [0]{\href }%
\providecommand \doibase [0]{http://dx.doi.org/}%
\providecommand \selectlanguage [0]{\@gobble}%
\providecommand \bibinfo  [0]{\@secondoftwo}%
\providecommand \bibfield  [0]{\@secondoftwo}%
\providecommand \translation [1]{[#1]}%
\providecommand \BibitemOpen [0]{}%
\providecommand \bibitemStop [0]{}%
\providecommand \bibitemNoStop [0]{.\EOS\space}%
\providecommand \EOS [0]{\spacefactor3000\relax}%
\providecommand \BibitemShut  [1]{\csname bibitem#1\endcsname}%
\let\auto@bib@innerbib\@empty
\bibitem [{\citenamefont {{Hulse}}\ and\ \citenamefont
  {{Taylor}}(1975)}]{hulsetaylor}%
  \BibitemOpen
  \bibfield  {author} {\bibinfo {author} {\bibfnamefont {R.~A.}\ \bibnamefont
  {{Hulse}}}\ and\ \bibinfo {author} {\bibfnamefont {J.~H.}\ \bibnamefont
  {{Taylor}}},\ }\bibfield  {title} {\enquote {\bibinfo {title} {{Discovery of
  a pulsar in a binary system.}}}\ }\href {\doibase 10.1086/181708} {\bibfield
  {journal} {\bibinfo  {journal} {\apjl}\ }\textbf {\bibinfo {volume} {195}},\
  \bibinfo {pages} {L51--L53} (\bibinfo {year} {1975})}\BibitemShut {NoStop}%
\bibitem [{\citenamefont {{Taylor}}\ and\ \citenamefont
  {{Weisberg}}(1982)}]{1982ApJ...253..908T}%
  \BibitemOpen
  \bibfield  {author} {\bibinfo {author} {\bibfnamefont {J.~H.}\ \bibnamefont
  {{Taylor}}}\ and\ \bibinfo {author} {\bibfnamefont {J.~M.}\ \bibnamefont
  {{Weisberg}}},\ }\bibfield  {title} {\enquote {\bibinfo {title} {{A new test
  of general relativity - Gravitational radiation and the binary pulsar PSR
  1913+16}},}\ }\href {\doibase 10.1086/159690} {\bibfield  {journal} {\bibinfo
   {journal} {\apj}\ }\textbf {\bibinfo {volume} {253}},\ \bibinfo {pages}
  {908--920} (\bibinfo {year} {1982})}\BibitemShut {NoStop}%
\bibitem [{\citenamefont {{The LIGO Scientific Collaboration}}\ \emph
  {et~al.}(2017{\natexlab{a}})\citenamefont {{The LIGO Scientific
  Collaboration}}, \citenamefont {{the Virgo Collaboration}}, \citenamefont
  {{Abbott}}, \citenamefont {{Abbott}}, \citenamefont {{Abbott}}, \citenamefont
  {{Acernese}}, \citenamefont {{Ackley}}, \citenamefont {{Adams}},
  \citenamefont {{Adams}}, \citenamefont {{Addesso}} \emph
  {et~al.}}]{LIGO-GW170817-bns}%
  \BibitemOpen
  \bibfield  {author} {\bibinfo {author} {\bibnamefont {{The LIGO Scientific
  Collaboration}}}, \bibinfo {author} {\bibnamefont {{the Virgo
  Collaboration}}}, \bibinfo {author} {\bibfnamefont {B.~P.}\ \bibnamefont
  {{Abbott}}}, \bibinfo {author} {\bibfnamefont {R.}~\bibnamefont {{Abbott}}},
  \bibinfo {author} {\bibfnamefont {T.~D.}\ \bibnamefont {{Abbott}}}, \bibinfo
  {author} {\bibfnamefont {F.}~\bibnamefont {{Acernese}}}, \bibinfo {author}
  {\bibfnamefont {K.}~\bibnamefont {{Ackley}}}, \bibinfo {author}
  {\bibfnamefont {C.}~\bibnamefont {{Adams}}}, \bibinfo {author} {\bibfnamefont
  {T.}~\bibnamefont {{Adams}}}, \bibinfo {author} {\bibfnamefont
  {P.}~\bibnamefont {{Addesso}}},  \emph {et~al.},\ }\bibfield  {title}
  {\enquote {\bibinfo {title} {{GW170817: Observation of gravitational waves
  from a binary neutron star inspiral}},}\ }\href {\doibase
  10.1103/PhysRevLett.119.161101} {\bibfield  {journal} {\bibinfo  {journal}
  {\prl}\ }\textbf {\bibinfo {volume} {119}},\ \bibinfo {pages} {161101}
  (\bibinfo {year} {2017}{\natexlab{a}})}\BibitemShut {NoStop}%
\bibitem [{\citenamefont {{The LIGO Scientific Collaboration}}\ \emph
  {et~al.}(2017{\natexlab{b}})\citenamefont {{The LIGO Scientific
  Collaboration}}, \citenamefont {{the Virgo Collaboration}}, \citenamefont
  {{Abbott}}, \citenamefont {{Abbott}}, \citenamefont {{Abbott}}, \citenamefont
  {{Acernese}}, \citenamefont {{Ackley}}, \citenamefont {{Adams}},
  \citenamefont {{Adams}}, \citenamefont {{Addesso}},\ and\ \citenamefont
  {et~al.}}]{LIGO-GW170817-kilonova}%
  \BibitemOpen
  \bibfield  {author} {\bibinfo {author} {\bibnamefont {{The LIGO Scientific
  Collaboration}}}, \bibinfo {author} {\bibnamefont {{the Virgo
  Collaboration}}}, \bibinfo {author} {\bibfnamefont {B.~P.}\ \bibnamefont
  {{Abbott}}}, \bibinfo {author} {\bibfnamefont {R.}~\bibnamefont {{Abbott}}},
  \bibinfo {author} {\bibfnamefont {T.~D.}\ \bibnamefont {{Abbott}}}, \bibinfo
  {author} {\bibfnamefont {F.}~\bibnamefont {{Acernese}}}, \bibinfo {author}
  {\bibfnamefont {K.}~\bibnamefont {{Ackley}}}, \bibinfo {author}
  {\bibfnamefont {C.}~\bibnamefont {{Adams}}}, \bibinfo {author} {\bibfnamefont
  {T.}~\bibnamefont {{Adams}}}, \bibinfo {author} {\bibfnamefont
  {P.}~\bibnamefont {{Addesso}}}, \ and\ \bibinfo {author} {\bibnamefont
  {et~al.}},\ }\bibfield  {title} {\enquote {\bibinfo {title} {{Estimating the
  Contribution of Dynamical Ejecta in the Kilonova Associated with
  GW170817}},}\ }\href {\doibase 10.3847/2041-8213/aa9478} {\bibfield
  {journal} {\bibinfo  {journal} {\apjl}\ }\textbf {\bibinfo {volume} {850}},\
  \bibinfo {eid} {L39} (\bibinfo {year} {2017}{\natexlab{b}})},\ \Eprint
  {http://arxiv.org/abs/1710.05836} {arXiv:1710.05836 [astro-ph.HE]}
  \BibitemShut {NoStop}%
\bibitem [{\citenamefont {{The LIGO Scientific Collaboration}}\ \emph
  {et~al.}(2017{\natexlab{c}})\citenamefont {{The LIGO Scientific
  Collaboration}}, \citenamefont {{the Virgo Collaboration}}, \citenamefont
  {{Abbott}}, \citenamefont {{Abbott}}, \citenamefont {{Abbott}}, \citenamefont
  {{Acernese}}, \citenamefont {{Ackley}}, \citenamefont {{Adams}},
  \citenamefont {{Adams}}, \citenamefont {{Addesso}},\ and\ \citenamefont
  {et~al.}}]{LIGO-GW170817-mma}%
  \BibitemOpen
  \bibfield  {author} {\bibinfo {author} {\bibnamefont {{The LIGO Scientific
  Collaboration}}}, \bibinfo {author} {\bibnamefont {{the Virgo
  Collaboration}}}, \bibinfo {author} {\bibfnamefont {B.~P.}\ \bibnamefont
  {{Abbott}}}, \bibinfo {author} {\bibfnamefont {R.}~\bibnamefont {{Abbott}}},
  \bibinfo {author} {\bibfnamefont {T.~D.}\ \bibnamefont {{Abbott}}}, \bibinfo
  {author} {\bibfnamefont {F.}~\bibnamefont {{Acernese}}}, \bibinfo {author}
  {\bibfnamefont {K.}~\bibnamefont {{Ackley}}}, \bibinfo {author}
  {\bibfnamefont {C.}~\bibnamefont {{Adams}}}, \bibinfo {author} {\bibfnamefont
  {T.}~\bibnamefont {{Adams}}}, \bibinfo {author} {\bibfnamefont
  {P.}~\bibnamefont {{Addesso}}}, \ and\ \bibinfo {author} {\bibnamefont
  {et~al.}},\ }\bibfield  {title} {\enquote {\bibinfo {title} {{Multi-messenger
  Observations of a Binary Neutron Star Merger}},}\ }\href {\doibase
  doi:10.3847/2041-8213/aa91c9} {\bibfield  {journal} {\bibinfo  {journal}
  {\apjl}\ } (\bibinfo {year} {2017}{\natexlab{c}}),\
  doi:10.3847/2041-8213/aa91c9}\BibitemShut {NoStop}%
\bibitem [{\citenamefont {{The LIGO Scientific Collaboration}}\ \emph
  {et~al.}(2017{\natexlab{d}})\citenamefont {{The LIGO Scientific
  Collaboration}}, \citenamefont {{the Virgo Collaboration}}, \citenamefont
  {{Abbott}}, \citenamefont {{Abbott}}, \citenamefont {{Abbott}}, \citenamefont
  {{Acernese}}, \citenamefont {{Ackley}}, \citenamefont {{Adams}},
  \citenamefont {{Adams}}, \citenamefont {{Addesso}},\ and\ \citenamefont
  {et~al.}}]{LIGO-GW170817-astro}%
  \BibitemOpen
  \bibfield  {author} {\bibinfo {author} {\bibnamefont {{The LIGO Scientific
  Collaboration}}}, \bibinfo {author} {\bibnamefont {{the Virgo
  Collaboration}}}, \bibinfo {author} {\bibfnamefont {B.~P.}\ \bibnamefont
  {{Abbott}}}, \bibinfo {author} {\bibfnamefont {R.}~\bibnamefont {{Abbott}}},
  \bibinfo {author} {\bibfnamefont {T.~D.}\ \bibnamefont {{Abbott}}}, \bibinfo
  {author} {\bibfnamefont {F.}~\bibnamefont {{Acernese}}}, \bibinfo {author}
  {\bibfnamefont {K.}~\bibnamefont {{Ackley}}}, \bibinfo {author}
  {\bibfnamefont {C.}~\bibnamefont {{Adams}}}, \bibinfo {author} {\bibfnamefont
  {T.}~\bibnamefont {{Adams}}}, \bibinfo {author} {\bibfnamefont
  {P.}~\bibnamefont {{Addesso}}}, \ and\ \bibinfo {author} {\bibnamefont
  {et~al.}},\ }\bibfield  {title} {\enquote {\bibinfo {title} {{On the
  Progenitor of Binary Neutron Star Merger GW170817}},}\ }\href {\doibase
  10.3847/2041-8213/aa93fc} {\bibfield  {journal} {\bibinfo  {journal} {\apjl}\
  }\textbf {\bibinfo {volume} {850}},\ \bibinfo {eid} {L40} (\bibinfo {year}
  {2017}{\natexlab{d}})},\ \Eprint {http://arxiv.org/abs/1710.05838}
  {arXiv:1710.05838 [astro-ph.HE]} \BibitemShut {NoStop}%
\bibitem [{\citenamefont {Tanvir}\ \emph {et~al.}(2017)\citenamefont {Tanvir},
  \citenamefont {Levan}, \citenamefont {Gonz{\'{a}}lez-Fern{\'{a}}ndez},
  \citenamefont {Korobkin}, \citenamefont {Mandel}, \citenamefont {Rosswog},
  \citenamefont {Hjorth}, \citenamefont {D'Avanzo}, \citenamefont {Fruchter},
  \citenamefont {Fryer} \emph {et~al.}}]{Tanvir_2017}%
  \BibitemOpen
  \bibfield  {author} {\bibinfo {author} {\bibfnamefont {N.~R.}\ \bibnamefont
  {Tanvir}}, \bibinfo {author} {\bibfnamefont {A.~J.}\ \bibnamefont {Levan}},
  \bibinfo {author} {\bibfnamefont {C.}~\bibnamefont
  {Gonz{\'{a}}lez-Fern{\'{a}}ndez}}, \bibinfo {author} {\bibfnamefont
  {O.}~\bibnamefont {Korobkin}}, \bibinfo {author} {\bibfnamefont
  {I.}~\bibnamefont {Mandel}}, \bibinfo {author} {\bibfnamefont
  {S.}~\bibnamefont {Rosswog}}, \bibinfo {author} {\bibfnamefont
  {J.}~\bibnamefont {Hjorth}}, \bibinfo {author} {\bibfnamefont
  {P.}~\bibnamefont {D'Avanzo}}, \bibinfo {author} {\bibfnamefont {A.~S.}\
  \bibnamefont {Fruchter}}, \bibinfo {author} {\bibfnamefont {C.~L.}\
  \bibnamefont {Fryer}},  \emph {et~al.},\ }\bibfield  {title} {\enquote
  {\bibinfo {title} {The emergence of a lanthanide-rich kilonova following the
  merger of two neutron stars},}\ }\href {\doibase 10.3847/2041-8213/aa90b6}
  {\bibfield  {journal} {\bibinfo  {journal} {The Astrophysical Journal}\
  }\textbf {\bibinfo {volume} {848}},\ \bibinfo {pages} {L27} (\bibinfo {year}
  {2017})}\BibitemShut {NoStop}%
\bibitem [{\citenamefont {{Lattimer}}\ and\ \citenamefont
  {{Schramm}}(1974)}]{1974ApJ...192L.145L}%
  \BibitemOpen
  \bibfield  {author} {\bibinfo {author} {\bibfnamefont {J.~M.}\ \bibnamefont
  {{Lattimer}}}\ and\ \bibinfo {author} {\bibfnamefont {D.~N.}\ \bibnamefont
  {{Schramm}}},\ }\bibfield  {title} {\enquote {\bibinfo {title}
  {{Black-Hole-Neutron-Star Collisions}},}\ }\href {\doibase 10.1086/181612}
  {\bibfield  {journal} {\bibinfo  {journal} {\apjl}\ }\textbf {\bibinfo
  {volume} {192}},\ \bibinfo {pages} {L145} (\bibinfo {year}
  {1974})}\BibitemShut {NoStop}%
\bibitem [{\citenamefont {{Lattimer}}\ and\ \citenamefont
  {{Schramm}}(1976)}]{1976ApJ...210..549L}%
  \BibitemOpen
  \bibfield  {author} {\bibinfo {author} {\bibfnamefont {J.~M.}\ \bibnamefont
  {{Lattimer}}}\ and\ \bibinfo {author} {\bibfnamefont {D.~N.}\ \bibnamefont
  {{Schramm}}},\ }\bibfield  {title} {\enquote {\bibinfo {title} {{The tidal
  disruption of neutron stars by black holes in close binaries.}}}\ }\href
  {\doibase 10.1086/154860} {\bibfield  {journal} {\bibinfo  {journal} {\apj}\
  }\textbf {\bibinfo {volume} {210}},\ \bibinfo {pages} {549--567} (\bibinfo
  {year} {1976})}\BibitemShut {NoStop}%
\bibitem [{\citenamefont {{Symbalisty}}\ and\ \citenamefont
  {{Schramm}}(1982)}]{1982ApL....22..143S}%
  \BibitemOpen
  \bibfield  {author} {\bibinfo {author} {\bibfnamefont {E.}~\bibnamefont
  {{Symbalisty}}}\ and\ \bibinfo {author} {\bibfnamefont {D.~N.}\ \bibnamefont
  {{Schramm}}},\ }\bibfield  {title} {\enquote {\bibinfo {title} {{Neutron Star
  Collisions and the r-Process}},}\ }\href@noop {} {\bibfield  {journal}
  {\bibinfo  {journal} {\apjl}\ }\textbf {\bibinfo {volume} {22}},\ \bibinfo
  {pages} {143} (\bibinfo {year} {1982})}\BibitemShut {NoStop}%
\bibitem [{\citenamefont {{Eichler}}\ \emph {et~al.}(1989)\citenamefont
  {{Eichler}}, \citenamefont {{Livio}}, \citenamefont {{Piran}},\ and\
  \citenamefont {{Schramm}}}]{1989Natur.340..126E}%
  \BibitemOpen
  \bibfield  {author} {\bibinfo {author} {\bibfnamefont {David}\ \bibnamefont
  {{Eichler}}}, \bibinfo {author} {\bibfnamefont {Mario}\ \bibnamefont
  {{Livio}}}, \bibinfo {author} {\bibfnamefont {Tsvi}\ \bibnamefont {{Piran}}},
  \ and\ \bibinfo {author} {\bibfnamefont {David~N.}\ \bibnamefont
  {{Schramm}}},\ }\bibfield  {title} {\enquote {\bibinfo {title}
  {{Nucleosynthesis, neutrino bursts and {\ensuremath{\gamma}}-rays from
  coalescing neutron stars}},}\ }\href {\doibase 10.1038/340126a0} {\bibfield
  {journal} {\bibinfo  {journal} {\nat}\ }\textbf {\bibinfo {volume} {340}},\
  \bibinfo {pages} {126--128} (\bibinfo {year} {1989})}\BibitemShut {NoStop}%
\bibitem [{\citenamefont {{Rosswog}}(2015)}]{2015IJMPD..2430012R}%
  \BibitemOpen
  \bibfield  {author} {\bibinfo {author} {\bibfnamefont {Stephan}\ \bibnamefont
  {{Rosswog}}},\ }\bibfield  {title} {\enquote {\bibinfo {title} {{The
  multi-messenger picture of compact binary mergers}},}\ }\href {\doibase
  10.1142/S0218271815300128} {\bibfield  {journal} {\bibinfo  {journal}
  {International Journal of Modern Physics D}\ }\textbf {\bibinfo {volume}
  {24}},\ \bibinfo {eid} {1530012-52} (\bibinfo {year} {2015})},\ \Eprint
  {http://arxiv.org/abs/1501.02081} {arXiv:1501.02081 [astro-ph.HE]}
  \BibitemShut {NoStop}%
\bibitem [{\citenamefont {{Li}}\ and\ \citenamefont
  {{Paczy{\'n}ski}}(1998)}]{LiLX1998}%
  \BibitemOpen
  \bibfield  {author} {\bibinfo {author} {\bibfnamefont {Li-Xin}\ \bibnamefont
  {{Li}}}\ and\ \bibinfo {author} {\bibfnamefont {Bohdan}\ \bibnamefont
  {{Paczy{\'n}ski}}},\ }\bibfield  {title} {\enquote {\bibinfo {title}
  {{Transient Events from Neutron Star Mergers}},}\ }\href {\doibase
  10.1086/311680} {\bibfield  {journal} {\bibinfo  {journal} {\apjl}\ }\textbf
  {\bibinfo {volume} {507}},\ \bibinfo {pages} {L59--L62} (\bibinfo {year}
  {1998})}\BibitemShut {NoStop}%
\bibitem [{\citenamefont {{Kulkarni}}(2005)}]{Kulkarni2005}%
  \BibitemOpen
  \bibfield  {author} {\bibinfo {author} {\bibfnamefont {S.~R.}\ \bibnamefont
  {{Kulkarni}}},\ }\bibfield  {title} {\enquote {\bibinfo {title} {{Modeling
  Supernova-like Explosions Associated with Gamma-ray Bursts with Short
  Durations}},}\ }\href@noop {} {\bibfield  {journal} {\bibinfo  {journal}
  {arXiv e-prints}\ ,\ \bibinfo {eid} {astro-ph/0510256}} (\bibinfo {year}
  {2005})},\ \Eprint {http://arxiv.org/abs/astro-ph/0510256}
  {arXiv:astro-ph/0510256 [astro-ph]} \BibitemShut {NoStop}%
\bibitem [{\citenamefont {{Metzger}}\ \emph {et~al.}(2010)\citenamefont
  {{Metzger}}, \citenamefont {{Mart{\'{\i}}nez-Pinedo}}, \citenamefont
  {{Darbha}}, \citenamefont {{Quataert}}, \citenamefont {{Arcones}},
  \citenamefont {{Kasen}}, \citenamefont {{Thomas}}, \citenamefont {{Nugent}},
  \citenamefont {{Panov}},\ and\ \citenamefont
  {{Zinner}}}]{short-grb-Metzger-EMCounterpartViaRProcess-2010}%
  \BibitemOpen
  \bibfield  {author} {\bibinfo {author} {\bibfnamefont {B.~D.}\ \bibnamefont
  {{Metzger}}}, \bibinfo {author} {\bibfnamefont {G.}~\bibnamefont
  {{Mart{\'{\i}}nez-Pinedo}}}, \bibinfo {author} {\bibfnamefont
  {S.}~\bibnamefont {{Darbha}}}, \bibinfo {author} {\bibfnamefont
  {E.}~\bibnamefont {{Quataert}}}, \bibinfo {author} {\bibfnamefont
  {A.}~\bibnamefont {{Arcones}}}, \bibinfo {author} {\bibfnamefont
  {D.}~\bibnamefont {{Kasen}}}, \bibinfo {author} {\bibfnamefont
  {R.}~\bibnamefont {{Thomas}}}, \bibinfo {author} {\bibfnamefont
  {P.}~\bibnamefont {{Nugent}}}, \bibinfo {author} {\bibfnamefont {I.~V.}\
  \bibnamefont {{Panov}}}, \ and\ \bibinfo {author} {\bibfnamefont {N.~T.}\
  \bibnamefont {{Zinner}}},\ }\bibfield  {title} {\enquote {\bibinfo {title}
  {{Electromagnetic counterparts of compact object mergers powered by the
  radioactive decay of r-process nuclei}},}\ }\href {\doibase
  10.1111/j.1365-2966.2010.16864.x} {\bibfield  {journal} {\bibinfo  {journal}
  {\mnras}\ ,\ \bibinfo {pages} {840--+}} (\bibinfo {year} {2010})}\BibitemShut
  {NoStop}%
\bibitem [{\citenamefont {Metzger}(2019)}]{Metzger_LRR_2019}%
  \BibitemOpen
  \bibfield  {author} {\bibinfo {author} {\bibfnamefont {Brian~D.}\
  \bibnamefont {Metzger}},\ }\bibfield  {title} {\enquote {\bibinfo {title}
  {Kilonovae},}\ }\href {\doibase 10.1007/s41114-019-0024-0} {\bibfield
  {journal} {\bibinfo  {journal} {Living Reviews in Relativity}\ }\textbf
  {\bibinfo {volume} {23}} (\bibinfo {year} {2019}),\
  10.1007/s41114-019-0024-0}\BibitemShut {NoStop}%
\bibitem [{\citenamefont {{Roberts}}\ \emph {et~al.}(2011)\citenamefont
  {{Roberts}}, \citenamefont {{Kasen}}, \citenamefont {{Lee}},\ and\
  \citenamefont {{Ramirez-Ruiz}}}]{Roberts2011}%
  \BibitemOpen
  \bibfield  {author} {\bibinfo {author} {\bibfnamefont {L.~F.}\ \bibnamefont
  {{Roberts}}}, \bibinfo {author} {\bibfnamefont {D.}~\bibnamefont {{Kasen}}},
  \bibinfo {author} {\bibfnamefont {W.~H.}\ \bibnamefont {{Lee}}}, \ and\
  \bibinfo {author} {\bibfnamefont {E.}~\bibnamefont {{Ramirez-Ruiz}}},\
  }\bibfield  {title} {\enquote {\bibinfo {title} {{Electromagnetic Transients
  Powered by Nuclear Decay in the Tidal Tails of Coalescing Compact
  Binaries}},}\ }\href {\doibase 10.1088/2041-8205/736/1/L21} {\bibfield
  {journal} {\bibinfo  {journal} {\apjl}\ }\textbf {\bibinfo {volume} {736}},\
  \bibinfo {eid} {L21} (\bibinfo {year} {2011})},\ \Eprint
  {http://arxiv.org/abs/1104.5504} {arXiv:1104.5504 [astro-ph.HE]} \BibitemShut
  {NoStop}%
\bibitem [{\citenamefont {{Goriely}}\ \emph {et~al.}(2011)\citenamefont
  {{Goriely}}, \citenamefont {{Bauswein}},\ and\ \citenamefont
  {{Janka}}}]{Goriely2011}%
  \BibitemOpen
  \bibfield  {author} {\bibinfo {author} {\bibfnamefont {Stephane}\
  \bibnamefont {{Goriely}}}, \bibinfo {author} {\bibfnamefont {Andreas}\
  \bibnamefont {{Bauswein}}}, \ and\ \bibinfo {author} {\bibfnamefont
  {Hans-Thomas}\ \bibnamefont {{Janka}}},\ }\bibfield  {title} {\enquote
  {\bibinfo {title} {{r-process Nucleosynthesis in Dynamically Ejected Matter
  of Neutron Star Mergers}},}\ }\href {\doibase 10.1088/2041-8205/738/2/L32}
  {\bibfield  {journal} {\bibinfo  {journal} {\apjl}\ }\textbf {\bibinfo
  {volume} {738}},\ \bibinfo {eid} {L32} (\bibinfo {year} {2011})},\ \Eprint
  {http://arxiv.org/abs/1107.0899} {arXiv:1107.0899 [astro-ph.SR]} \BibitemShut
  {NoStop}%
\bibitem [{\citenamefont {{Metzger}}\ and\ \citenamefont
  {{Berger}}(2012)}]{short-grb-GWCoincidenceEM-MetzgerBerger2011}%
  \BibitemOpen
  \bibfield  {author} {\bibinfo {author} {\bibfnamefont {B.~D.}\ \bibnamefont
  {{Metzger}}}\ and\ \bibinfo {author} {\bibfnamefont {E.}~\bibnamefont
  {{Berger}}},\ }\bibfield  {title} {\enquote {\bibinfo {title} {{What is the
  Most Promising Electromagnetic Counterpart of a Neutron Star Binary
  Merger?}}}\ }\href {\doibase 10.1088/0004-637X/746/1/48} {\bibfield
  {journal} {\bibinfo  {journal} {\apj}\ }\textbf {\bibinfo {volume} {746}},\
  \bibinfo {eid} {48} (\bibinfo {year} {2012})}\BibitemShut {NoStop}%
\bibitem [{\citenamefont {{Korobkin}}\ \emph {et~al.}(2012)\citenamefont
  {{Korobkin}}, \citenamefont {{Rosswog}}, \citenamefont {{Arcones}},\ and\
  \citenamefont {{Winteler}}}]{Korobkin_2012}%
  \BibitemOpen
  \bibfield  {author} {\bibinfo {author} {\bibfnamefont {O.}~\bibnamefont
  {{Korobkin}}}, \bibinfo {author} {\bibfnamefont {S.}~\bibnamefont
  {{Rosswog}}}, \bibinfo {author} {\bibfnamefont {A.}~\bibnamefont
  {{Arcones}}}, \ and\ \bibinfo {author} {\bibfnamefont {C.}~\bibnamefont
  {{Winteler}}},\ }\bibfield  {title} {\enquote {\bibinfo {title} {{On the
  astrophysical robustness of the neutron star merger r-process}},}\ }\href
  {\doibase 10.1111/j.1365-2966.2012.21859.x} {\bibfield  {journal} {\bibinfo
  {journal} {Monthly Notices of the Royal Astronomical Society}\ }\textbf
  {\bibinfo {volume} {426}},\ \bibinfo {pages} {1940--1949} (\bibinfo {year}
  {2012})}\BibitemShut {NoStop}%
\bibitem [{\citenamefont {{Cowan}}\ \emph {et~al.}(2021)\citenamefont
  {{Cowan}}, \citenamefont {{Sneden}}, \citenamefont {{Lawler}}, \citenamefont
  {{Aprahamian}}, \citenamefont {{Wiescher}}, \citenamefont {{Langanke}},
  \citenamefont {{Mart{\'\i}nez-Pinedo}},\ and\ \citenamefont
  {{Thielemann}}}]{2021RvMP...93a5002C}%
  \BibitemOpen
  \bibfield  {author} {\bibinfo {author} {\bibfnamefont {John~J.}\ \bibnamefont
  {{Cowan}}}, \bibinfo {author} {\bibfnamefont {Christopher}\ \bibnamefont
  {{Sneden}}}, \bibinfo {author} {\bibfnamefont {James~E.}\ \bibnamefont
  {{Lawler}}}, \bibinfo {author} {\bibfnamefont {Ani}\ \bibnamefont
  {{Aprahamian}}}, \bibinfo {author} {\bibfnamefont {Michael}\ \bibnamefont
  {{Wiescher}}}, \bibinfo {author} {\bibfnamefont {Karlheinz}\ \bibnamefont
  {{Langanke}}}, \bibinfo {author} {\bibfnamefont {Gabriel}\ \bibnamefont
  {{Mart{\'\i}nez-Pinedo}}}, \ and\ \bibinfo {author} {\bibfnamefont
  {Friedrich-Karl}\ \bibnamefont {{Thielemann}}},\ }\bibfield  {title}
  {\enquote {\bibinfo {title} {{Origin of the heaviest elements: The rapid
  neutron-capture process}},}\ }\href {\doibase 10.1103/RevModPhys.93.015002}
  {\bibfield  {journal} {\bibinfo  {journal} {Reviews of Modern Physics}\
  }\textbf {\bibinfo {volume} {93}},\ \bibinfo {eid} {015002} (\bibinfo {year}
  {2021})},\ \Eprint {http://arxiv.org/abs/1901.01410} {arXiv:1901.01410
  [astro-ph.HE]} \BibitemShut {NoStop}%
\bibitem [{\citenamefont {{Abbott}}\ \emph {et~al.}(2017)\citenamefont
  {{Abbott}}, \citenamefont {{Abbott}}, \citenamefont {{Abbott}}, \citenamefont
  {{Acernese}}, \citenamefont {{Ackley}}, \citenamefont {{Adams}},
  \citenamefont {{Adams}}, \citenamefont {{Addesso}}, \citenamefont
  {{Adhikari}}, \citenamefont {{Adya}},\ and\ \citenamefont
  {et~al.}}]{2017ApJ...848L..13A}%
  \BibitemOpen
  \bibfield  {author} {\bibinfo {author} {\bibfnamefont {B.~P.}\ \bibnamefont
  {{Abbott}}}, \bibinfo {author} {\bibfnamefont {R.}~\bibnamefont {{Abbott}}},
  \bibinfo {author} {\bibfnamefont {T.~D.}\ \bibnamefont {{Abbott}}}, \bibinfo
  {author} {\bibfnamefont {F.}~\bibnamefont {{Acernese}}}, \bibinfo {author}
  {\bibfnamefont {K.}~\bibnamefont {{Ackley}}}, \bibinfo {author}
  {\bibfnamefont {C.}~\bibnamefont {{Adams}}}, \bibinfo {author} {\bibfnamefont
  {T.}~\bibnamefont {{Adams}}}, \bibinfo {author} {\bibfnamefont
  {P.}~\bibnamefont {{Addesso}}}, \bibinfo {author} {\bibfnamefont {R.~X.}\
  \bibnamefont {{Adhikari}}}, \bibinfo {author} {\bibfnamefont {V.~B.}\
  \bibnamefont {{Adya}}}, \ and\ \bibinfo {author} {\bibnamefont {et~al.}},\
  }\bibfield  {title} {\enquote {\bibinfo {title} {{Gravitational Waves and
  Gamma-Rays from a Binary Neutron Star Merger: GW170817 and GRB 170817A}},}\
  }\href {\doibase 10.3847/2041-8213/aa920c} {\bibfield  {journal} {\bibinfo
  {journal} {\apjl}\ }\textbf {\bibinfo {volume} {848}},\ \bibinfo {eid} {L13}
  (\bibinfo {year} {2017})},\ \Eprint {http://arxiv.org/abs/1710.05834}
  {arXiv:1710.05834 [astro-ph.HE]} \BibitemShut {NoStop}%
\bibitem [{\citenamefont {{Savchenko}}\ \emph {et~al.}(2017)\citenamefont
  {{Savchenko}}, \citenamefont {{Ferrigno}}, \citenamefont {{Kuulkers}},
  \citenamefont {{Bazzano}}, \citenamefont {{Bozzo}}, \citenamefont {{Brandt}},
  \citenamefont {{Chenevez}}, \citenamefont {{Courvoisier}}, \citenamefont
  {{Diehl}}, \citenamefont {{Domingo}}, \citenamefont {{Hanlon}}, \citenamefont
  {{Jourdain}}, \citenamefont {{von Kienlin}}, \citenamefont {{Laurent}},
  \citenamefont {{Lebrun}}, \citenamefont {{Lutovinov}}, \citenamefont
  {{Martin-Carrillo}}, \citenamefont {{Mereghetti}}, \citenamefont
  {{Natalucci}}, \citenamefont {{Rodi}}, \citenamefont {{Roques}},
  \citenamefont {{Sunyaev}},\ and\ \citenamefont
  {{Ubertini}}}]{2017ApJ...848L..15S}%
  \BibitemOpen
  \bibfield  {author} {\bibinfo {author} {\bibfnamefont {V.}~\bibnamefont
  {{Savchenko}}}, \bibinfo {author} {\bibfnamefont {C.}~\bibnamefont
  {{Ferrigno}}}, \bibinfo {author} {\bibfnamefont {E.}~\bibnamefont
  {{Kuulkers}}}, \bibinfo {author} {\bibfnamefont {A.}~\bibnamefont
  {{Bazzano}}}, \bibinfo {author} {\bibfnamefont {E.}~\bibnamefont {{Bozzo}}},
  \bibinfo {author} {\bibfnamefont {S.}~\bibnamefont {{Brandt}}}, \bibinfo
  {author} {\bibfnamefont {J.}~\bibnamefont {{Chenevez}}}, \bibinfo {author}
  {\bibfnamefont {T.~J.~L.}\ \bibnamefont {{Courvoisier}}}, \bibinfo {author}
  {\bibfnamefont {R.}~\bibnamefont {{Diehl}}}, \bibinfo {author} {\bibfnamefont
  {A.}~\bibnamefont {{Domingo}}}, \bibinfo {author} {\bibfnamefont
  {L.}~\bibnamefont {{Hanlon}}}, \bibinfo {author} {\bibfnamefont
  {E.}~\bibnamefont {{Jourdain}}}, \bibinfo {author} {\bibfnamefont
  {A.}~\bibnamefont {{von Kienlin}}}, \bibinfo {author} {\bibfnamefont
  {P.}~\bibnamefont {{Laurent}}}, \bibinfo {author} {\bibfnamefont
  {F.}~\bibnamefont {{Lebrun}}}, \bibinfo {author} {\bibfnamefont
  {A.}~\bibnamefont {{Lutovinov}}}, \bibinfo {author} {\bibfnamefont
  {A.}~\bibnamefont {{Martin-Carrillo}}}, \bibinfo {author} {\bibfnamefont
  {S.}~\bibnamefont {{Mereghetti}}}, \bibinfo {author} {\bibfnamefont
  {L.}~\bibnamefont {{Natalucci}}}, \bibinfo {author} {\bibfnamefont
  {J.}~\bibnamefont {{Rodi}}}, \bibinfo {author} {\bibfnamefont {J.~P.}\
  \bibnamefont {{Roques}}}, \bibinfo {author} {\bibfnamefont {R.}~\bibnamefont
  {{Sunyaev}}}, \ and\ \bibinfo {author} {\bibfnamefont {P.}~\bibnamefont
  {{Ubertini}}},\ }\bibfield  {title} {\enquote {\bibinfo {title} {{INTEGRAL
  Detection of the First Prompt Gamma-Ray Signal Coincident with the
  Gravitational-wave Event GW170817}},}\ }\href {\doibase
  10.3847/2041-8213/aa8f94} {\bibfield  {journal} {\bibinfo  {journal} {\apjl}\
  }\textbf {\bibinfo {volume} {848}},\ \bibinfo {eid} {L15} (\bibinfo {year}
  {2017})},\ \Eprint {http://arxiv.org/abs/1710.05449} {arXiv:1710.05449
  [astro-ph.HE]} \BibitemShut {NoStop}%
\bibitem [{\citenamefont {{Cowperthwaite}}\ \emph {et~al.}(2017)\citenamefont
  {{Cowperthwaite}}, \citenamefont {{Berger}}, \citenamefont {{Villar}},
  \citenamefont {{Metzger}}, \citenamefont {{Nicholl}}, \citenamefont
  {{Chornock}}, \citenamefont {{Blanchard}}, \citenamefont {{Fong}},
  \citenamefont {{Margutti}}, \citenamefont {{Soares-Santos}} \emph
  {et~al.}}]{Cowperthwaite2017}%
  \BibitemOpen
  \bibfield  {author} {\bibinfo {author} {\bibfnamefont {P.~S.}\ \bibnamefont
  {{Cowperthwaite}}}, \bibinfo {author} {\bibfnamefont {E.}~\bibnamefont
  {{Berger}}}, \bibinfo {author} {\bibfnamefont {V.~A.}\ \bibnamefont
  {{Villar}}}, \bibinfo {author} {\bibfnamefont {B.~D.}\ \bibnamefont
  {{Metzger}}}, \bibinfo {author} {\bibfnamefont {M.}~\bibnamefont
  {{Nicholl}}}, \bibinfo {author} {\bibfnamefont {R.}~\bibnamefont
  {{Chornock}}}, \bibinfo {author} {\bibfnamefont {P.~K.}\ \bibnamefont
  {{Blanchard}}}, \bibinfo {author} {\bibfnamefont {W.}~\bibnamefont {{Fong}}},
  \bibinfo {author} {\bibfnamefont {R.}~\bibnamefont {{Margutti}}}, \bibinfo
  {author} {\bibfnamefont {M.}~\bibnamefont {{Soares-Santos}}},  \emph
  {et~al.},\ }\bibfield  {title} {\enquote {\bibinfo {title} {{The
  Electromagnetic Counterpart of the Binary Neutron Star Merger LIGO/Virgo
  GW170817. II. UV, Optical, and Near-infrared Light Curves and Comparison to
  Kilonova Models}},}\ }\href {\doibase 10.3847/2041-8213/aa8fc7} {\bibfield
  {journal} {\bibinfo  {journal} {\apjl}\ }\textbf {\bibinfo {volume} {848}},\
  \bibinfo {eid} {L17} (\bibinfo {year} {2017})}\BibitemShut {NoStop}%
\bibitem [{\citenamefont {{Troja}}\ \emph {et~al.}(2017)\citenamefont
  {{Troja}}, \citenamefont {{Piro}}, \citenamefont {{van Eerten}},
  \citenamefont {{Wollaeger}}, \citenamefont {{Im}}, \citenamefont {{Fox}},
  \citenamefont {{Butler}}, \citenamefont {{Cenko}}, \citenamefont
  {{Sakamoto}}, \citenamefont {{Fryer}}, \citenamefont {{Ricci}}, \citenamefont
  {{Lien}}, \citenamefont {{Ryan}}, \citenamefont {{Korobkin}}, \citenamefont
  {{Lee}}, \citenamefont {{Burgess}}, \citenamefont {{Lee}}, \citenamefont
  {{Watson}}, \citenamefont {{Choi}}, \citenamefont {{Covino}}, \citenamefont
  {{D'Avanzo}}, \citenamefont {{Fontes}}, \citenamefont {{Gonz{\'a}lez}},
  \citenamefont {{Khandrika}}, \citenamefont {{Kim}}, \citenamefont {{Kim}},
  \citenamefont {{Lee}}, \citenamefont {{Lee}}, \citenamefont {{Kutyrev}},
  \citenamefont {{Lim}}, \citenamefont {{S{\'a}nchez-Ram{\'\i}rez}},
  \citenamefont {{Veilleux}}, \citenamefont {{Wieringa}},\ and\ \citenamefont
  {{Yoon}}}]{Troja_2017}%
  \BibitemOpen
  \bibfield  {author} {\bibinfo {author} {\bibfnamefont {E.}~\bibnamefont
  {{Troja}}}, \bibinfo {author} {\bibfnamefont {L.}~\bibnamefont {{Piro}}},
  \bibinfo {author} {\bibfnamefont {H.}~\bibnamefont {{van Eerten}}}, \bibinfo
  {author} {\bibfnamefont {R.~T.}\ \bibnamefont {{Wollaeger}}}, \bibinfo
  {author} {\bibfnamefont {M.}~\bibnamefont {{Im}}}, \bibinfo {author}
  {\bibfnamefont {O.~D.}\ \bibnamefont {{Fox}}}, \bibinfo {author}
  {\bibfnamefont {N.~R.}\ \bibnamefont {{Butler}}}, \bibinfo {author}
  {\bibfnamefont {S.~B.}\ \bibnamefont {{Cenko}}}, \bibinfo {author}
  {\bibfnamefont {T.}~\bibnamefont {{Sakamoto}}}, \bibinfo {author}
  {\bibfnamefont {C.~L.}\ \bibnamefont {{Fryer}}}, \bibinfo {author}
  {\bibfnamefont {R.}~\bibnamefont {{Ricci}}}, \bibinfo {author} {\bibfnamefont
  {A.}~\bibnamefont {{Lien}}}, \bibinfo {author} {\bibfnamefont {R.~E.}\
  \bibnamefont {{Ryan}}}, \bibinfo {author} {\bibfnamefont {O.}~\bibnamefont
  {{Korobkin}}}, \bibinfo {author} {\bibfnamefont {S.~K.}\ \bibnamefont
  {{Lee}}}, \bibinfo {author} {\bibfnamefont {J.~M.}\ \bibnamefont
  {{Burgess}}}, \bibinfo {author} {\bibfnamefont {W.~H.}\ \bibnamefont
  {{Lee}}}, \bibinfo {author} {\bibfnamefont {A.~M.}\ \bibnamefont {{Watson}}},
  \bibinfo {author} {\bibfnamefont {C.}~\bibnamefont {{Choi}}}, \bibinfo
  {author} {\bibfnamefont {S.}~\bibnamefont {{Covino}}}, \bibinfo {author}
  {\bibfnamefont {P.}~\bibnamefont {{D'Avanzo}}}, \bibinfo {author}
  {\bibfnamefont {C.~J.}\ \bibnamefont {{Fontes}}}, \bibinfo {author}
  {\bibfnamefont {J.~Becerra}\ \bibnamefont {{Gonz{\'a}lez}}}, \bibinfo
  {author} {\bibfnamefont {H.~G.}\ \bibnamefont {{Khandrika}}}, \bibinfo
  {author} {\bibfnamefont {J.}~\bibnamefont {{Kim}}}, \bibinfo {author}
  {\bibfnamefont {S.~L.}\ \bibnamefont {{Kim}}}, \bibinfo {author}
  {\bibfnamefont {C.~U.}\ \bibnamefont {{Lee}}}, \bibinfo {author}
  {\bibfnamefont {H.~M.}\ \bibnamefont {{Lee}}}, \bibinfo {author}
  {\bibfnamefont {A.}~\bibnamefont {{Kutyrev}}}, \bibinfo {author}
  {\bibfnamefont {G.}~\bibnamefont {{Lim}}}, \bibinfo {author} {\bibfnamefont
  {R.}~\bibnamefont {{S{\'a}nchez-Ram{\'\i}rez}}}, \bibinfo {author}
  {\bibfnamefont {S.}~\bibnamefont {{Veilleux}}}, \bibinfo {author}
  {\bibfnamefont {M.~H.}\ \bibnamefont {{Wieringa}}}, \ and\ \bibinfo {author}
  {\bibfnamefont {Y.}~\bibnamefont {{Yoon}}},\ }\bibfield  {title} {\enquote
  {\bibinfo {title} {{The X-ray counterpart to the gravitational-wave event
  GW170817}},}\ }\href {\doibase 10.1038/nature24290} {\bibfield  {journal}
  {\bibinfo  {journal} {\nat}\ }\textbf {\bibinfo {volume} {551}},\ \bibinfo
  {pages} {71--74} (\bibinfo {year} {2017})},\ \Eprint
  {http://arxiv.org/abs/1710.05433} {arXiv:1710.05433 [astro-ph.HE]}
  \BibitemShut {NoStop}%
\bibitem [{\citenamefont {{Villar}}\ \emph {et~al.}(2017)\citenamefont
  {{Villar}}, \citenamefont {{Guillochon}}, \citenamefont {{Berger}},
  \citenamefont {{Metzger}}, \citenamefont {{Cowperthwaite}}, \citenamefont
  {{Nicholl}}, \citenamefont {{Alexand er}}, \citenamefont {{Blanchard}},
  \citenamefont {{Chornock}}, \citenamefont {{Eftekhari}}, \citenamefont
  {{Fong}}, \citenamefont {{Margutti}},\ and\ \citenamefont
  {{Williams}}}]{2017ApJ...851L..21V}%
  \BibitemOpen
  \bibfield  {author} {\bibinfo {author} {\bibfnamefont {V.~A.}\ \bibnamefont
  {{Villar}}}, \bibinfo {author} {\bibfnamefont {J.}~\bibnamefont
  {{Guillochon}}}, \bibinfo {author} {\bibfnamefont {E.}~\bibnamefont
  {{Berger}}}, \bibinfo {author} {\bibfnamefont {B.~D.}\ \bibnamefont
  {{Metzger}}}, \bibinfo {author} {\bibfnamefont {P.~S.}\ \bibnamefont
  {{Cowperthwaite}}}, \bibinfo {author} {\bibfnamefont {M.}~\bibnamefont
  {{Nicholl}}}, \bibinfo {author} {\bibfnamefont {K.~D.}\ \bibnamefont
  {{Alexand er}}}, \bibinfo {author} {\bibfnamefont {P.~K.}\ \bibnamefont
  {{Blanchard}}}, \bibinfo {author} {\bibfnamefont {R.}~\bibnamefont
  {{Chornock}}}, \bibinfo {author} {\bibfnamefont {T.}~\bibnamefont
  {{Eftekhari}}}, \bibinfo {author} {\bibfnamefont {W.}~\bibnamefont {{Fong}}},
  \bibinfo {author} {\bibfnamefont {R.}~\bibnamefont {{Margutti}}}, \ and\
  \bibinfo {author} {\bibfnamefont {P.~K.~G.}\ \bibnamefont {{Williams}}},\
  }\bibfield  {title} {\enquote {\bibinfo {title} {{The Combined Ultraviolet,
  Optical, and Near-infrared Light Curves of the Kilonova Associated with the
  Binary Neutron Star Merger GW170817: Unified Data Set, Analytic Models, and
  Physical Implications}},}\ }\href {\doibase 10.3847/2041-8213/aa9c84}
  {\bibfield  {journal} {\bibinfo  {journal} {\apjl}\ }\textbf {\bibinfo
  {volume} {851}},\ \bibinfo {eid} {L21} (\bibinfo {year} {2017})}\BibitemShut
  {NoStop}%
\bibitem [{\citenamefont {{Pian}}\ \emph {et~al.}(2017)\citenamefont {{Pian}},
  \citenamefont {{D'Avanzo}}, \citenamefont {{Benetti}}, \citenamefont
  {{Branchesi}}, \citenamefont {{Brocato}}, \citenamefont {{Campana}},
  \citenamefont {{Cappellaro}}, \citenamefont {{Covino}}, \citenamefont
  {{D'Elia}}, \citenamefont {{Fynbo}} \emph {et~al.}}]{2017Natur.551...67P}%
  \BibitemOpen
  \bibfield  {author} {\bibinfo {author} {\bibfnamefont {E.}~\bibnamefont
  {{Pian}}}, \bibinfo {author} {\bibfnamefont {P.}~\bibnamefont {{D'Avanzo}}},
  \bibinfo {author} {\bibfnamefont {S.}~\bibnamefont {{Benetti}}}, \bibinfo
  {author} {\bibfnamefont {M.}~\bibnamefont {{Branchesi}}}, \bibinfo {author}
  {\bibfnamefont {E.}~\bibnamefont {{Brocato}}}, \bibinfo {author}
  {\bibfnamefont {S.}~\bibnamefont {{Campana}}}, \bibinfo {author}
  {\bibfnamefont {E.}~\bibnamefont {{Cappellaro}}}, \bibinfo {author}
  {\bibfnamefont {S.}~\bibnamefont {{Covino}}}, \bibinfo {author}
  {\bibfnamefont {V.}~\bibnamefont {{D'Elia}}}, \bibinfo {author}
  {\bibfnamefont {J.~P.~U.}\ \bibnamefont {{Fynbo}}},  \emph {et~al.},\
  }\bibfield  {title} {\enquote {\bibinfo {title} {{Spectroscopic
  identification of r-process nucleosynthesis in a double neutron-star
  merger}},}\ }\href {\doibase 10.1038/nature24298} {\bibfield  {journal}
  {\bibinfo  {journal} {\nat}\ }\textbf {\bibinfo {volume} {551}},\ \bibinfo
  {pages} {67--70} (\bibinfo {year} {2017})}\BibitemShut {NoStop}%
\bibitem [{\citenamefont {{Nicholl}}\ \emph {et~al.}(2017)\citenamefont
  {{Nicholl}}, \citenamefont {{Berger}}, \citenamefont {{Kasen}}, \citenamefont
  {{Metzger}}, \citenamefont {{Elias}}, \citenamefont {{Brice{\~n}o}},
  \citenamefont {{Alexander}}, \citenamefont {{Blanchard}}, \citenamefont
  {{Chornock}}, \citenamefont {{Cowperthwaite}}, \citenamefont {{Eftekhari}},
  \citenamefont {{Fong}}, \citenamefont {{Margutti}}, \citenamefont {{Villar}},
  \citenamefont {{Williams}}, \citenamefont {{Brown}}, \citenamefont {{Annis}},
  \citenamefont {{Bahramian}}, \citenamefont {{Brout}}, \citenamefont
  {{Brown}}, \citenamefont {{Chen}}, \citenamefont {{Clemens}}, \citenamefont
  {{Dennihy}}, \citenamefont {{Dunlap}}, \citenamefont {{Holz}}, \citenamefont
  {{Marchesini}}, \citenamefont {{Massaro}}, \citenamefont {{Moskowitz}},
  \citenamefont {{Pelisoli}}, \citenamefont {{Rest}}, \citenamefont {{Ricci}},
  \citenamefont {{Sako}}, \citenamefont {{Soares-Santos}},\ and\ \citenamefont
  {{Strader}}}]{2017ApJ...848L..18N}%
  \BibitemOpen
  \bibfield  {author} {\bibinfo {author} {\bibfnamefont {M.}~\bibnamefont
  {{Nicholl}}}, \bibinfo {author} {\bibfnamefont {E.}~\bibnamefont {{Berger}}},
  \bibinfo {author} {\bibfnamefont {D.}~\bibnamefont {{Kasen}}}, \bibinfo
  {author} {\bibfnamefont {B.~D.}\ \bibnamefont {{Metzger}}}, \bibinfo {author}
  {\bibfnamefont {J.}~\bibnamefont {{Elias}}}, \bibinfo {author} {\bibfnamefont
  {C.}~\bibnamefont {{Brice{\~n}o}}}, \bibinfo {author} {\bibfnamefont {K.~D.}\
  \bibnamefont {{Alexander}}}, \bibinfo {author} {\bibfnamefont {P.~K.}\
  \bibnamefont {{Blanchard}}}, \bibinfo {author} {\bibfnamefont
  {R.}~\bibnamefont {{Chornock}}}, \bibinfo {author} {\bibfnamefont {P.~S.}\
  \bibnamefont {{Cowperthwaite}}}, \bibinfo {author} {\bibfnamefont
  {T.}~\bibnamefont {{Eftekhari}}}, \bibinfo {author} {\bibfnamefont
  {W.}~\bibnamefont {{Fong}}}, \bibinfo {author} {\bibfnamefont
  {R.}~\bibnamefont {{Margutti}}}, \bibinfo {author} {\bibfnamefont {V.~A.}\
  \bibnamefont {{Villar}}}, \bibinfo {author} {\bibfnamefont {P.~K.~G.}\
  \bibnamefont {{Williams}}}, \bibinfo {author} {\bibfnamefont
  {W.}~\bibnamefont {{Brown}}}, \bibinfo {author} {\bibfnamefont
  {J.}~\bibnamefont {{Annis}}}, \bibinfo {author} {\bibfnamefont
  {A.}~\bibnamefont {{Bahramian}}}, \bibinfo {author} {\bibfnamefont
  {D.}~\bibnamefont {{Brout}}}, \bibinfo {author} {\bibfnamefont {D.~A.}\
  \bibnamefont {{Brown}}}, \bibinfo {author} {\bibfnamefont {H.-Y.}\
  \bibnamefont {{Chen}}}, \bibinfo {author} {\bibfnamefont {J.~C.}\
  \bibnamefont {{Clemens}}}, \bibinfo {author} {\bibfnamefont {E.}~\bibnamefont
  {{Dennihy}}}, \bibinfo {author} {\bibfnamefont {B.}~\bibnamefont {{Dunlap}}},
  \bibinfo {author} {\bibfnamefont {D.~E.}\ \bibnamefont {{Holz}}}, \bibinfo
  {author} {\bibfnamefont {E.}~\bibnamefont {{Marchesini}}}, \bibinfo {author}
  {\bibfnamefont {F.}~\bibnamefont {{Massaro}}}, \bibinfo {author}
  {\bibfnamefont {N.}~\bibnamefont {{Moskowitz}}}, \bibinfo {author}
  {\bibfnamefont {I.}~\bibnamefont {{Pelisoli}}}, \bibinfo {author}
  {\bibfnamefont {A.}~\bibnamefont {{Rest}}}, \bibinfo {author} {\bibfnamefont
  {F.}~\bibnamefont {{Ricci}}}, \bibinfo {author} {\bibfnamefont
  {M.}~\bibnamefont {{Sako}}}, \bibinfo {author} {\bibfnamefont
  {M.}~\bibnamefont {{Soares-Santos}}}, \ and\ \bibinfo {author} {\bibfnamefont
  {J.}~\bibnamefont {{Strader}}},\ }\bibfield  {title} {\enquote {\bibinfo
  {title} {{The Electromagnetic Counterpart of the Binary Neutron Star Merger
  LIGO/Virgo GW170817. III. Optical and UV Spectra of a Blue Kilonova from Fast
  Polar Ejecta}},}\ }\href {\doibase 10.3847/2041-8213/aa9029} {\bibfield
  {journal} {\bibinfo  {journal} {\apjl}\ }\textbf {\bibinfo {volume} {848}},\
  \bibinfo {eid} {L18} (\bibinfo {year} {2017})},\ \Eprint
  {http://arxiv.org/abs/1710.05456} {arXiv:1710.05456 [astro-ph.HE]}
  \BibitemShut {NoStop}%
\bibitem [{\citenamefont {{Miller}}\ \emph {et~al.}(2019)\citenamefont
  {{Miller}}, \citenamefont {{Ryan}}, \citenamefont {{Dolence}}, \citenamefont
  {{Burrows}}, \citenamefont {{Fontes}}, \citenamefont {{Fryer}}, \citenamefont
  {{Korobkin}}, \citenamefont {{Lippuner}}, \citenamefont {{Mumpower}},\ and\
  \citenamefont {{Wollaeger}}}]{2019PhRvD.100b3008M}%
  \BibitemOpen
  \bibfield  {author} {\bibinfo {author} {\bibfnamefont {Jonah~M.}\
  \bibnamefont {{Miller}}}, \bibinfo {author} {\bibfnamefont {Benjamin~R.}\
  \bibnamefont {{Ryan}}}, \bibinfo {author} {\bibfnamefont {Joshua~C.}\
  \bibnamefont {{Dolence}}}, \bibinfo {author} {\bibfnamefont {Adam}\
  \bibnamefont {{Burrows}}}, \bibinfo {author} {\bibfnamefont {Christopher~J.}\
  \bibnamefont {{Fontes}}}, \bibinfo {author} {\bibfnamefont {Christopher~L.}\
  \bibnamefont {{Fryer}}}, \bibinfo {author} {\bibfnamefont {Oleg}\
  \bibnamefont {{Korobkin}}}, \bibinfo {author} {\bibfnamefont {Jonas}\
  \bibnamefont {{Lippuner}}}, \bibinfo {author} {\bibfnamefont {Matthew~R.}\
  \bibnamefont {{Mumpower}}}, \ and\ \bibinfo {author} {\bibfnamefont
  {Ryan~T.}\ \bibnamefont {{Wollaeger}}},\ }\bibfield  {title} {\enquote
  {\bibinfo {title} {{Full transport model of GW170817-like disk produces a
  blue kilonova}},}\ }\href {\doibase 10.1103/PhysRevD.100.023008} {\bibfield
  {journal} {\bibinfo  {journal} {\prd}\ }\textbf {\bibinfo {volume} {100}},\
  \bibinfo {eid} {023008} (\bibinfo {year} {2019})}\BibitemShut {NoStop}%
\bibitem [{\citenamefont {{Tanaka}}\ \emph {et~al.}(2020)\citenamefont
  {{Tanaka}}, \citenamefont {{Kato}}, \citenamefont {{Gaigalas}},\ and\
  \citenamefont {{Kawaguchi}}}]{2020MNRAS.496.1369T}%
  \BibitemOpen
  \bibfield  {author} {\bibinfo {author} {\bibfnamefont {Masaomi}\ \bibnamefont
  {{Tanaka}}}, \bibinfo {author} {\bibfnamefont {Daiji}\ \bibnamefont
  {{Kato}}}, \bibinfo {author} {\bibfnamefont {Gediminas}\ \bibnamefont
  {{Gaigalas}}}, \ and\ \bibinfo {author} {\bibfnamefont {Kyohei}\ \bibnamefont
  {{Kawaguchi}}},\ }\bibfield  {title} {\enquote {\bibinfo {title} {{Systematic
  opacity calculations for kilonovae}},}\ }\href {\doibase
  10.1093/mnras/staa1576} {\bibfield  {journal} {\bibinfo  {journal} {\mnras}\
  }\textbf {\bibinfo {volume} {496}},\ \bibinfo {pages} {1369--1392} (\bibinfo
  {year} {2020})},\ \Eprint {http://arxiv.org/abs/1906.08914} {arXiv:1906.08914
  [astro-ph.HE]} \BibitemShut {NoStop}%
\bibitem [{\citenamefont {{Barnes}}\ \emph {et~al.}(2021)\citenamefont
  {{Barnes}}, \citenamefont {{Zhu}}, \citenamefont {{Lund}}, \citenamefont
  {{Sprouse}}, \citenamefont {{Vassh}}, \citenamefont {{McLaughlin}},
  \citenamefont {{Mumpower}},\ and\ \citenamefont
  {{Surman}}}]{2021ApJ...918...44B}%
  \BibitemOpen
  \bibfield  {author} {\bibinfo {author} {\bibfnamefont {Jennifer}\
  \bibnamefont {{Barnes}}}, \bibinfo {author} {\bibfnamefont {Y.~L.}\
  \bibnamefont {{Zhu}}}, \bibinfo {author} {\bibfnamefont {K.~A.}\ \bibnamefont
  {{Lund}}}, \bibinfo {author} {\bibfnamefont {T.~M.}\ \bibnamefont
  {{Sprouse}}}, \bibinfo {author} {\bibfnamefont {N.}~\bibnamefont {{Vassh}}},
  \bibinfo {author} {\bibfnamefont {G.~C.}\ \bibnamefont {{McLaughlin}}},
  \bibinfo {author} {\bibfnamefont {M.~R.}\ \bibnamefont {{Mumpower}}}, \ and\
  \bibinfo {author} {\bibfnamefont {R.}~\bibnamefont {{Surman}}},\ }\bibfield
  {title} {\enquote {\bibinfo {title} {{Kilonovae Across the Nuclear Physics
  Landscape: The Impact of Nuclear Physics Uncertainties on r-process-powered
  Emission}},}\ }\href {\doibase 10.3847/1538-4357/ac0aec} {\bibfield
  {journal} {\bibinfo  {journal} {\apj}\ }\textbf {\bibinfo {volume} {918}},\
  \bibinfo {eid} {44} (\bibinfo {year} {2021})},\ \Eprint
  {http://arxiv.org/abs/2010.11182} {arXiv:2010.11182 [astro-ph.HE]}
  \BibitemShut {NoStop}%
\bibitem [{\citenamefont {{Zhu}}\ \emph {et~al.}(2021)\citenamefont {{Zhu}},
  \citenamefont {{Lund}}, \citenamefont {{Barnes}}, \citenamefont {{Sprouse}},
  \citenamefont {{Vassh}}, \citenamefont {{McLaughlin}}, \citenamefont
  {{Mumpower}},\ and\ \citenamefont {{Surman}}}]{2021ApJ...906...94Z}%
  \BibitemOpen
  \bibfield  {author} {\bibinfo {author} {\bibfnamefont {Y.~L.}\ \bibnamefont
  {{Zhu}}}, \bibinfo {author} {\bibfnamefont {K.~A.}\ \bibnamefont {{Lund}}},
  \bibinfo {author} {\bibfnamefont {J.}~\bibnamefont {{Barnes}}}, \bibinfo
  {author} {\bibfnamefont {T.~M.}\ \bibnamefont {{Sprouse}}}, \bibinfo {author}
  {\bibfnamefont {N.}~\bibnamefont {{Vassh}}}, \bibinfo {author} {\bibfnamefont
  {G.~C.}\ \bibnamefont {{McLaughlin}}}, \bibinfo {author} {\bibfnamefont
  {M.~R.}\ \bibnamefont {{Mumpower}}}, \ and\ \bibinfo {author} {\bibfnamefont
  {R.}~\bibnamefont {{Surman}}},\ }\bibfield  {title} {\enquote {\bibinfo
  {title} {{Modeling Kilonova Light Curves: Dependence on Nuclear Inputs}},}\
  }\href {\doibase 10.3847/1538-4357/abc69e} {\bibfield  {journal} {\bibinfo
  {journal} {\apj}\ }\textbf {\bibinfo {volume} {906}},\ \bibinfo {eid} {94}
  (\bibinfo {year} {2021})}\BibitemShut {NoStop}%
\bibitem [{\citenamefont {{Sneden}}\ \emph {et~al.}(1994)\citenamefont
  {{Sneden}}, \citenamefont {{Preston}}, \citenamefont {{McWilliam}},\ and\
  \citenamefont {{Searle}}}]{1994ApJ...431L..27S}%
  \BibitemOpen
  \bibfield  {author} {\bibinfo {author} {\bibfnamefont {Christopher}\
  \bibnamefont {{Sneden}}}, \bibinfo {author} {\bibfnamefont {George~W.}\
  \bibnamefont {{Preston}}}, \bibinfo {author} {\bibfnamefont {Andrew}\
  \bibnamefont {{McWilliam}}}, \ and\ \bibinfo {author} {\bibfnamefont
  {Leonard}\ \bibnamefont {{Searle}}},\ }\bibfield  {title} {\enquote {\bibinfo
  {title} {{Ultra--Metal-poor Halo Stars: The Remarkable Spectrum of CS
  22892-052}},}\ }\href {\doibase 10.1086/187464} {\bibfield  {journal}
  {\bibinfo  {journal} {\apjl}\ }\textbf {\bibinfo {volume} {431}},\ \bibinfo
  {pages} {L27} (\bibinfo {year} {1994})}\BibitemShut {NoStop}%
\bibitem [{\citenamefont {{Hill}}\ \emph
  {et~al.}(2002{\natexlab{a}})\citenamefont {{Hill}}, \citenamefont {{Plez}},
  \citenamefont {{Cayrel}}, \citenamefont {{Beers}}, \citenamefont
  {{Nordstr{\"o}m}}, \citenamefont {{Andersen}}, \citenamefont {{Spite}},
  \citenamefont {{Spite}}, \citenamefont {{Barbuy}}, \citenamefont
  {{Bonifacio}}, \citenamefont {{Depagne}}, \citenamefont {{Fran{\c{c}}ois}},\
  and\ \citenamefont {{Primas}}}]{2002A&A...387..560H}%
  \BibitemOpen
  \bibfield  {author} {\bibinfo {author} {\bibfnamefont {V.}~\bibnamefont
  {{Hill}}}, \bibinfo {author} {\bibfnamefont {B.}~\bibnamefont {{Plez}}},
  \bibinfo {author} {\bibfnamefont {R.}~\bibnamefont {{Cayrel}}}, \bibinfo
  {author} {\bibfnamefont {T.~C.}\ \bibnamefont {{Beers}}}, \bibinfo {author}
  {\bibfnamefont {B.}~\bibnamefont {{Nordstr{\"o}m}}}, \bibinfo {author}
  {\bibfnamefont {J.}~\bibnamefont {{Andersen}}}, \bibinfo {author}
  {\bibfnamefont {M.}~\bibnamefont {{Spite}}}, \bibinfo {author} {\bibfnamefont
  {F.}~\bibnamefont {{Spite}}}, \bibinfo {author} {\bibfnamefont
  {B.}~\bibnamefont {{Barbuy}}}, \bibinfo {author} {\bibfnamefont
  {P.}~\bibnamefont {{Bonifacio}}}, \bibinfo {author} {\bibfnamefont
  {E.}~\bibnamefont {{Depagne}}}, \bibinfo {author} {\bibfnamefont
  {P.}~\bibnamefont {{Fran{\c{c}}ois}}}, \ and\ \bibinfo {author}
  {\bibfnamefont {F.}~\bibnamefont {{Primas}}},\ }\bibfield  {title} {\enquote
  {\bibinfo {title} {{First stars. I. The extreme r-element rich, iron-poor
  halo giant CS 31082-001. Implications for the r-process site(s) and
  radioactive cosmochronology}},}\ }\href {\doibase 10.1051/0004-6361:20020434}
  {\bibfield  {journal} {\bibinfo  {journal} {\aap}\ }\textbf {\bibinfo
  {volume} {387}},\ \bibinfo {pages} {560--579} (\bibinfo {year}
  {2002}{\natexlab{a}})},\ \Eprint {http://arxiv.org/abs/astro-ph/0203462}
  {arXiv:astro-ph/0203462 [astro-ph]} \BibitemShut {NoStop}%
\bibitem [{\citenamefont {{Sneden}}\ \emph
  {et~al.}(2003{\natexlab{a}})\citenamefont {{Sneden}}, \citenamefont
  {{Cowan}}, \citenamefont {{Lawler}}, \citenamefont {{Ivans}}, \citenamefont
  {{Burles}}, \citenamefont {{Beers}}, \citenamefont {{Primas}}, \citenamefont
  {{Hill}}, \citenamefont {{Truran}}, \citenamefont {{Fuller}}, \citenamefont
  {{Pfeiffer}},\ and\ \citenamefont {{Kratz}}}]{2003ApJ...591..936S}%
  \BibitemOpen
  \bibfield  {author} {\bibinfo {author} {\bibfnamefont {Christopher}\
  \bibnamefont {{Sneden}}}, \bibinfo {author} {\bibfnamefont {John~J.}\
  \bibnamefont {{Cowan}}}, \bibinfo {author} {\bibfnamefont {James~E.}\
  \bibnamefont {{Lawler}}}, \bibinfo {author} {\bibfnamefont {Inese~I.}\
  \bibnamefont {{Ivans}}}, \bibinfo {author} {\bibfnamefont {Scott}\
  \bibnamefont {{Burles}}}, \bibinfo {author} {\bibfnamefont {Timothy~C.}\
  \bibnamefont {{Beers}}}, \bibinfo {author} {\bibfnamefont {Francesca}\
  \bibnamefont {{Primas}}}, \bibinfo {author} {\bibfnamefont {Vanessa}\
  \bibnamefont {{Hill}}}, \bibinfo {author} {\bibfnamefont {James~W.}\
  \bibnamefont {{Truran}}}, \bibinfo {author} {\bibfnamefont {George~M.}\
  \bibnamefont {{Fuller}}}, \bibinfo {author} {\bibfnamefont {Bernd}\
  \bibnamefont {{Pfeiffer}}}, \ and\ \bibinfo {author} {\bibfnamefont
  {Karl-Ludwig}\ \bibnamefont {{Kratz}}},\ }\bibfield  {title} {\enquote
  {\bibinfo {title} {{The Extremely Metal-poor, Neutron Capture-rich Star CS
  22892-052: A Comprehensive Abundance Analysis}},}\ }\href {\doibase
  10.1086/375491} {\bibfield  {journal} {\bibinfo  {journal} {\apj}\ }\textbf
  {\bibinfo {volume} {591}},\ \bibinfo {pages} {936--953} (\bibinfo {year}
  {2003}{\natexlab{a}})},\ \Eprint {http://arxiv.org/abs/astro-ph/0303542}
  {arXiv:astro-ph/0303542 [astro-ph]} \BibitemShut {NoStop}%
\bibitem [{\citenamefont {{Frebel}}\ \emph {et~al.}(2007)\citenamefont
  {{Frebel}}, \citenamefont {{Christlieb}}, \citenamefont {{Norris}},
  \citenamefont {{Thom}}, \citenamefont {{Beers}},\ and\ \citenamefont
  {{Rhee}}}]{2007ApJ...660L.117F}%
  \BibitemOpen
  \bibfield  {author} {\bibinfo {author} {\bibfnamefont {Anna}\ \bibnamefont
  {{Frebel}}}, \bibinfo {author} {\bibfnamefont {Norbert}\ \bibnamefont
  {{Christlieb}}}, \bibinfo {author} {\bibfnamefont {John~E.}\ \bibnamefont
  {{Norris}}}, \bibinfo {author} {\bibfnamefont {Christopher}\ \bibnamefont
  {{Thom}}}, \bibinfo {author} {\bibfnamefont {Timothy~C.}\ \bibnamefont
  {{Beers}}}, \ and\ \bibinfo {author} {\bibfnamefont {Jaehyon}\ \bibnamefont
  {{Rhee}}},\ }\bibfield  {title} {\enquote {\bibinfo {title} {{Discovery of HE
  1523-0901, a Strongly r-Process-enhanced Metal-poor Star with Detected
  Uranium}},}\ }\href {\doibase 10.1086/518122} {\bibfield  {journal} {\bibinfo
   {journal} {\apjl}\ }\textbf {\bibinfo {volume} {660}},\ \bibinfo {pages}
  {L117--L120} (\bibinfo {year} {2007})},\ \Eprint
  {http://arxiv.org/abs/astro-ph/0703414} {arXiv:astro-ph/0703414 [astro-ph]}
  \BibitemShut {NoStop}%
\bibitem [{\citenamefont {{Siqueira Mello}}\ \emph {et~al.}(2013)\citenamefont
  {{Siqueira Mello}}, \citenamefont {{Spite}}, \citenamefont {{Barbuy}},
  \citenamefont {{Spite}}, \citenamefont {{Caffau}}, \citenamefont {{Hill}},
  \citenamefont {{Wanajo}}, \citenamefont {{Primas}}, \citenamefont {{Plez}},
  \citenamefont {{Cayrel}}, \citenamefont {{Andersen}}, \citenamefont
  {{Nordstr{\"o}m}}, \citenamefont {{Sneden}}, \citenamefont {{Beers}},
  \citenamefont {{Bonifacio}}, \citenamefont {{Fran{\c{c}}ois}},\ and\
  \citenamefont {{Molaro}}}]{2013A&A...550A.122S}%
  \BibitemOpen
  \bibfield  {author} {\bibinfo {author} {\bibfnamefont {C.}~\bibnamefont
  {{Siqueira Mello}}}, \bibinfo {author} {\bibfnamefont {M.}~\bibnamefont
  {{Spite}}}, \bibinfo {author} {\bibfnamefont {B.}~\bibnamefont {{Barbuy}}},
  \bibinfo {author} {\bibfnamefont {F.}~\bibnamefont {{Spite}}}, \bibinfo
  {author} {\bibfnamefont {E.}~\bibnamefont {{Caffau}}}, \bibinfo {author}
  {\bibfnamefont {V.}~\bibnamefont {{Hill}}}, \bibinfo {author} {\bibfnamefont
  {S.}~\bibnamefont {{Wanajo}}}, \bibinfo {author} {\bibfnamefont
  {F.}~\bibnamefont {{Primas}}}, \bibinfo {author} {\bibfnamefont
  {B.}~\bibnamefont {{Plez}}}, \bibinfo {author} {\bibfnamefont
  {R.}~\bibnamefont {{Cayrel}}}, \bibinfo {author} {\bibfnamefont
  {J.}~\bibnamefont {{Andersen}}}, \bibinfo {author} {\bibfnamefont
  {B.}~\bibnamefont {{Nordstr{\"o}m}}}, \bibinfo {author} {\bibfnamefont
  {C.}~\bibnamefont {{Sneden}}}, \bibinfo {author} {\bibfnamefont {T.~C.}\
  \bibnamefont {{Beers}}}, \bibinfo {author} {\bibfnamefont {P.}~\bibnamefont
  {{Bonifacio}}}, \bibinfo {author} {\bibfnamefont {P.}~\bibnamefont
  {{Fran{\c{c}}ois}}}, \ and\ \bibinfo {author} {\bibfnamefont
  {P.}~\bibnamefont {{Molaro}}},\ }\bibfield  {title} {\enquote {\bibinfo
  {title} {{First stars. XVI. HST/STIS abundances of heavy elements in the
  uranium-rich metal-poor star CS 31082-001}},}\ }\href {\doibase
  10.1051/0004-6361/201219949} {\bibfield  {journal} {\bibinfo  {journal}
  {\aap}\ }\textbf {\bibinfo {volume} {550}},\ \bibinfo {eid} {A122} (\bibinfo
  {year} {2013})},\ \Eprint {http://arxiv.org/abs/1212.0211} {arXiv:1212.0211
  [astro-ph.SR]} \BibitemShut {NoStop}%
\bibitem [{\citenamefont {Lorusso}\ \emph {et~al.}(2015)\citenamefont
  {Lorusso}, \citenamefont {Nishimura}, \citenamefont {Xu}, \citenamefont
  {Jungclaus}, \citenamefont {Shimizu}, \citenamefont {Simpson}, \citenamefont
  {S\"oderstr\"om}, \citenamefont {Watanabe}, \citenamefont {Browne},
  \citenamefont {Doornenbal}, \citenamefont {Gey}, \citenamefont {Jung},
  \citenamefont {Meyer}, \citenamefont {Sumikama}, \citenamefont {Taprogge},
  \citenamefont {Vajta}, \citenamefont {Wu}, \citenamefont {Baba},
  \citenamefont {Benzoni}, \citenamefont {Chae}, \citenamefont {Crespi},
  \citenamefont {Fukuda}, \citenamefont {Gernh\"auser}, \citenamefont {Inabe},
  \citenamefont {Isobe}, \citenamefont {Kajino}, \citenamefont {Kameda},
  \citenamefont {Kim}, \citenamefont {Kim}, \citenamefont {Kojouharov},
  \citenamefont {Kondev}, \citenamefont {Kubo}, \citenamefont {Kurz},
  \citenamefont {Kwon}, \citenamefont {Lane}, \citenamefont {Li}, \citenamefont
  {Montaner-Piz\'a}, \citenamefont {Moschner}, \citenamefont {Naqvi},
  \citenamefont {Niikura}, \citenamefont {Nishibata}, \citenamefont {Odahara},
  \citenamefont {Orlandi}, \citenamefont {Patel}, \citenamefont {Podoly\'ak},
  \citenamefont {Sakurai}, \citenamefont {Schaffner}, \citenamefont {Schury},
  \citenamefont {Shibagaki}, \citenamefont {Steiger}, \citenamefont {Suzuki},
  \citenamefont {Takeda}, \citenamefont {Wendt}, \citenamefont {Yagi},\ and\
  \citenamefont {Yoshinaga}}]{PhysRevLett.114.192501}%
  \BibitemOpen
  \bibfield  {author} {\bibinfo {author} {\bibfnamefont {G.}~\bibnamefont
  {Lorusso}}, \bibinfo {author} {\bibfnamefont {S.}~\bibnamefont {Nishimura}},
  \bibinfo {author} {\bibfnamefont {Z.~Y.}\ \bibnamefont {Xu}}, \bibinfo
  {author} {\bibfnamefont {A.}~\bibnamefont {Jungclaus}}, \bibinfo {author}
  {\bibfnamefont {Y.}~\bibnamefont {Shimizu}}, \bibinfo {author} {\bibfnamefont
  {G.~S.}\ \bibnamefont {Simpson}}, \bibinfo {author} {\bibfnamefont {P.-A.}\
  \bibnamefont {S\"oderstr\"om}}, \bibinfo {author} {\bibfnamefont
  {H.}~\bibnamefont {Watanabe}}, \bibinfo {author} {\bibfnamefont
  {F.}~\bibnamefont {Browne}}, \bibinfo {author} {\bibfnamefont
  {P.}~\bibnamefont {Doornenbal}}, \bibinfo {author} {\bibfnamefont
  {G.}~\bibnamefont {Gey}}, \bibinfo {author} {\bibfnamefont {H.~S.}\
  \bibnamefont {Jung}}, \bibinfo {author} {\bibfnamefont {B.}~\bibnamefont
  {Meyer}}, \bibinfo {author} {\bibfnamefont {T.}~\bibnamefont {Sumikama}},
  \bibinfo {author} {\bibfnamefont {J.}~\bibnamefont {Taprogge}}, \bibinfo
  {author} {\bibfnamefont {Zs.}\ \bibnamefont {Vajta}}, \bibinfo {author}
  {\bibfnamefont {J.}~\bibnamefont {Wu}}, \bibinfo {author} {\bibfnamefont
  {H.}~\bibnamefont {Baba}}, \bibinfo {author} {\bibfnamefont {G.}~\bibnamefont
  {Benzoni}}, \bibinfo {author} {\bibfnamefont {K.~Y.}\ \bibnamefont {Chae}},
  \bibinfo {author} {\bibfnamefont {F.~C.~L.}\ \bibnamefont {Crespi}}, \bibinfo
  {author} {\bibfnamefont {N.}~\bibnamefont {Fukuda}}, \bibinfo {author}
  {\bibfnamefont {R.}~\bibnamefont {Gernh\"auser}}, \bibinfo {author}
  {\bibfnamefont {N.}~\bibnamefont {Inabe}}, \bibinfo {author} {\bibfnamefont
  {T.}~\bibnamefont {Isobe}}, \bibinfo {author} {\bibfnamefont
  {T.}~\bibnamefont {Kajino}}, \bibinfo {author} {\bibfnamefont
  {D.}~\bibnamefont {Kameda}}, \bibinfo {author} {\bibfnamefont {G.~D.}\
  \bibnamefont {Kim}}, \bibinfo {author} {\bibfnamefont {Y.-K.}\ \bibnamefont
  {Kim}}, \bibinfo {author} {\bibfnamefont {I.}~\bibnamefont {Kojouharov}},
  \bibinfo {author} {\bibfnamefont {F.~G.}\ \bibnamefont {Kondev}}, \bibinfo
  {author} {\bibfnamefont {T.}~\bibnamefont {Kubo}}, \bibinfo {author}
  {\bibfnamefont {N.}~\bibnamefont {Kurz}}, \bibinfo {author} {\bibfnamefont
  {Y.~K.}\ \bibnamefont {Kwon}}, \bibinfo {author} {\bibfnamefont {G.~J.}\
  \bibnamefont {Lane}}, \bibinfo {author} {\bibfnamefont {Z.}~\bibnamefont
  {Li}}, \bibinfo {author} {\bibfnamefont {A.}~\bibnamefont {Montaner-Piz\'a}},
  \bibinfo {author} {\bibfnamefont {K.}~\bibnamefont {Moschner}}, \bibinfo
  {author} {\bibfnamefont {F.}~\bibnamefont {Naqvi}}, \bibinfo {author}
  {\bibfnamefont {M.}~\bibnamefont {Niikura}}, \bibinfo {author} {\bibfnamefont
  {H.}~\bibnamefont {Nishibata}}, \bibinfo {author} {\bibfnamefont
  {A.}~\bibnamefont {Odahara}}, \bibinfo {author} {\bibfnamefont
  {R.}~\bibnamefont {Orlandi}}, \bibinfo {author} {\bibfnamefont
  {Z.}~\bibnamefont {Patel}}, \bibinfo {author} {\bibfnamefont {Zs.}\
  \bibnamefont {Podoly\'ak}}, \bibinfo {author} {\bibfnamefont
  {H.}~\bibnamefont {Sakurai}}, \bibinfo {author} {\bibfnamefont
  {H.}~\bibnamefont {Schaffner}}, \bibinfo {author} {\bibfnamefont
  {P.}~\bibnamefont {Schury}}, \bibinfo {author} {\bibfnamefont
  {S.}~\bibnamefont {Shibagaki}}, \bibinfo {author} {\bibfnamefont
  {K.}~\bibnamefont {Steiger}}, \bibinfo {author} {\bibfnamefont
  {H.}~\bibnamefont {Suzuki}}, \bibinfo {author} {\bibfnamefont
  {H.}~\bibnamefont {Takeda}}, \bibinfo {author} {\bibfnamefont
  {A.}~\bibnamefont {Wendt}}, \bibinfo {author} {\bibfnamefont
  {A.}~\bibnamefont {Yagi}}, \ and\ \bibinfo {author} {\bibfnamefont
  {K.}~\bibnamefont {Yoshinaga}},\ }\bibfield  {title} {\enquote {\bibinfo
  {title} {$\ensuremath{\beta}$-decay half-lives of 110 neutron-rich nuclei
  across the $n=82$ shell gap: Implications for the mechanism and universality
  of the astrophysical $r$ process},}\ }\href {\doibase
  10.1103/PhysRevLett.114.192501} {\bibfield  {journal} {\bibinfo  {journal}
  {Phys. Rev. Lett.}\ }\textbf {\bibinfo {volume} {114}},\ \bibinfo {pages}
  {192501} (\bibinfo {year} {2015})}\BibitemShut {NoStop}%
\bibitem [{\citenamefont {{Korobkin}}\ \emph {et~al.}(2021)\citenamefont
  {{Korobkin}}, \citenamefont {{Wollaeger}}, \citenamefont {{Fryer}},
  \citenamefont {{Hungerford}}, \citenamefont {{Rosswog}}, \citenamefont
  {{Fontes}}, \citenamefont {{Mumpower}}, \citenamefont {{Chase}},
  \citenamefont {{Even}}, \citenamefont {{Miller}}, \citenamefont {{Misch}},\
  and\ \citenamefont {{Lippuner}}}]{2021ApJ...910..116K}%
  \BibitemOpen
  \bibfield  {author} {\bibinfo {author} {\bibfnamefont {Oleg}\ \bibnamefont
  {{Korobkin}}}, \bibinfo {author} {\bibfnamefont {Ryan~T.}\ \bibnamefont
  {{Wollaeger}}}, \bibinfo {author} {\bibfnamefont {Christopher~L.}\
  \bibnamefont {{Fryer}}}, \bibinfo {author} {\bibfnamefont {Aimee~L.}\
  \bibnamefont {{Hungerford}}}, \bibinfo {author} {\bibfnamefont {Stephan}\
  \bibnamefont {{Rosswog}}}, \bibinfo {author} {\bibfnamefont {Christopher~J.}\
  \bibnamefont {{Fontes}}}, \bibinfo {author} {\bibfnamefont {Matthew~R.}\
  \bibnamefont {{Mumpower}}}, \bibinfo {author} {\bibfnamefont {Eve~A.}\
  \bibnamefont {{Chase}}}, \bibinfo {author} {\bibfnamefont {Wesley~P.}\
  \bibnamefont {{Even}}}, \bibinfo {author} {\bibfnamefont {Jonah}\
  \bibnamefont {{Miller}}}, \bibinfo {author} {\bibfnamefont {G.~Wendell}\
  \bibnamefont {{Misch}}}, \ and\ \bibinfo {author} {\bibfnamefont {Jonas}\
  \bibnamefont {{Lippuner}}},\ }\bibfield  {title} {\enquote {\bibinfo {title}
  {{Axisymmetric Radiative Transfer Models of Kilonovae}},}\ }\href {\doibase
  10.3847/1538-4357/abe1b5} {\bibfield  {journal} {\bibinfo  {journal} {\apj}\
  }\textbf {\bibinfo {volume} {910}},\ \bibinfo {eid} {116} (\bibinfo {year}
  {2021})}\BibitemShut {NoStop}%
\bibitem [{\citenamefont {{Perego}}\ \emph {et~al.}(2014)\citenamefont
  {{Perego}}, \citenamefont {{Rosswog}}, \citenamefont {{Cabez{\'o}n}},
  \citenamefont {{Korobkin}}, \citenamefont {{K{\"a}ppeli}}, \citenamefont
  {{Arcones}},\ and\ \citenamefont {{Liebend{\"o}rfer}}}]{2014MNRAS.443.3134P}%
  \BibitemOpen
  \bibfield  {author} {\bibinfo {author} {\bibfnamefont {A.}~\bibnamefont
  {{Perego}}}, \bibinfo {author} {\bibfnamefont {S.}~\bibnamefont {{Rosswog}}},
  \bibinfo {author} {\bibfnamefont {R.~M.}\ \bibnamefont {{Cabez{\'o}n}}},
  \bibinfo {author} {\bibfnamefont {O.}~\bibnamefont {{Korobkin}}}, \bibinfo
  {author} {\bibfnamefont {R.}~\bibnamefont {{K{\"a}ppeli}}}, \bibinfo {author}
  {\bibfnamefont {A.}~\bibnamefont {{Arcones}}}, \ and\ \bibinfo {author}
  {\bibfnamefont {M.}~\bibnamefont {{Liebend{\"o}rfer}}},\ }\bibfield  {title}
  {\enquote {\bibinfo {title} {{Neutrino-driven winds from neutron star merger
  remnants}},}\ }\href {\doibase 10.1093/mnras/stu1352} {\bibfield  {journal}
  {\bibinfo  {journal} {\mnras}\ }\textbf {\bibinfo {volume} {443}},\ \bibinfo
  {pages} {3134--3156} (\bibinfo {year} {2014})},\ \Eprint
  {http://arxiv.org/abs/1405.6730} {arXiv:1405.6730 [astro-ph.HE]} \BibitemShut
  {NoStop}%
\bibitem [{\citenamefont {{Rosswog}}\ \emph {et~al.}(2014)\citenamefont
  {{Rosswog}}, \citenamefont {{Korobkin}}, \citenamefont {{Arcones}},
  \citenamefont {{Thielemann}},\ and\ \citenamefont
  {{Piran}}}]{2014MNRAS.439..744R}%
  \BibitemOpen
  \bibfield  {author} {\bibinfo {author} {\bibfnamefont {S.}~\bibnamefont
  {{Rosswog}}}, \bibinfo {author} {\bibfnamefont {O.}~\bibnamefont
  {{Korobkin}}}, \bibinfo {author} {\bibfnamefont {A.}~\bibnamefont
  {{Arcones}}}, \bibinfo {author} {\bibfnamefont {F.-K.}\ \bibnamefont
  {{Thielemann}}}, \ and\ \bibinfo {author} {\bibfnamefont {T.}~\bibnamefont
  {{Piran}}},\ }\bibfield  {title} {\enquote {\bibinfo {title} {{The long-term
  evolution of neutron star merger remnants - I. The impact of r-process
  nucleosynthesis}},}\ }\href {\doibase 10.1093/mnras/stt2502} {\bibfield
  {journal} {\bibinfo  {journal} {\mnras}\ }\textbf {\bibinfo {volume} {439}},\
  \bibinfo {pages} {744--756} (\bibinfo {year} {2014})}\BibitemShut {NoStop}%
\bibitem [{\citenamefont {{Ristic}}\ \emph {et~al.}(2022)\citenamefont
  {{Ristic}}, \citenamefont {{Champion}}, \citenamefont {{O'Shaughnessy}},
  \citenamefont {{Wollaeger}}, \citenamefont {{Korobkin}}, \citenamefont
  {{Chase}}, \citenamefont {{Fryer}}, \citenamefont {{Hungerford}},\ and\
  \citenamefont {{Fontes}}}]{Ristic22}%
  \BibitemOpen
  \bibfield  {author} {\bibinfo {author} {\bibfnamefont {M.}~\bibnamefont
  {{Ristic}}}, \bibinfo {author} {\bibfnamefont {E.}~\bibnamefont
  {{Champion}}}, \bibinfo {author} {\bibfnamefont {R.}~\bibnamefont
  {{O'Shaughnessy}}}, \bibinfo {author} {\bibfnamefont {R.}~\bibnamefont
  {{Wollaeger}}}, \bibinfo {author} {\bibfnamefont {O.}~\bibnamefont
  {{Korobkin}}}, \bibinfo {author} {\bibfnamefont {E.~A.}\ \bibnamefont
  {{Chase}}}, \bibinfo {author} {\bibfnamefont {C.~L.}\ \bibnamefont
  {{Fryer}}}, \bibinfo {author} {\bibfnamefont {A.~L.}\ \bibnamefont
  {{Hungerford}}}, \ and\ \bibinfo {author} {\bibfnamefont {C.~J.}\
  \bibnamefont {{Fontes}}},\ }\bibfield  {title} {\enquote {\bibinfo {title}
  {{Interpolating detailed simulations of kilonovae: Adaptive learning and
  parameter inference applications}},}\ }\href {\doibase
  10.1103/PhysRevResearch.4.013046} {\bibfield  {journal} {\bibinfo  {journal}
  {Physical Review Research}\ }\textbf {\bibinfo {volume} {4}},\ \bibinfo {eid}
  {013046} (\bibinfo {year} {2022})},\ \Eprint
  {http://arxiv.org/abs/2105.07013} {arXiv:2105.07013 [astro-ph.HE]}
  \BibitemShut {NoStop}%
\bibitem [{\citenamefont {Wollaeger}\ and\ \citenamefont {van
  Rossum}(2014)}]{SuperNu}%
  \BibitemOpen
  \bibfield  {author} {\bibinfo {author} {\bibfnamefont {Ryan~T.}\ \bibnamefont
  {Wollaeger}}\ and\ \bibinfo {author} {\bibfnamefont {Daniel~R.}\ \bibnamefont
  {van Rossum}},\ }\bibfield  {title} {\enquote {\bibinfo {title} {{RADIATION}
  {TRANSPORT} {FOR} {EXPLOSIVE} {OUTFLOWS}: {OPACITY} {REGROUPING}},}\ }\href
  {\doibase 10.1088/0067-0049/214/2/28} {\bibfield  {journal} {\bibinfo
  {journal} {The Astrophysical Journal Supplement Series}\ }\textbf {\bibinfo
  {volume} {214}},\ \bibinfo {pages} {28} (\bibinfo {year} {2014})}\BibitemShut
  {NoStop}%
\bibitem [{\citenamefont {{Winteler}}\ \emph {et~al.}(2012)\citenamefont
  {{Winteler}}, \citenamefont {{K{\"a}ppeli}}, \citenamefont {{Perego}},
  \citenamefont {{Arcones}}, \citenamefont {{Vasset}}, \citenamefont
  {{Nishimura}}, \citenamefont {{Liebend{\"o}rfer}},\ and\ \citenamefont
  {{Thielemann}}}]{2012ApJ...750L..22W}%
  \BibitemOpen
  \bibfield  {author} {\bibinfo {author} {\bibfnamefont {C.}~\bibnamefont
  {{Winteler}}}, \bibinfo {author} {\bibfnamefont {R.}~\bibnamefont
  {{K{\"a}ppeli}}}, \bibinfo {author} {\bibfnamefont {A.}~\bibnamefont
  {{Perego}}}, \bibinfo {author} {\bibfnamefont {A.}~\bibnamefont {{Arcones}}},
  \bibinfo {author} {\bibfnamefont {N.}~\bibnamefont {{Vasset}}}, \bibinfo
  {author} {\bibfnamefont {N.}~\bibnamefont {{Nishimura}}}, \bibinfo {author}
  {\bibfnamefont {M.}~\bibnamefont {{Liebend{\"o}rfer}}}, \ and\ \bibinfo
  {author} {\bibfnamefont {F.~K.}\ \bibnamefont {{Thielemann}}},\ }\bibfield
  {title} {\enquote {\bibinfo {title} {{Magnetorotationally Driven Supernovae
  as the Origin of Early Galaxy r-process Elements?}}}\ }\href {\doibase
  10.1088/2041-8205/750/1/L22} {\bibfield  {journal} {\bibinfo  {journal}
  {\apjl}\ }\textbf {\bibinfo {volume} {750}},\ \bibinfo {eid} {L22} (\bibinfo
  {year} {2012})}\BibitemShut {NoStop}%
\bibitem [{\citenamefont {{Barnes}}\ \emph {et~al.}(2016)\citenamefont
  {{Barnes}}, \citenamefont {{Kasen}}, \citenamefont {{Wu}},\ and\
  \citenamefont {{Mart{\'\i}nez-Pinedo}}}]{Barnes_2016}%
  \BibitemOpen
  \bibfield  {author} {\bibinfo {author} {\bibfnamefont {Jennifer}\
  \bibnamefont {{Barnes}}}, \bibinfo {author} {\bibfnamefont {Daniel}\
  \bibnamefont {{Kasen}}}, \bibinfo {author} {\bibfnamefont {Meng-Ru}\
  \bibnamefont {{Wu}}}, \ and\ \bibinfo {author} {\bibfnamefont {Gabriel}\
  \bibnamefont {{Mart{\'\i}nez-Pinedo}}},\ }\bibfield  {title} {\enquote
  {\bibinfo {title} {{Radioactivity and Thermalization in the Ejecta of Compact
  Object Mergers and Their Impact on Kilonova Light Curves}},}\ }\href
  {\doibase 10.3847/0004-637X/829/2/110} {\bibfield  {journal} {\bibinfo
  {journal} {The Astrophysical Journal}\ }\textbf {\bibinfo {volume} {829}},\
  \bibinfo {eid} {110} (\bibinfo {year} {2016})}\BibitemShut {NoStop}%
\bibitem [{\citenamefont {Wollaeger}\ \emph {et~al.}(2018)\citenamefont
  {Wollaeger}, \citenamefont {Korobkin}, \citenamefont {Fontes}, \citenamefont
  {Rosswog}, \citenamefont {Even}, \citenamefont {Fryer}, \citenamefont
  {Sollerman}, \citenamefont {Hungerford}, \citenamefont {van Rossum},\ and\
  \citenamefont {Wollaber}}]{Wollaeger2018}%
  \BibitemOpen
  \bibfield  {author} {\bibinfo {author} {\bibfnamefont {Ryan~T}\ \bibnamefont
  {Wollaeger}}, \bibinfo {author} {\bibfnamefont {Oleg}\ \bibnamefont
  {Korobkin}}, \bibinfo {author} {\bibfnamefont {Christopher~J}\ \bibnamefont
  {Fontes}}, \bibinfo {author} {\bibfnamefont {Stephan~K}\ \bibnamefont
  {Rosswog}}, \bibinfo {author} {\bibfnamefont {Wesley~P}\ \bibnamefont
  {Even}}, \bibinfo {author} {\bibfnamefont {Christopher~L}\ \bibnamefont
  {Fryer}}, \bibinfo {author} {\bibfnamefont {Jesper}\ \bibnamefont
  {Sollerman}}, \bibinfo {author} {\bibfnamefont {Aimee~L}\ \bibnamefont
  {Hungerford}}, \bibinfo {author} {\bibfnamefont {Daniel~R}\ \bibnamefont
  {van Rossum}}, \ and\ \bibinfo {author} {\bibfnamefont {Allan~B}\
  \bibnamefont {Wollaber}},\ }\bibfield  {title} {\enquote {\bibinfo {title}
  {{Impact of ejecta morphology and composition on the electromagnetic
  signatures of neutron star mergers}},}\ }\href {\doibase
  10.1093/mnras/sty1018} {\bibfield  {journal} {\bibinfo  {journal} {Monthly
  Notices of the Royal Astronomical Society}\ }\textbf {\bibinfo {volume}
  {478}},\ \bibinfo {pages} {3298--3334} (\bibinfo {year} {2018})},\ \Eprint
  {http://arxiv.org/abs/https://academic.oup.com/mnras/article-pdf/478/3/3298/25067894/sty1018.pdf}
  {https://academic.oup.com/mnras/article-pdf/478/3/3298/25067894/sty1018.pdf}
  \BibitemShut {NoStop}%
\bibitem [{\citenamefont {{Fontes}}\ \emph {et~al.}(2015)\citenamefont
  {{Fontes}}, \citenamefont {{Zhang}}, \citenamefont {{Abdallah Jr.}},
  \citenamefont {{Clark}}, \citenamefont {{Kilcrease}}, \citenamefont
  {{Colgan}}, \citenamefont {{Cunningham}}, \citenamefont {{Hakel}},
  \citenamefont {{Magee}},\ and\ \citenamefont
  {{Sherrill}}}]{2015JPhB...48n4014F}%
  \BibitemOpen
  \bibfield  {author} {\bibinfo {author} {\bibfnamefont {C.~J.}\ \bibnamefont
  {{Fontes}}}, \bibinfo {author} {\bibfnamefont {H.~L.}\ \bibnamefont
  {{Zhang}}}, \bibinfo {author} {\bibfnamefont {J.}~\bibnamefont {{Abdallah
  Jr.}}}, \bibinfo {author} {\bibfnamefont {R.~E.~H.}\ \bibnamefont {{Clark}}},
  \bibinfo {author} {\bibfnamefont {D.~P.}\ \bibnamefont {{Kilcrease}}},
  \bibinfo {author} {\bibfnamefont {J.}~\bibnamefont {{Colgan}}}, \bibinfo
  {author} {\bibfnamefont {R.~T.}\ \bibnamefont {{Cunningham}}}, \bibinfo
  {author} {\bibfnamefont {P.}~\bibnamefont {{Hakel}}}, \bibinfo {author}
  {\bibfnamefont {N.~H.}\ \bibnamefont {{Magee}}}, \ and\ \bibinfo {author}
  {\bibfnamefont {M.~E.}\ \bibnamefont {{Sherrill}}},\ }\bibfield  {title}
  {\enquote {\bibinfo {title} {{The Los Alamos suite of relativistic atomic
  physics codes}},}\ }\href {\doibase 10.1088/0953-4075/48/14/144014}
  {\bibfield  {journal} {\bibinfo  {journal} {Journal of Physics B Atomic
  Molecular Physics}\ }\textbf {\bibinfo {volume} {48}},\ \bibinfo {eid}
  {144014} (\bibinfo {year} {2015})}\BibitemShut {NoStop}%
\bibitem [{\citenamefont {{Fontes}}\ \emph {et~al.}(2020)\citenamefont
  {{Fontes}}, \citenamefont {{Fryer}}, \citenamefont {{Hungerford}},
  \citenamefont {{Wollaeger}},\ and\ \citenamefont
  {{Korobkin}}}]{2020MNRAS.493.4143F}%
  \BibitemOpen
  \bibfield  {author} {\bibinfo {author} {\bibfnamefont {C.~J.}\ \bibnamefont
  {{Fontes}}}, \bibinfo {author} {\bibfnamefont {C.~L.}\ \bibnamefont
  {{Fryer}}}, \bibinfo {author} {\bibfnamefont {A.~L.}\ \bibnamefont
  {{Hungerford}}}, \bibinfo {author} {\bibfnamefont {R.~T.}\ \bibnamefont
  {{Wollaeger}}}, \ and\ \bibinfo {author} {\bibfnamefont {O.}~\bibnamefont
  {{Korobkin}}},\ }\bibfield  {title} {\enquote {\bibinfo {title} {{A
  line-binned treatment of opacities for the spectra and light curves from
  neutron star mergers}},}\ }\href {\doibase 10.1093/mnras/staa485} {\bibfield
  {journal} {\bibinfo  {journal} {\mnras}\ }\textbf {\bibinfo {volume} {493}},\
  \bibinfo {pages} {4143--4171} (\bibinfo {year} {2020})}\BibitemShut {NoStop}%
\bibitem [{\citenamefont {{Wofford}}\ \emph {et~al.}(2022)\citenamefont
  {{Wofford}}, \citenamefont {{Yelikar}}, \citenamefont {{Gallagher}},
  \citenamefont {{Champion}}, \citenamefont {{Wysocki}}, \citenamefont
  {{Delfavero}}, \citenamefont {{Lange}}, \citenamefont {{Rose}}, \citenamefont
  {{Valsan}}, \citenamefont {{Morisaki}}, \citenamefont {{Read}}, \citenamefont
  {{Henshaw}},\ and\ \citenamefont {{O'Shaughnessy}}}]{Wofford22}%
  \BibitemOpen
  \bibfield  {author} {\bibinfo {author} {\bibfnamefont {J.}~\bibnamefont
  {{Wofford}}}, \bibinfo {author} {\bibfnamefont {A.}~\bibnamefont
  {{Yelikar}}}, \bibinfo {author} {\bibfnamefont {H.}~\bibnamefont
  {{Gallagher}}}, \bibinfo {author} {\bibfnamefont {E.}~\bibnamefont
  {{Champion}}}, \bibinfo {author} {\bibfnamefont {D.}~\bibnamefont
  {{Wysocki}}}, \bibinfo {author} {\bibfnamefont {V.}~\bibnamefont
  {{Delfavero}}}, \bibinfo {author} {\bibfnamefont {J.}~\bibnamefont
  {{Lange}}}, \bibinfo {author} {\bibfnamefont {C.}~\bibnamefont {{Rose}}},
  \bibinfo {author} {\bibfnamefont {V.}~\bibnamefont {{Valsan}}}, \bibinfo
  {author} {\bibfnamefont {S.}~\bibnamefont {{Morisaki}}}, \bibinfo {author}
  {\bibfnamefont {J.}~\bibnamefont {{Read}}}, \bibinfo {author} {\bibfnamefont
  {C.}~\bibnamefont {{Henshaw}}}, \ and\ \bibinfo {author} {\bibfnamefont
  {R.}~\bibnamefont {{O'Shaughnessy}}},\ }\bibfield  {title} {\enquote
  {\bibinfo {title} {{Expanding RIFT: Improving performance for GW parameter
  inference}},}\ }\href {\doibase 10.48550/arXiv.2210.07912} {\bibfield
  {journal} {\bibinfo  {journal} {arXiv e-prints}\ ,\ \bibinfo {eid}
  {arXiv:2210.07912}} (\bibinfo {year} {2022})},\ \Eprint
  {http://arxiv.org/abs/2210.07912} {arXiv:2210.07912 [gr-qc]} \BibitemShut
  {NoStop}%
\bibitem [{\citenamefont {{Cameron}}(1957)}]{1957PASP...69..201C}%
  \BibitemOpen
  \bibfield  {author} {\bibinfo {author} {\bibfnamefont {A.~G.~W.}\
  \bibnamefont {{Cameron}}},\ }\bibfield  {title} {\enquote {\bibinfo {title}
  {{Nuclear Reactions in Stars and Nucleogenesis}},}\ }\href {\doibase
  10.1086/127051} {\bibfield  {journal} {\bibinfo  {journal} {\pasp}\ }\textbf
  {\bibinfo {volume} {69}},\ \bibinfo {pages} {201} (\bibinfo {year}
  {1957})}\BibitemShut {NoStop}%
\bibitem [{\citenamefont {Burbidge}\ \emph {et~al.}(1957)\citenamefont
  {Burbidge}, \citenamefont {Burbidge}, \citenamefont {Fowler},\ and\
  \citenamefont {Hoyle}}]{RevModPhys.29.547}%
  \BibitemOpen
  \bibfield  {author} {\bibinfo {author} {\bibfnamefont {E.~Margaret}\
  \bibnamefont {Burbidge}}, \bibinfo {author} {\bibfnamefont {G.~R.}\
  \bibnamefont {Burbidge}}, \bibinfo {author} {\bibfnamefont {William~A.}\
  \bibnamefont {Fowler}}, \ and\ \bibinfo {author} {\bibfnamefont
  {F.}~\bibnamefont {Hoyle}},\ }\bibfield  {title} {\enquote {\bibinfo {title}
  {Synthesis of the elements in stars},}\ }\href {\doibase
  10.1103/RevModPhys.29.547} {\bibfield  {journal} {\bibinfo  {journal} {Rev.
  Mod. Phys.}\ }\textbf {\bibinfo {volume} {29}},\ \bibinfo {pages} {547--650}
  (\bibinfo {year} {1957})}\BibitemShut {NoStop}%
\bibitem [{\citenamefont {{M{\"o}ller}}\ \emph {et~al.}(2016)\citenamefont
  {{M{\"o}ller}}, \citenamefont {{Sierk}}, \citenamefont {{Ichikawa}},\ and\
  \citenamefont {{Sagawa}}}]{2016ADNDT.109....1M}%
  \BibitemOpen
  \bibfield  {author} {\bibinfo {author} {\bibfnamefont {P.}~\bibnamefont
  {{M{\"o}ller}}}, \bibinfo {author} {\bibfnamefont {A.~J.}\ \bibnamefont
  {{Sierk}}}, \bibinfo {author} {\bibfnamefont {T.}~\bibnamefont {{Ichikawa}}},
  \ and\ \bibinfo {author} {\bibfnamefont {H.}~\bibnamefont {{Sagawa}}},\
  }\bibfield  {title} {\enquote {\bibinfo {title} {{Nuclear ground-state masses
  and deformations: FRDM(2012)}},}\ }\href {\doibase 10.1016/j.adt.2015.10.002}
  {\bibfield  {journal} {\bibinfo  {journal} {Atomic Data and Nuclear Data
  Tables}\ }\textbf {\bibinfo {volume} {109}},\ \bibinfo {pages} {1--204}
  (\bibinfo {year} {2016})},\ \Eprint {http://arxiv.org/abs/1508.06294}
  {arXiv:1508.06294 [nucl-th]} \BibitemShut {NoStop}%
\bibitem [{\citenamefont {{Pearson}}\ \emph {et~al.}(2014)\citenamefont
  {{Pearson}}, \citenamefont {{Chamel}}, \citenamefont {{Fantina}},\ and\
  \citenamefont {{Goriely}}}]{2014EPJA...50...43P}%
  \BibitemOpen
  \bibfield  {author} {\bibinfo {author} {\bibfnamefont {J.~M.}\ \bibnamefont
  {{Pearson}}}, \bibinfo {author} {\bibfnamefont {N.}~\bibnamefont {{Chamel}}},
  \bibinfo {author} {\bibfnamefont {A.~F.}\ \bibnamefont {{Fantina}}}, \ and\
  \bibinfo {author} {\bibfnamefont {S.}~\bibnamefont {{Goriely}}},\ }\bibfield
  {title} {\enquote {\bibinfo {title} {{Symmetry energy: nuclear masses and
  neutron stars}},}\ }\href {\doibase 10.1140/epja/i2014-14043-8} {\bibfield
  {journal} {\bibinfo  {journal} {European Physical Journal A}\ }\textbf
  {\bibinfo {volume} {50}},\ \bibinfo {eid} {43} (\bibinfo {year} {2014})},\
  \Eprint {http://arxiv.org/abs/1309.2783} {arXiv:1309.2783 [nucl-th]}
  \BibitemShut {NoStop}%
\bibitem [{\citenamefont {{Mumpower}}\ \emph {et~al.}(2020)\citenamefont
  {{Mumpower}}, \citenamefont {{Jaffke}}, \citenamefont {{Verriere}},\ and\
  \citenamefont {{Randrup}}}]{2020PhRvC.101e4607M}%
  \BibitemOpen
  \bibfield  {author} {\bibinfo {author} {\bibfnamefont {M.~R.}\ \bibnamefont
  {{Mumpower}}}, \bibinfo {author} {\bibfnamefont {P.}~\bibnamefont
  {{Jaffke}}}, \bibinfo {author} {\bibfnamefont {M.}~\bibnamefont
  {{Verriere}}}, \ and\ \bibinfo {author} {\bibfnamefont {J.}~\bibnamefont
  {{Randrup}}},\ }\bibfield  {title} {\enquote {\bibinfo {title} {{Primary
  fission fragment mass yields across the chart of nuclides}},}\ }\href
  {\doibase 10.1103/PhysRevC.101.054607} {\bibfield  {journal} {\bibinfo
  {journal} {\prc}\ }\textbf {\bibinfo {volume} {101}},\ \bibinfo {eid}
  {054607} (\bibinfo {year} {2020})},\ \Eprint
  {http://arxiv.org/abs/1911.06344} {arXiv:1911.06344 [nucl-th]} \BibitemShut
  {NoStop}%
\bibitem [{\citenamefont {{Panov}}\ \emph {et~al.}(2010)\citenamefont
  {{Panov}}, \citenamefont {{Korneev}}, \citenamefont {{Rauscher}},
  \citenamefont {{Mart{\'\i}nez-Pinedo}}, \citenamefont {{Keli{\'c}-Heil}},
  \citenamefont {{Zinner}},\ and\ \citenamefont
  {{Thielemann}}}]{2010A&A...513A..61P}%
  \BibitemOpen
  \bibfield  {author} {\bibinfo {author} {\bibfnamefont {I.~V.}\ \bibnamefont
  {{Panov}}}, \bibinfo {author} {\bibfnamefont {I.~Yu.}\ \bibnamefont
  {{Korneev}}}, \bibinfo {author} {\bibfnamefont {T.}~\bibnamefont
  {{Rauscher}}}, \bibinfo {author} {\bibfnamefont {G.}~\bibnamefont
  {{Mart{\'\i}nez-Pinedo}}}, \bibinfo {author} {\bibfnamefont {A.}~\bibnamefont
  {{Keli{\'c}-Heil}}}, \bibinfo {author} {\bibfnamefont {N.~T.}\ \bibnamefont
  {{Zinner}}}, \ and\ \bibinfo {author} {\bibfnamefont {F.~K.}\ \bibnamefont
  {{Thielemann}}},\ }\bibfield  {title} {\enquote {\bibinfo {title}
  {{Neutron-induced astrophysical reaction rates for translead nuclei}},}\
  }\href {\doibase 10.1051/0004-6361/200911967} {\bibfield  {journal} {\bibinfo
   {journal} {\aap}\ }\textbf {\bibinfo {volume} {513}},\ \bibinfo {eid} {A61}
  (\bibinfo {year} {2010})},\ \Eprint {http://arxiv.org/abs/0911.2181}
  {arXiv:0911.2181 [astro-ph.SR]} \BibitemShut {NoStop}%
\bibitem [{\citenamefont {{Kiuchi}}\ \emph {et~al.}(2022)\citenamefont
  {{Kiuchi}}, \citenamefont {{Fujibayashi}}, \citenamefont {{Hayashi}},
  \citenamefont {{Kyutoku}}, \citenamefont {{Sekiguchi}},\ and\ \citenamefont
  {{Shibata}}}]{2022arXiv221107637K}%
  \BibitemOpen
  \bibfield  {author} {\bibinfo {author} {\bibfnamefont {Kenta}\ \bibnamefont
  {{Kiuchi}}}, \bibinfo {author} {\bibfnamefont {Sho}\ \bibnamefont
  {{Fujibayashi}}}, \bibinfo {author} {\bibfnamefont {Kota}\ \bibnamefont
  {{Hayashi}}}, \bibinfo {author} {\bibfnamefont {Koutarou}\ \bibnamefont
  {{Kyutoku}}}, \bibinfo {author} {\bibfnamefont {Yuichiro}\ \bibnamefont
  {{Sekiguchi}}}, \ and\ \bibinfo {author} {\bibfnamefont {Masaru}\
  \bibnamefont {{Shibata}}},\ }\bibfield  {title} {\enquote {\bibinfo {title}
  {{Self-consistent picture of the mass ejection from a one second-long binary
  neutron star merger leaving a short-lived remnant in general-relativistic
  neutrino-radiation magnetohydrodynamic simulation}},}\ }\href {\doibase
  10.48550/arXiv.2211.07637} {\bibfield  {journal} {\bibinfo  {journal} {arXiv
  e-prints}\ ,\ \bibinfo {eid} {arXiv:2211.07637}} (\bibinfo {year} {2022})},\
  \Eprint {http://arxiv.org/abs/2211.07637} {arXiv:2211.07637 [astro-ph.HE]}
  \BibitemShut {NoStop}%
\bibitem [{\citenamefont {{Sprouse}}\ \emph {et~al.}(2015)\citenamefont
  {{Sprouse}}, \citenamefont {{Mumpower}}, \citenamefont {{Surman}},\ and\
  \citenamefont {{Aprahamian}}}]{2015APS..DNP.EA097S}%
  \BibitemOpen
  \bibfield  {author} {\bibinfo {author} {\bibfnamefont {Trevor}\ \bibnamefont
  {{Sprouse}}}, \bibinfo {author} {\bibfnamefont {Matthew}\ \bibnamefont
  {{Mumpower}}}, \bibinfo {author} {\bibfnamefont {Rebecca}\ \bibnamefont
  {{Surman}}}, \ and\ \bibinfo {author} {\bibfnamefont {Ani}\ \bibnamefont
  {{Aprahamian}}},\ }\bibfield  {title} {\enquote {\bibinfo {title} {{A
  generalized framework for nucleosynthesis calculations}},}\ }in\ \href@noop
  {} {\emph {\bibinfo {booktitle} {APS Division of Nuclear Physics Meeting
  Abstracts}}},\ \bibinfo {series} {APS Meeting Abstracts}, Vol.\ \bibinfo
  {volume} {2015}\ (\bibinfo {year} {2015})\ p.\ \bibinfo {pages}
  {EA.097}\BibitemShut {NoStop}%
\bibitem [{\citenamefont {{Arlandini}}\ \emph {et~al.}(1999)\citenamefont
  {{Arlandini}}, \citenamefont {{K{\"a}ppeler}}, \citenamefont {{Wisshak}},
  \citenamefont {{Gallino}}, \citenamefont {{Lugaro}}, \citenamefont
  {{Busso}},\ and\ \citenamefont {{Straniero}}}]{Arlandini1999}%
  \BibitemOpen
  \bibfield  {author} {\bibinfo {author} {\bibfnamefont {Claudio}\ \bibnamefont
  {{Arlandini}}}, \bibinfo {author} {\bibfnamefont {Franz}\ \bibnamefont
  {{K{\"a}ppeler}}}, \bibinfo {author} {\bibfnamefont {Klaus}\ \bibnamefont
  {{Wisshak}}}, \bibinfo {author} {\bibfnamefont {Roberto}\ \bibnamefont
  {{Gallino}}}, \bibinfo {author} {\bibfnamefont {Maria}\ \bibnamefont
  {{Lugaro}}}, \bibinfo {author} {\bibfnamefont {Maurizio}\ \bibnamefont
  {{Busso}}}, \ and\ \bibinfo {author} {\bibfnamefont {Oscar}\ \bibnamefont
  {{Straniero}}},\ }\bibfield  {title} {\enquote {\bibinfo {title} {{Neutron
  Capture in Low-Mass Asymptotic Giant Branch Stars: Cross Sections and
  Abundance Signatures}},}\ }\href {\doibase 10.1086/307938} {\bibfield
  {journal} {\bibinfo  {journal} {\apj}\ }\textbf {\bibinfo {volume} {525}},\
  \bibinfo {pages} {886--900} (\bibinfo {year} {1999})},\ \Eprint
  {http://arxiv.org/abs/astro-ph/9906266} {arXiv:astro-ph/9906266 [astro-ph]}
  \BibitemShut {NoStop}%
\bibitem [{\citenamefont {{Sneden}}\ \emph {et~al.}(2008)\citenamefont
  {{Sneden}}, \citenamefont {{Cowan}},\ and\ \citenamefont
  {{Gallino}}}]{Sneden2008}%
  \BibitemOpen
  \bibfield  {author} {\bibinfo {author} {\bibfnamefont {C.}~\bibnamefont
  {{Sneden}}}, \bibinfo {author} {\bibfnamefont {J.~J.}\ \bibnamefont
  {{Cowan}}}, \ and\ \bibinfo {author} {\bibfnamefont {R.}~\bibnamefont
  {{Gallino}}},\ }\bibfield  {title} {\enquote {\bibinfo {title}
  {{Neutron-capture elements in the early galaxy.}}}\ }\href {\doibase
  10.1146/annurev.astro.46.060407.145207} {\bibfield  {journal} {\bibinfo
  {journal} {\araa}\ }\textbf {\bibinfo {volume} {46}},\ \bibinfo {pages}
  {241--288} (\bibinfo {year} {2008})}\BibitemShut {NoStop}%
\bibitem [{\citenamefont {Rosswog}\ \emph {et~al.}(2017)\citenamefont
  {Rosswog}, \citenamefont {Feindt}, \citenamefont {Korobkin}, \citenamefont
  {Wu}, \citenamefont {Sollerman}, \citenamefont {Goobar},\ and\ \citenamefont
  {Martinez-Pinedo}}]{Rosswog_2017}%
  \BibitemOpen
  \bibfield  {author} {\bibinfo {author} {\bibfnamefont {S}~\bibnamefont
  {Rosswog}}, \bibinfo {author} {\bibfnamefont {U}~\bibnamefont {Feindt}},
  \bibinfo {author} {\bibfnamefont {O}~\bibnamefont {Korobkin}}, \bibinfo
  {author} {\bibfnamefont {M-R}\ \bibnamefont {Wu}}, \bibinfo {author}
  {\bibfnamefont {J}~\bibnamefont {Sollerman}}, \bibinfo {author}
  {\bibfnamefont {A}~\bibnamefont {Goobar}}, \ and\ \bibinfo {author}
  {\bibfnamefont {G}~\bibnamefont {Martinez-Pinedo}},\ }\bibfield  {title}
  {\enquote {\bibinfo {title} {Detectability of compact binary merger
  macronovae},}\ }\href {\doibase 10.1088/1361-6382/aa68a9} {\bibfield
  {journal} {\bibinfo  {journal} {Classical and Quantum Gravity}\ }\textbf
  {\bibinfo {volume} {34}},\ \bibinfo {pages} {104001} (\bibinfo {year}
  {2017})}\BibitemShut {NoStop}%
\bibitem [{\citenamefont {{Radice}}\ \emph {et~al.}(2018)\citenamefont
  {{Radice}}, \citenamefont {{Perego}}, \citenamefont {{Hotokezaka}},
  \citenamefont {{Bernuzzi}}, \citenamefont {{Fromm}},\ and\ \citenamefont
  {{Roberts}}}]{2018ApJ...869L..35R}%
  \BibitemOpen
  \bibfield  {author} {\bibinfo {author} {\bibfnamefont {David}\ \bibnamefont
  {{Radice}}}, \bibinfo {author} {\bibfnamefont {Albino}\ \bibnamefont
  {{Perego}}}, \bibinfo {author} {\bibfnamefont {Kenta}\ \bibnamefont
  {{Hotokezaka}}}, \bibinfo {author} {\bibfnamefont {Sebastiano}\ \bibnamefont
  {{Bernuzzi}}}, \bibinfo {author} {\bibfnamefont {Steven~A.}\ \bibnamefont
  {{Fromm}}}, \ and\ \bibinfo {author} {\bibfnamefont {Luke~F.}\ \bibnamefont
  {{Roberts}}},\ }\bibfield  {title} {\enquote {\bibinfo {title}
  {{Viscous-dynamical Ejecta from Binary Neutron Star Mergers}},}\ }\href
  {\doibase 10.3847/2041-8213/aaf053} {\bibfield  {journal} {\bibinfo
  {journal} {\apjl}\ }\textbf {\bibinfo {volume} {869}},\ \bibinfo {eid} {L35}
  (\bibinfo {year} {2018})},\ \Eprint {http://arxiv.org/abs/1809.11163}
  {arXiv:1809.11163 [astro-ph.HE]} \BibitemShut {NoStop}%
\bibitem [{\citenamefont {{Fern{\'a}ndez}}\ \emph {et~al.}(2019)\citenamefont
  {{Fern{\'a}ndez}}, \citenamefont {{Tchekhovskoy}}, \citenamefont
  {{Quataert}}, \citenamefont {{Foucart}},\ and\ \citenamefont
  {{Kasen}}}]{2019MNRAS.482.3373F}%
  \BibitemOpen
  \bibfield  {author} {\bibinfo {author} {\bibfnamefont {Rodrigo}\ \bibnamefont
  {{Fern{\'a}ndez}}}, \bibinfo {author} {\bibfnamefont {Alexander}\
  \bibnamefont {{Tchekhovskoy}}}, \bibinfo {author} {\bibfnamefont {Eliot}\
  \bibnamefont {{Quataert}}}, \bibinfo {author} {\bibfnamefont {Francois}\
  \bibnamefont {{Foucart}}}, \ and\ \bibinfo {author} {\bibfnamefont {Daniel}\
  \bibnamefont {{Kasen}}},\ }\bibfield  {title} {\enquote {\bibinfo {title}
  {{Long-term GRMHD simulations of neutron star merger accretion discs:
  implications for electromagnetic counterparts}},}\ }\href {\doibase
  10.1093/mnras/sty2932} {\bibfield  {journal} {\bibinfo  {journal} {\mnras}\
  }\textbf {\bibinfo {volume} {482}},\ \bibinfo {pages} {3373--3393} (\bibinfo
  {year} {2019})},\ \Eprint {http://arxiv.org/abs/1808.00461} {arXiv:1808.00461
  [astro-ph.HE]} \BibitemShut {NoStop}%
\bibitem [{\citenamefont {{Nedora}}\ \emph {et~al.}(2021)\citenamefont
  {{Nedora}}, \citenamefont {{Bernuzzi}}, \citenamefont {{Radice}},
  \citenamefont {{Daszuta}}, \citenamefont {{Endrizzi}}, \citenamefont
  {{Perego}}, \citenamefont {{Prakash}}, \citenamefont {{Safarzadeh}},
  \citenamefont {{Schianchi}},\ and\ \citenamefont
  {{Logoteta}}}]{2021ApJ...906...98N}%
  \BibitemOpen
  \bibfield  {author} {\bibinfo {author} {\bibfnamefont {Vsevolod}\
  \bibnamefont {{Nedora}}}, \bibinfo {author} {\bibfnamefont {Sebastiano}\
  \bibnamefont {{Bernuzzi}}}, \bibinfo {author} {\bibfnamefont {David}\
  \bibnamefont {{Radice}}}, \bibinfo {author} {\bibfnamefont {Boris}\
  \bibnamefont {{Daszuta}}}, \bibinfo {author} {\bibfnamefont {Andrea}\
  \bibnamefont {{Endrizzi}}}, \bibinfo {author} {\bibfnamefont {Albino}\
  \bibnamefont {{Perego}}}, \bibinfo {author} {\bibfnamefont {Aviral}\
  \bibnamefont {{Prakash}}}, \bibinfo {author} {\bibfnamefont {Mohammadtaher}\
  \bibnamefont {{Safarzadeh}}}, \bibinfo {author} {\bibfnamefont {Federico}\
  \bibnamefont {{Schianchi}}}, \ and\ \bibinfo {author} {\bibfnamefont
  {Domenico}\ \bibnamefont {{Logoteta}}},\ }\bibfield  {title} {\enquote
  {\bibinfo {title} {{Numerical Relativity Simulations of the Neutron Star
  Merger GW170817: Long-term Remnant Evolutions, Winds, Remnant Disks, and
  Nucleosynthesis}},}\ }\href {\doibase 10.3847/1538-4357/abc9be} {\bibfield
  {journal} {\bibinfo  {journal} {\apj}\ }\textbf {\bibinfo {volume} {906}},\
  \bibinfo {eid} {98} (\bibinfo {year} {2021})},\ \Eprint
  {http://arxiv.org/abs/2008.04333} {arXiv:2008.04333 [astro-ph.HE]}
  \BibitemShut {NoStop}%
\bibitem [{\citenamefont {{Abohalima}}\ and\ \citenamefont
  {{Frebel}}(2018)}]{2018ApJS..238...36A}%
  \BibitemOpen
  \bibfield  {author} {\bibinfo {author} {\bibfnamefont {Abdu}\ \bibnamefont
  {{Abohalima}}}\ and\ \bibinfo {author} {\bibfnamefont {Anna}\ \bibnamefont
  {{Frebel}}},\ }\bibfield  {title} {\enquote {\bibinfo {title}
  {{JINAbase{\textemdash}A Database for Chemical Abundances of Metal-poor
  Stars}},}\ }\href {\doibase 10.3847/1538-4365/aadfe9} {\bibfield  {journal}
  {\bibinfo  {journal} {\apjs}\ }\textbf {\bibinfo {volume} {238}},\ \bibinfo
  {eid} {36} (\bibinfo {year} {2018})},\ \Eprint
  {http://arxiv.org/abs/1711.04410} {arXiv:1711.04410 [astro-ph.SR]}
  \BibitemShut {NoStop}%
\bibitem [{\citenamefont {{Sneden}}\ \emph
  {et~al.}(2003{\natexlab{b}})\citenamefont {{Sneden}}, \citenamefont
  {{Cowan}}, \citenamefont {{Lawler}}, \citenamefont {{Ivans}}, \citenamefont
  {{Burles}}, \citenamefont {{Beers}}, \citenamefont {{Primas}}, \citenamefont
  {{Hill}}, \citenamefont {{Truran}}, \citenamefont {{Fuller}}, \citenamefont
  {{Pfeiffer}},\ and\ \citenamefont {{Kratz}}}]{SNE03}%
  \BibitemOpen
  \bibfield  {author} {\bibinfo {author} {\bibfnamefont {Christopher}\
  \bibnamefont {{Sneden}}}, \bibinfo {author} {\bibfnamefont {John~J.}\
  \bibnamefont {{Cowan}}}, \bibinfo {author} {\bibfnamefont {James~E.}\
  \bibnamefont {{Lawler}}}, \bibinfo {author} {\bibfnamefont {Inese~I.}\
  \bibnamefont {{Ivans}}}, \bibinfo {author} {\bibfnamefont {Scott}\
  \bibnamefont {{Burles}}}, \bibinfo {author} {\bibfnamefont {Timothy~C.}\
  \bibnamefont {{Beers}}}, \bibinfo {author} {\bibfnamefont {Francesca}\
  \bibnamefont {{Primas}}}, \bibinfo {author} {\bibfnamefont {Vanessa}\
  \bibnamefont {{Hill}}}, \bibinfo {author} {\bibfnamefont {James~W.}\
  \bibnamefont {{Truran}}}, \bibinfo {author} {\bibfnamefont {George~M.}\
  \bibnamefont {{Fuller}}}, \bibinfo {author} {\bibfnamefont {Bernd}\
  \bibnamefont {{Pfeiffer}}}, \ and\ \bibinfo {author} {\bibfnamefont
  {Karl-Ludwig}\ \bibnamefont {{Kratz}}},\ }\bibfield  {title} {\enquote
  {\bibinfo {title} {{The Extremely Metal-poor, Neutron Capture-rich Star CS
  22892-052: A Comprehensive Abundance Analysis}},}\ }\href {\doibase
  10.1086/375491} {\bibfield  {journal} {\bibinfo  {journal} {\apj}\ }\textbf
  {\bibinfo {volume} {591}},\ \bibinfo {pages} {936--953} (\bibinfo {year}
  {2003}{\natexlab{b}})},\ \Eprint {http://arxiv.org/abs/astro-ph/0303542}
  {arXiv:astro-ph/0303542 [astro-ph]} \BibitemShut {NoStop}%
\bibitem [{\citenamefont {{Hayek}}\ \emph {et~al.}(2009)\citenamefont
  {{Hayek}}, \citenamefont {{Wiesendahl}}, \citenamefont {{Christlieb}},
  \citenamefont {{Eriksson}}, \citenamefont {{Korn}}, \citenamefont
  {{Barklem}}, \citenamefont {{Hill}}, \citenamefont {{Beers}}, \citenamefont
  {{Farouqi}}, \citenamefont {{Pfeiffer}},\ and\ \citenamefont
  {{Kratz}}}]{HAY09}%
  \BibitemOpen
  \bibfield  {author} {\bibinfo {author} {\bibfnamefont {W.}~\bibnamefont
  {{Hayek}}}, \bibinfo {author} {\bibfnamefont {U.}~\bibnamefont
  {{Wiesendahl}}}, \bibinfo {author} {\bibfnamefont {N.}~\bibnamefont
  {{Christlieb}}}, \bibinfo {author} {\bibfnamefont {K.}~\bibnamefont
  {{Eriksson}}}, \bibinfo {author} {\bibfnamefont {A.~J.}\ \bibnamefont
  {{Korn}}}, \bibinfo {author} {\bibfnamefont {P.~S.}\ \bibnamefont
  {{Barklem}}}, \bibinfo {author} {\bibfnamefont {V.}~\bibnamefont {{Hill}}},
  \bibinfo {author} {\bibfnamefont {T.~C.}\ \bibnamefont {{Beers}}}, \bibinfo
  {author} {\bibfnamefont {K.}~\bibnamefont {{Farouqi}}}, \bibinfo {author}
  {\bibfnamefont {B.}~\bibnamefont {{Pfeiffer}}}, \ and\ \bibinfo {author}
  {\bibfnamefont {K.-L.}\ \bibnamefont {{Kratz}}},\ }\bibfield  {title}
  {\enquote {\bibinfo {title} {{The Hamburg/ESO R-process enhanced star survey
  (HERES). IV. Detailed abundance analysis and age dating of the strongly
  r-process enhanced stars CS 29491-069 and HE 1219-0312}},}\ }\href {\doibase
  10.1051/0004-6361/200811121} {\bibfield  {journal} {\bibinfo  {journal}
  {\aap}\ }\textbf {\bibinfo {volume} {504}},\ \bibinfo {pages} {511--524}
  (\bibinfo {year} {2009})},\ \Eprint {http://arxiv.org/abs/0910.0707}
  {arXiv:0910.0707 [astro-ph.SR]} \BibitemShut {NoStop}%
\bibitem [{\citenamefont {{Hill}}\ \emph
  {et~al.}(2002{\natexlab{b}})\citenamefont {{Hill}}, \citenamefont {{Plez}},
  \citenamefont {{Cayrel}}, \citenamefont {{Beers}}, \citenamefont
  {{Nordstr{\"o}m}}, \citenamefont {{Andersen}}, \citenamefont {{Spite}},
  \citenamefont {{Spite}}, \citenamefont {{Barbuy}}, \citenamefont
  {{Bonifacio}}, \citenamefont {{Depagne}}, \citenamefont {{Fran{\c c}ois}},\
  and\ \citenamefont {{Primas}}}]{HIL02}%
  \BibitemOpen
  \bibfield  {author} {\bibinfo {author} {\bibfnamefont {V.}~\bibnamefont
  {{Hill}}}, \bibinfo {author} {\bibfnamefont {B.}~\bibnamefont {{Plez}}},
  \bibinfo {author} {\bibfnamefont {R.}~\bibnamefont {{Cayrel}}}, \bibinfo
  {author} {\bibfnamefont {T.~C.}\ \bibnamefont {{Beers}}}, \bibinfo {author}
  {\bibfnamefont {B.}~\bibnamefont {{Nordstr{\"o}m}}}, \bibinfo {author}
  {\bibfnamefont {J.}~\bibnamefont {{Andersen}}}, \bibinfo {author}
  {\bibfnamefont {M.}~\bibnamefont {{Spite}}}, \bibinfo {author} {\bibfnamefont
  {F.}~\bibnamefont {{Spite}}}, \bibinfo {author} {\bibfnamefont
  {B.}~\bibnamefont {{Barbuy}}}, \bibinfo {author} {\bibfnamefont
  {P.}~\bibnamefont {{Bonifacio}}}, \bibinfo {author} {\bibfnamefont
  {E.}~\bibnamefont {{Depagne}}}, \bibinfo {author} {\bibfnamefont
  {P.}~\bibnamefont {{Fran{\c c}ois}}}, \ and\ \bibinfo {author} {\bibfnamefont
  {F.}~\bibnamefont {{Primas}}},\ }\bibfield  {title} {\enquote {\bibinfo
  {title} {{First stars. I. The extreme r-element rich, iron-poor halo giant CS
  31082-001. Implications for the r-process site(s) and radioactive
  cosmochronology}},}\ }\href {\doibase 10.1051/0004-6361:20020434} {\bibfield
  {journal} {\bibinfo  {journal} {\aap}\ }\textbf {\bibinfo {volume} {387}},\
  \bibinfo {pages} {560--579} (\bibinfo {year} {2002}{\natexlab{b}})},\ \Eprint
  {http://arxiv.org/abs/astro-ph/0203462} {astro-ph/0203462} \BibitemShut
  {NoStop}%
\bibitem [{\citenamefont {{Westin}}\ \emph {et~al.}(2000)\citenamefont
  {{Westin}}, \citenamefont {{Sneden}}, \citenamefont {{Gustafsson}},\ and\
  \citenamefont {{Cowan}}}]{WES00}%
  \BibitemOpen
  \bibfield  {author} {\bibinfo {author} {\bibfnamefont {J.}~\bibnamefont
  {{Westin}}}, \bibinfo {author} {\bibfnamefont {C.}~\bibnamefont {{Sneden}}},
  \bibinfo {author} {\bibfnamefont {B.}~\bibnamefont {{Gustafsson}}}, \ and\
  \bibinfo {author} {\bibfnamefont {J.~J.}\ \bibnamefont {{Cowan}}},\
  }\bibfield  {title} {\enquote {\bibinfo {title} {{The r-Process-enriched
  Low-Metallicity Giant HD 115444}},}\ }\href {\doibase 10.1086/308407}
  {\bibfield  {journal} {\bibinfo  {journal} {\apj}\ }\textbf {\bibinfo
  {volume} {530}},\ \bibinfo {pages} {783--799} (\bibinfo {year} {2000})},\
  \Eprint {http://arxiv.org/abs/astro-ph/9910376} {astro-ph/9910376}
  \BibitemShut {NoStop}%
\bibitem [{\citenamefont {{Ivans}}\ \emph {et~al.}(2006)\citenamefont
  {{Ivans}}, \citenamefont {{Simmerer}}, \citenamefont {{Sneden}},
  \citenamefont {{Lawler}}, \citenamefont {{Cowan}}, \citenamefont
  {{Gallino}},\ and\ \citenamefont {{Bisterzo}}}]{IVA06}%
  \BibitemOpen
  \bibfield  {author} {\bibinfo {author} {\bibfnamefont {I.~I.}\ \bibnamefont
  {{Ivans}}}, \bibinfo {author} {\bibfnamefont {J.}~\bibnamefont {{Simmerer}}},
  \bibinfo {author} {\bibfnamefont {C.}~\bibnamefont {{Sneden}}}, \bibinfo
  {author} {\bibfnamefont {J.~E.}\ \bibnamefont {{Lawler}}}, \bibinfo {author}
  {\bibfnamefont {J.~J.}\ \bibnamefont {{Cowan}}}, \bibinfo {author}
  {\bibfnamefont {R.}~\bibnamefont {{Gallino}}}, \ and\ \bibinfo {author}
  {\bibfnamefont {S.}~\bibnamefont {{Bisterzo}}},\ }\bibfield  {title}
  {\enquote {\bibinfo {title} {{Near-Ultraviolet Observations of HD 221170: New
  Insights into the Nature of r-Process-rich Stars}},}\ }\href {\doibase
  10.1086/504069} {\bibfield  {journal} {\bibinfo  {journal} {\apj}\ }\textbf
  {\bibinfo {volume} {645}},\ \bibinfo {pages} {613--633} (\bibinfo {year}
  {2006})},\ \Eprint {http://arxiv.org/abs/astro-ph/0604180} {astro-ph/0604180}
  \BibitemShut {NoStop}%
\bibitem [{\citenamefont {{Cowan}}\ \emph {et~al.}(2002)\citenamefont
  {{Cowan}}, \citenamefont {{Sneden}}, \citenamefont {{Burles}}, \citenamefont
  {{Ivans}}, \citenamefont {{Beers}}, \citenamefont {{Truran}}, \citenamefont
  {{Lawler}}, \citenamefont {{Primas}}, \citenamefont {{Fuller}}, \citenamefont
  {{Pfeiffer}},\ and\ \citenamefont {{Kratz}}}]{COW02}%
  \BibitemOpen
  \bibfield  {author} {\bibinfo {author} {\bibfnamefont {J.~J.}\ \bibnamefont
  {{Cowan}}}, \bibinfo {author} {\bibfnamefont {C.}~\bibnamefont {{Sneden}}},
  \bibinfo {author} {\bibfnamefont {S.}~\bibnamefont {{Burles}}}, \bibinfo
  {author} {\bibfnamefont {I.~I.}\ \bibnamefont {{Ivans}}}, \bibinfo {author}
  {\bibfnamefont {T.~C.}\ \bibnamefont {{Beers}}}, \bibinfo {author}
  {\bibfnamefont {J.~W.}\ \bibnamefont {{Truran}}}, \bibinfo {author}
  {\bibfnamefont {J.~E.}\ \bibnamefont {{Lawler}}}, \bibinfo {author}
  {\bibfnamefont {F.}~\bibnamefont {{Primas}}}, \bibinfo {author}
  {\bibfnamefont {G.~M.}\ \bibnamefont {{Fuller}}}, \bibinfo {author}
  {\bibfnamefont {B.}~\bibnamefont {{Pfeiffer}}}, \ and\ \bibinfo {author}
  {\bibfnamefont {K.-L.}\ \bibnamefont {{Kratz}}},\ }\bibfield  {title}
  {\enquote {\bibinfo {title} {{The Chemical Composition and Age of the
  Metal-poor Halo Star BD +17$\deg$3248}},}\ }\href {\doibase 10.1086/340347}
  {\bibfield  {journal} {\bibinfo  {journal} {\apj}\ }\textbf {\bibinfo
  {volume} {572}},\ \bibinfo {pages} {861--879} (\bibinfo {year} {2002})},\
  \Eprint {http://arxiv.org/abs/astro-ph/0202429} {astro-ph/0202429}
  \BibitemShut {NoStop}%
\bibitem [{\citenamefont {{Most}}\ and\ \citenamefont
  {{Raithel}}(2021)}]{2021PhRvD.104l4012M}%
  \BibitemOpen
  \bibfield  {author} {\bibinfo {author} {\bibfnamefont {Elias~R.}\
  \bibnamefont {{Most}}}\ and\ \bibinfo {author} {\bibfnamefont {Carolyn~A.}\
  \bibnamefont {{Raithel}}},\ }\bibfield  {title} {\enquote {\bibinfo {title}
  {{Impact of the nuclear symmetry energy on the post-merger phase of a binary
  neutron star coalescence}},}\ }\href {\doibase 10.1103/PhysRevD.104.124012}
  {\bibfield  {journal} {\bibinfo  {journal} {\prd}\ }\textbf {\bibinfo
  {volume} {104}},\ \bibinfo {eid} {124012} (\bibinfo {year} {2021})},\ \Eprint
  {http://arxiv.org/abs/2107.06804} {arXiv:2107.06804 [astro-ph.HE]}
  \BibitemShut {NoStop}%
\bibitem [{\citenamefont {Heinzel}\ \emph {et~al.}(2021)\citenamefont
  {Heinzel}, \citenamefont {Coughlin}, \citenamefont {Dietrich}, \citenamefont
  {Bulla}, \citenamefont {Antier}, \citenamefont {Christensen}, \citenamefont
  {Coulter}, \citenamefont {Foley}, \citenamefont {Issa},\ and\ \citenamefont
  {Khetan}}]{gwastro-mergers-em-CoughlinGPKilonova-2020}%
  \BibitemOpen
  \bibfield  {author} {\bibinfo {author} {\bibfnamefont {J}~\bibnamefont
  {Heinzel}}, \bibinfo {author} {\bibfnamefont {M~W}\ \bibnamefont {Coughlin}},
  \bibinfo {author} {\bibfnamefont {T}~\bibnamefont {Dietrich}}, \bibinfo
  {author} {\bibfnamefont {M}~\bibnamefont {Bulla}}, \bibinfo {author}
  {\bibfnamefont {S}~\bibnamefont {Antier}}, \bibinfo {author} {\bibfnamefont
  {N}~\bibnamefont {Christensen}}, \bibinfo {author} {\bibfnamefont {D~A}\
  \bibnamefont {Coulter}}, \bibinfo {author} {\bibfnamefont {R~J}\ \bibnamefont
  {Foley}}, \bibinfo {author} {\bibfnamefont {L}~\bibnamefont {Issa}}, \ and\
  \bibinfo {author} {\bibfnamefont {N}~\bibnamefont {Khetan}},\ }\bibfield
  {title} {\enquote {\bibinfo {title} {Comparing inclination-dependent analyses
  of kilonova transients},}\ }\href {\doibase 10.1093/mnras/stab221} {\bibfield
   {journal} {\bibinfo  {journal} {Monthly Notices of the Royal Astronomical
  Society}\ }\textbf {\bibinfo {volume} {502}},\ \bibinfo {pages} {3057--3065}
  (\bibinfo {year} {2021})}\BibitemShut {NoStop}%
\bibitem [{\citenamefont {{Coughlin}}\ \emph {et~al.}(2018)\citenamefont
  {{Coughlin}}, \citenamefont {{Dietrich}}, \citenamefont {{Doctor}},
  \citenamefont {{Kasen}}, \citenamefont {{Coughlin}}, \citenamefont
  {{Jerkstrand}}, \citenamefont {{Leloudas}}, \citenamefont {{McBrien}},
  \citenamefont {{Metzger}}, \citenamefont {{O'Shaughnessy}},\ and\
  \citenamefont {{Smartt}}}]{2018MNRAS.480.3871C}%
  \BibitemOpen
  \bibfield  {author} {\bibinfo {author} {\bibfnamefont {Michael~W.}\
  \bibnamefont {{Coughlin}}}, \bibinfo {author} {\bibfnamefont {Tim}\
  \bibnamefont {{Dietrich}}}, \bibinfo {author} {\bibfnamefont {Zoheyr}\
  \bibnamefont {{Doctor}}}, \bibinfo {author} {\bibfnamefont {Daniel}\
  \bibnamefont {{Kasen}}}, \bibinfo {author} {\bibfnamefont {Scott}\
  \bibnamefont {{Coughlin}}}, \bibinfo {author} {\bibfnamefont {Anders}\
  \bibnamefont {{Jerkstrand}}}, \bibinfo {author} {\bibfnamefont {Giorgos}\
  \bibnamefont {{Leloudas}}}, \bibinfo {author} {\bibfnamefont {Owen}\
  \bibnamefont {{McBrien}}}, \bibinfo {author} {\bibfnamefont {Brian~D.}\
  \bibnamefont {{Metzger}}}, \bibinfo {author} {\bibfnamefont {Richard}\
  \bibnamefont {{O'Shaughnessy}}}, \ and\ \bibinfo {author} {\bibfnamefont
  {Stephen~J.}\ \bibnamefont {{Smartt}}},\ }\bibfield  {title} {\enquote
  {\bibinfo {title} {{Constraints on the neutron star equation of state from
  AT2017gfo using radiative transfer simulations}},}\ }\href {\doibase
  10.1093/mnras/sty2174} {\bibfield  {journal} {\bibinfo  {journal} {\mnras}\
  }\textbf {\bibinfo {volume} {480}},\ \bibinfo {pages} {3871--3878} (\bibinfo
  {year} {2018})}\BibitemShut {NoStop}%
\bibitem [{\citenamefont {{Coughlin}}\ \emph {et~al.}(2019)\citenamefont
  {{Coughlin}}, \citenamefont {{Dietrich}}, \citenamefont {{Margalit}},\ and\
  \citenamefont {{Metzger}}}]{2019MNRAS.489L..91C}%
  \BibitemOpen
  \bibfield  {author} {\bibinfo {author} {\bibfnamefont {Michael~W.}\
  \bibnamefont {{Coughlin}}}, \bibinfo {author} {\bibfnamefont {Tim}\
  \bibnamefont {{Dietrich}}}, \bibinfo {author} {\bibfnamefont {Ben}\
  \bibnamefont {{Margalit}}}, \ and\ \bibinfo {author} {\bibfnamefont
  {Brian~D.}\ \bibnamefont {{Metzger}}},\ }\bibfield  {title} {\enquote
  {\bibinfo {title} {{Multimessenger Bayesian parameter inference of a binary
  neutron star merger}},}\ }\href {\doibase 10.1093/mnrasl/slz133} {\bibfield
  {journal} {\bibinfo  {journal} {\mnras}\ }\textbf {\bibinfo {volume} {489}},\
  \bibinfo {pages} {L91--L96} (\bibinfo {year} {2019})}\BibitemShut {NoStop}%
\bibitem [{\citenamefont {{Smartt}}\ \emph {et~al.}(2017)\citenamefont
  {{Smartt}}, \citenamefont {{Chen}}, \citenamefont {{Jerkstrand}},
  \citenamefont {{Coughlin}}, \citenamefont {{Kankare}}, \citenamefont {{Sim}},
  \citenamefont {{Fraser}}, \citenamefont {{Inserra}}, \citenamefont
  {{Maguire}}, \citenamefont {{Chambers}} \emph
  {et~al.}}]{2017Natur.551...75S}%
  \BibitemOpen
  \bibfield  {author} {\bibinfo {author} {\bibfnamefont {S.~J.}\ \bibnamefont
  {{Smartt}}}, \bibinfo {author} {\bibfnamefont {T.~W.}\ \bibnamefont
  {{Chen}}}, \bibinfo {author} {\bibfnamefont {A.}~\bibnamefont
  {{Jerkstrand}}}, \bibinfo {author} {\bibfnamefont {M.}~\bibnamefont
  {{Coughlin}}}, \bibinfo {author} {\bibfnamefont {E.}~\bibnamefont
  {{Kankare}}}, \bibinfo {author} {\bibfnamefont {S.~A.}\ \bibnamefont
  {{Sim}}}, \bibinfo {author} {\bibfnamefont {M.}~\bibnamefont {{Fraser}}},
  \bibinfo {author} {\bibfnamefont {C.}~\bibnamefont {{Inserra}}}, \bibinfo
  {author} {\bibfnamefont {K.}~\bibnamefont {{Maguire}}}, \bibinfo {author}
  {\bibfnamefont {K.~C.}\ \bibnamefont {{Chambers}}},  \emph {et~al.},\
  }\bibfield  {title} {\enquote {\bibinfo {title} {{A kilonova as the
  electromagnetic counterpart to a gravitational-wave source}},}\ }\href
  {\doibase 10.1038/nature24303} {\bibfield  {journal} {\bibinfo  {journal}
  {\nat}\ }\textbf {\bibinfo {volume} {551}},\ \bibinfo {pages} {75--79}
  (\bibinfo {year} {2017})}\BibitemShut {NoStop}%
\bibitem [{\citenamefont {Breschi}\ \emph {et~al.}(2021)\citenamefont
  {Breschi}, \citenamefont {Perego}, \citenamefont {Bernuzzi}, \citenamefont
  {Del Pozzo}, \citenamefont {Nedora}, \citenamefont {Radice},\ and\
  \citenamefont {Vescovi}}]{2021arXiv210101201B}%
  \BibitemOpen
  \bibfield  {author} {\bibinfo {author} {\bibfnamefont {Matteo}\ \bibnamefont
  {Breschi}}, \bibinfo {author} {\bibfnamefont {Albino}\ \bibnamefont
  {Perego}}, \bibinfo {author} {\bibfnamefont {Sebastiano}\ \bibnamefont
  {Bernuzzi}}, \bibinfo {author} {\bibfnamefont {Walter}\ \bibnamefont
  {Del Pozzo}}, \bibinfo {author} {\bibfnamefont {Vsevolod}\ \bibnamefont
  {Nedora}}, \bibinfo {author} {\bibfnamefont {David}\ \bibnamefont {Radice}},
  \ and\ \bibinfo {author} {\bibfnamefont {Diego}\ \bibnamefont {Vescovi}},\
  }\bibfield  {title} {\enquote {\bibinfo {title} {At2017gfo: Bayesian
  inference and model selection of multicomponent kilonovae and constraints on
  the neutron star equation of state},}\ }\href {\doibase
  10.1093/mnras/stab1287} {\bibfield  {journal} {\bibinfo  {journal} {Monthly
  Notices of the Royal Astronomical Society}\ }\textbf {\bibinfo {volume}
  {505}},\ \bibinfo {pages} {1661--1677} (\bibinfo {year} {2021})}\BibitemShut
  {NoStop}%
\bibitem [{\citenamefont {Nicholl}\ \emph {et~al.}(2021)\citenamefont
  {Nicholl}, \citenamefont {Margalit}, \citenamefont {Schmidt}, \citenamefont
  {Smith}, \citenamefont {Ridley},\ and\ \citenamefont {Nuttall}}]{Nicholl21}%
  \BibitemOpen
  \bibfield  {author} {\bibinfo {author} {\bibfnamefont {Matt}\ \bibnamefont
  {Nicholl}}, \bibinfo {author} {\bibfnamefont {Ben}\ \bibnamefont {Margalit}},
  \bibinfo {author} {\bibfnamefont {Patricia}\ \bibnamefont {Schmidt}},
  \bibinfo {author} {\bibfnamefont {Graham~P}\ \bibnamefont {Smith}}, \bibinfo
  {author} {\bibfnamefont {Evan~J}\ \bibnamefont {Ridley}}, \ and\ \bibinfo
  {author} {\bibfnamefont {James}\ \bibnamefont {Nuttall}},\ }\bibfield
  {title} {\enquote {\bibinfo {title} {Tight multimessenger constraints on the
  neutron star equation of state from gw170817 and a forward model for kilonova
  light-curve synthesis},}\ }\href {\doibase 10.1093/mnras/stab1523} {\bibfield
   {journal} {\bibinfo  {journal} {Monthly Notices of the Royal Astronomical
  Society}\ }\textbf {\bibinfo {volume} {505}},\ \bibinfo {pages} {3016--3032}
  (\bibinfo {year} {2021})}\BibitemShut {NoStop}%
\bibitem [{\citenamefont {{Luko{\v{s}}iute}}\ \emph {et~al.}(2022)\citenamefont
  {{Luko{\v{s}}iute}}, \citenamefont {{Raaijmakers}}, \citenamefont {{Doctor}},
  \citenamefont {{Soares-Santos}},\ and\ \citenamefont
  {{Nord}}}]{2022MNRAS.516.1137L}%
  \BibitemOpen
  \bibfield  {author} {\bibinfo {author} {\bibfnamefont {K.}~\bibnamefont
  {{Luko{\v{s}}iute}}}, \bibinfo {author} {\bibfnamefont {G.}~\bibnamefont
  {{Raaijmakers}}}, \bibinfo {author} {\bibfnamefont {Z.}~\bibnamefont
  {{Doctor}}}, \bibinfo {author} {\bibfnamefont {M.}~\bibnamefont
  {{Soares-Santos}}}, \ and\ \bibinfo {author} {\bibfnamefont {B.}~\bibnamefont
  {{Nord}}},\ }\bibfield  {title} {\enquote {\bibinfo {title} {{KilonovaNet:
  Surrogate models of kilonova spectra with conditional variational
  autoencoders}},}\ }\href {\doibase 10.1093/mnras/stac2342} {\bibfield
  {journal} {\bibinfo  {journal} {\mnras}\ }\textbf {\bibinfo {volume} {516}},\
  \bibinfo {pages} {1137--1148} (\bibinfo {year} {2022})},\ \Eprint
  {http://arxiv.org/abs/2204.00285} {arXiv:2204.00285 [astro-ph.IM]}
  \BibitemShut {NoStop}%
\bibitem [{\citenamefont {{Almualla}}\ \emph {et~al.}(2021)\citenamefont
  {{Almualla}}, \citenamefont {{Ning}}, \citenamefont {{Bulla}}, \citenamefont
  {{Dietrich}}, \citenamefont {{Coughlin}},\ and\ \citenamefont
  {{Guessoum}}}]{2021arXiv211215470A}%
  \BibitemOpen
  \bibfield  {author} {\bibinfo {author} {\bibfnamefont {Mouza}\ \bibnamefont
  {{Almualla}}}, \bibinfo {author} {\bibfnamefont {Yuhong}\ \bibnamefont
  {{Ning}}}, \bibinfo {author} {\bibfnamefont {Mattia}\ \bibnamefont
  {{Bulla}}}, \bibinfo {author} {\bibfnamefont {Tim}\ \bibnamefont
  {{Dietrich}}}, \bibinfo {author} {\bibfnamefont {Michael~W.}\ \bibnamefont
  {{Coughlin}}}, \ and\ \bibinfo {author} {\bibfnamefont {Nidhal}\ \bibnamefont
  {{Guessoum}}},\ }\bibfield  {title} {\enquote {\bibinfo {title} {{Using
  Neural Networks to Perform Rapid High-Dimensional Kilonova Parameter
  Inference}},}\ }\href@noop {} {\bibfield  {journal} {\bibinfo  {journal}
  {arXiv e-prints}\ ,\ \bibinfo {eid} {arXiv:2112.15470}} (\bibinfo {year}
  {2021})},\ \Eprint {http://arxiv.org/abs/2112.15470} {arXiv:2112.15470
  [astro-ph.HE]} \BibitemShut {NoStop}%
\bibitem [{\citenamefont {{Kawaguchi}}\ \emph {et~al.}(2020)\citenamefont
  {{Kawaguchi}}, \citenamefont {{Shibata}},\ and\ \citenamefont
  {{Tanaka}}}]{2020ApJ...889..171K}%
  \BibitemOpen
  \bibfield  {author} {\bibinfo {author} {\bibfnamefont {Kyohei}\ \bibnamefont
  {{Kawaguchi}}}, \bibinfo {author} {\bibfnamefont {Masaru}\ \bibnamefont
  {{Shibata}}}, \ and\ \bibinfo {author} {\bibfnamefont {Masaomi}\ \bibnamefont
  {{Tanaka}}},\ }\bibfield  {title} {\enquote {\bibinfo {title} {{Diversity of
  Kilonova Light Curves}},}\ }\href {\doibase 10.3847/1538-4357/ab61f6}
  {\bibfield  {journal} {\bibinfo  {journal} {\apj}\ }\textbf {\bibinfo
  {volume} {889}},\ \bibinfo {eid} {171} (\bibinfo {year} {2020})}\BibitemShut
  {NoStop}%
\bibitem [{\citenamefont {{Radice}}\ \emph {et~al.}(2016)\citenamefont
  {{Radice}}, \citenamefont {{Galeazzi}}, \citenamefont {{Lippuner}},
  \citenamefont {{Roberts}}, \citenamefont {{Ott}},\ and\ \citenamefont
  {{Rezzolla}}}]{2016MNRAS.460.3255R}%
  \BibitemOpen
  \bibfield  {author} {\bibinfo {author} {\bibfnamefont {David}\ \bibnamefont
  {{Radice}}}, \bibinfo {author} {\bibfnamefont {Filippo}\ \bibnamefont
  {{Galeazzi}}}, \bibinfo {author} {\bibfnamefont {Jonas}\ \bibnamefont
  {{Lippuner}}}, \bibinfo {author} {\bibfnamefont {Luke~F.}\ \bibnamefont
  {{Roberts}}}, \bibinfo {author} {\bibfnamefont {Christian~D.}\ \bibnamefont
  {{Ott}}}, \ and\ \bibinfo {author} {\bibfnamefont {Luciano}\ \bibnamefont
  {{Rezzolla}}},\ }\bibfield  {title} {\enquote {\bibinfo {title} {{Dynamical
  mass ejection from binary neutron star mergers}},}\ }\href {\doibase
  10.1093/mnras/stw1227} {\bibfield  {journal} {\bibinfo  {journal} {\mnras}\
  }\textbf {\bibinfo {volume} {460}},\ \bibinfo {pages} {3255--3271} (\bibinfo
  {year} {2016})},\ \Eprint {http://arxiv.org/abs/1601.02426} {arXiv:1601.02426
  [astro-ph.HE]} \BibitemShut {NoStop}%
\bibitem [{\citenamefont {Dietrich}\ \emph {et~al.}(2017)\citenamefont
  {Dietrich}, \citenamefont {Bernuzzi}, \citenamefont {Ujevic},\ and\
  \citenamefont {Tichy}}]{PhysRevD.95.044045}%
  \BibitemOpen
  \bibfield  {author} {\bibinfo {author} {\bibfnamefont {Tim}\ \bibnamefont
  {Dietrich}}, \bibinfo {author} {\bibfnamefont {Sebastiano}\ \bibnamefont
  {Bernuzzi}}, \bibinfo {author} {\bibfnamefont {Maximiliano}\ \bibnamefont
  {Ujevic}}, \ and\ \bibinfo {author} {\bibfnamefont {Wolfgang}\ \bibnamefont
  {Tichy}},\ }\bibfield  {title} {\enquote {\bibinfo {title} {Gravitational
  waves and mass ejecta from binary neutron star mergers: Effect of the stars'
  rotation},}\ }\href {\doibase 10.1103/PhysRevD.95.044045} {\bibfield
  {journal} {\bibinfo  {journal} {Phys. Rev. D}\ }\textbf {\bibinfo {volume}
  {95}},\ \bibinfo {pages} {044045} (\bibinfo {year} {2017})}\BibitemShut
  {NoStop}%
\bibitem [{\citenamefont {{Shibata}}\ and\ \citenamefont
  {{Hotokezaka}}(2019)}]{2019ARNPS..69...41S}%
  \BibitemOpen
  \bibfield  {author} {\bibinfo {author} {\bibfnamefont {Masaru}\ \bibnamefont
  {{Shibata}}}\ and\ \bibinfo {author} {\bibfnamefont {Kenta}\ \bibnamefont
  {{Hotokezaka}}},\ }\bibfield  {title} {\enquote {\bibinfo {title} {{Merger
  and Mass Ejection of Neutron Star Binaries}},}\ }\href {\doibase
  10.1146/annurev-nucl-101918-023625} {\bibfield  {journal} {\bibinfo
  {journal} {Annual Review of Nuclear and Particle Science}\ }\textbf {\bibinfo
  {volume} {69}},\ \bibinfo {pages} {41--64} (\bibinfo {year} {2019})},\
  \Eprint {http://arxiv.org/abs/1908.02350} {arXiv:1908.02350 [astro-ph.HE]}
  \BibitemShut {NoStop}%
\bibitem [{\citenamefont {Troja}\ \emph {et~al.}(2020)\citenamefont {Troja},
  \citenamefont {van Eerten}, \citenamefont {Zhang}, \citenamefont {Ryan},
  \citenamefont {Piro}, \citenamefont {Ricci}, \citenamefont {O’Connor},
  \citenamefont {Wieringa}, \citenamefont {Cenko},\ and\ \citenamefont
  {Sakamoto}}]{Troja_2020}%
  \BibitemOpen
  \bibfield  {author} {\bibinfo {author} {\bibfnamefont {E}~\bibnamefont
  {Troja}}, \bibinfo {author} {\bibfnamefont {H}~\bibnamefont {van Eerten}},
  \bibinfo {author} {\bibfnamefont {B}~\bibnamefont {Zhang}}, \bibinfo {author}
  {\bibfnamefont {G}~\bibnamefont {Ryan}}, \bibinfo {author} {\bibfnamefont
  {L}~\bibnamefont {Piro}}, \bibinfo {author} {\bibfnamefont {R}~\bibnamefont
  {Ricci}}, \bibinfo {author} {\bibfnamefont {B}~\bibnamefont {O’Connor}},
  \bibinfo {author} {\bibfnamefont {M~H}\ \bibnamefont {Wieringa}}, \bibinfo
  {author} {\bibfnamefont {S~B}\ \bibnamefont {Cenko}}, \ and\ \bibinfo
  {author} {\bibfnamefont {T}~\bibnamefont {Sakamoto}},\ }\bibfield  {title}
  {\enquote {\bibinfo {title} {{A thousand days after the merger: Continued
  X-ray emission from GW170817}},}\ }\href {\doibase 10.1093/mnras/staa2626}
  {\bibfield  {journal} {\bibinfo  {journal} {Monthly Notices of the Royal
  Astronomical Society}\ }\textbf {\bibinfo {volume} {498}},\ \bibinfo {pages}
  {5643--5651} (\bibinfo {year} {2020})}\BibitemShut {NoStop}%
\bibitem [{\citenamefont {Evans}\ \emph {et~al.}(2017)\citenamefont {Evans},
  \citenamefont {Cenko}, \citenamefont {Kennea}, \citenamefont {Emery},
  \citenamefont {Kuin}, \citenamefont {Korobkin}, \citenamefont {Wollaeger},
  \citenamefont {Fryer}, \citenamefont {Madsen}, \citenamefont {Harrison} \emph
  {et~al.}}]{Evans_2017}%
  \BibitemOpen
  \bibfield  {author} {\bibinfo {author} {\bibfnamefont {P.~A.}\ \bibnamefont
  {Evans}}, \bibinfo {author} {\bibfnamefont {S.~B.}\ \bibnamefont {Cenko}},
  \bibinfo {author} {\bibfnamefont {J.~A.}\ \bibnamefont {Kennea}}, \bibinfo
  {author} {\bibfnamefont {S.~W.~K.}\ \bibnamefont {Emery}}, \bibinfo {author}
  {\bibfnamefont {N.~P.~M.}\ \bibnamefont {Kuin}}, \bibinfo {author}
  {\bibfnamefont {O.}~\bibnamefont {Korobkin}}, \bibinfo {author}
  {\bibfnamefont {R.~T.}\ \bibnamefont {Wollaeger}}, \bibinfo {author}
  {\bibfnamefont {C.~L.}\ \bibnamefont {Fryer}}, \bibinfo {author}
  {\bibfnamefont {K.~K.}\ \bibnamefont {Madsen}}, \bibinfo {author}
  {\bibfnamefont {F.~A.}\ \bibnamefont {Harrison}},  \emph {et~al.},\
  }\bibfield  {title} {\enquote {\bibinfo {title} {<i>swift</i> and
  <i>nustar</i> observations of gw170817: Detection of a blue kilonova},}\
  }\href {\doibase 10.1126/science.aap9580} {\bibfield  {journal} {\bibinfo
  {journal} {Science}\ }\textbf {\bibinfo {volume} {358}},\ \bibinfo {pages}
  {1565--1570} (\bibinfo {year} {2017})},\ \Eprint
  {http://arxiv.org/abs/https://www.science.org/doi/pdf/10.1126/science.aap9580}
  {https://www.science.org/doi/pdf/10.1126/science.aap9580} \BibitemShut
  {NoStop}%
\bibitem [{\citenamefont {{Ji}}\ \emph {et~al.}(2019)\citenamefont {{Ji}},
  \citenamefont {{Drout}},\ and\ \citenamefont
  {{Hansen}}}]{2019ApJ...882...40J}%
  \BibitemOpen
  \bibfield  {author} {\bibinfo {author} {\bibfnamefont {Alexander~P.}\
  \bibnamefont {{Ji}}}, \bibinfo {author} {\bibfnamefont {Maria~R.}\
  \bibnamefont {{Drout}}}, \ and\ \bibinfo {author} {\bibfnamefont {Terese~T.}\
  \bibnamefont {{Hansen}}},\ }\bibfield  {title} {\enquote {\bibinfo {title}
  {{The Lanthanide Fraction Distribution in Metal-poor Stars: A Test of Neutron
  Star Mergers as the Dominant r-process Site}},}\ }\href {\doibase
  10.3847/1538-4357/ab3291} {\bibfield  {journal} {\bibinfo  {journal} {\apj}\
  }\textbf {\bibinfo {volume} {882}},\ \bibinfo {eid} {40} (\bibinfo {year}
  {2019})},\ \Eprint {http://arxiv.org/abs/1905.01814} {arXiv:1905.01814
  [astro-ph.HE]} \BibitemShut {NoStop}%
\bibitem [{\citenamefont {{Farouqi}}\ \emph {et~al.}(2022)\citenamefont
  {{Farouqi}}, \citenamefont {{Thielemann}}, \citenamefont {{Rosswog}},\ and\
  \citenamefont {{Kratz}}}]{2022A&A...663A..70F}%
  \BibitemOpen
  \bibfield  {author} {\bibinfo {author} {\bibfnamefont {K.}~\bibnamefont
  {{Farouqi}}}, \bibinfo {author} {\bibfnamefont {F.~K.}\ \bibnamefont
  {{Thielemann}}}, \bibinfo {author} {\bibfnamefont {S.}~\bibnamefont
  {{Rosswog}}}, \ and\ \bibinfo {author} {\bibfnamefont {K.~L.}\ \bibnamefont
  {{Kratz}}},\ }\bibfield  {title} {\enquote {\bibinfo {title} {{Correlations
  of r-process elements in very metal-poor stars as clues to their
  nucleosynthesis sites}},}\ }\href {\doibase 10.1051/0004-6361/202141038}
  {\bibfield  {journal} {\bibinfo  {journal} {\aap}\ }\textbf {\bibinfo
  {volume} {663}},\ \bibinfo {eid} {A70} (\bibinfo {year} {2022})},\ \Eprint
  {http://arxiv.org/abs/2107.03486} {arXiv:2107.03486 [astro-ph.SR]}
  \BibitemShut {NoStop}%
\bibitem [{\citenamefont {{Fujibayashi}}\ \emph {et~al.}(2023)\citenamefont
  {{Fujibayashi}}, \citenamefont {{Kiuchi}}, \citenamefont {{Wanajo}},
  \citenamefont {{Kyutoku}}, \citenamefont {{Sekiguchi}},\ and\ \citenamefont
  {{Shibata}}}]{2023ApJ...942...39F}%
  \BibitemOpen
  \bibfield  {author} {\bibinfo {author} {\bibfnamefont {Sho}\ \bibnamefont
  {{Fujibayashi}}}, \bibinfo {author} {\bibfnamefont {Kenta}\ \bibnamefont
  {{Kiuchi}}}, \bibinfo {author} {\bibfnamefont {Shinya}\ \bibnamefont
  {{Wanajo}}}, \bibinfo {author} {\bibfnamefont {Koutarou}\ \bibnamefont
  {{Kyutoku}}}, \bibinfo {author} {\bibfnamefont {Yuichiro}\ \bibnamefont
  {{Sekiguchi}}}, \ and\ \bibinfo {author} {\bibfnamefont {Masaru}\
  \bibnamefont {{Shibata}}},\ }\bibfield  {title} {\enquote {\bibinfo {title}
  {{Comprehensive Study of Mass Ejection and Nucleosynthesis in Binary Neutron
  Star Mergers Leaving Short-lived Massive Neutron Stars}},}\ }\href {\doibase
  10.3847/1538-4357/ac9ce0} {\bibfield  {journal} {\bibinfo  {journal} {\apj}\
  }\textbf {\bibinfo {volume} {942}},\ \bibinfo {eid} {39} (\bibinfo {year}
  {2023})},\ \Eprint {http://arxiv.org/abs/2205.05557} {arXiv:2205.05557
  [astro-ph.HE]} \BibitemShut {NoStop}%
\bibitem [{\citenamefont {{Vassh}}\ \emph {et~al.}(2022)\citenamefont
  {{Vassh}}, \citenamefont {{McLaughlin}}, \citenamefont {{Mumpower}},\ and\
  \citenamefont {{Surman}}}]{2022arXiv220209437V}%
  \BibitemOpen
  \bibfield  {author} {\bibinfo {author} {\bibfnamefont {Nicole}\ \bibnamefont
  {{Vassh}}}, \bibinfo {author} {\bibfnamefont {Gail~C.}\ \bibnamefont
  {{McLaughlin}}}, \bibinfo {author} {\bibfnamefont {Matthew~R.}\ \bibnamefont
  {{Mumpower}}}, \ and\ \bibinfo {author} {\bibfnamefont {Rebecca}\
  \bibnamefont {{Surman}}},\ }\bibfield  {title} {\enquote {\bibinfo {title}
  {{The need for a local nuclear physics feature in the neutron-rich
  rare-earths to explain solar $r$-process abundances}},}\ }\href@noop {}
  {\bibfield  {journal} {\bibinfo  {journal} {arXiv e-prints}\ ,\ \bibinfo
  {eid} {arXiv:2202.09437}} (\bibinfo {year} {2022})},\ \Eprint
  {http://arxiv.org/abs/2202.09437} {arXiv:2202.09437 [nucl-th]} \BibitemShut
  {NoStop}%
\bibitem [{\citenamefont {{Kullmann}}\ \emph {et~al.}(2022)\citenamefont
  {{Kullmann}}, \citenamefont {{Goriely}}, \citenamefont {{Just}},
  \citenamefont {{Ardevol-Pulpillo}}, \citenamefont {{Bauswein}},\ and\
  \citenamefont {{Janka}}}]{2022MNRAS.510.2804K}%
  \BibitemOpen
  \bibfield  {author} {\bibinfo {author} {\bibfnamefont {I.}~\bibnamefont
  {{Kullmann}}}, \bibinfo {author} {\bibfnamefont {S.}~\bibnamefont
  {{Goriely}}}, \bibinfo {author} {\bibfnamefont {O.}~\bibnamefont {{Just}}},
  \bibinfo {author} {\bibfnamefont {R.}~\bibnamefont {{Ardevol-Pulpillo}}},
  \bibinfo {author} {\bibfnamefont {A.}~\bibnamefont {{Bauswein}}}, \ and\
  \bibinfo {author} {\bibfnamefont {H.~T.}\ \bibnamefont {{Janka}}},\
  }\bibfield  {title} {\enquote {\bibinfo {title} {{Dynamical ejecta of neutron
  star mergers with nucleonic weak processes I: nucleosynthesis}},}\ }\href
  {\doibase 10.1093/mnras/stab3393} {\bibfield  {journal} {\bibinfo  {journal}
  {\mnras}\ }\textbf {\bibinfo {volume} {510}},\ \bibinfo {pages} {2804--2819}
  (\bibinfo {year} {2022})},\ \Eprint {http://arxiv.org/abs/2109.02509}
  {arXiv:2109.02509 [astro-ph.HE]} \BibitemShut {NoStop}%
\end{thebibliography}%

\end{document}